\documentclass[journal,comsoc]{IEEEtran}

\usepackage[utf8]{inputenc}
\usepackage{enumerate}
\usepackage{xcolor}
\usepackage[backend=biber,style=ieee,defernumbers=true]{biblatex}

\makeatletter
\let\blx@rerun@biber\relax
\makeatother

\usepackage{balance}
\usepackage{graphicx}
\usepackage{subfigure}
\usepackage{amsmath,amssymb,amsthm}
\usepackage{multirow}
\usepackage{float}
\usepackage{caption}
\usepackage{mathtools}
\usepackage{cases}
\usepackage{censor}
\usepackage{acronym}
\usepackage{placeins}
\usepackage{verbatim}
\usepackage{array}
\usepackage{newtxmath}
\newcolumntype{P}[1]{>{\centering\arraybackslash}p{#1}}
\newcolumntype{M}[1]{>{\centering\arraybackslash}m{#1}}

\newcommand\fig[1]{Figure~\ref{fig:#1}}

\newcommand\subfigwidth{0.95\columnwidth}
\newcommand\subfiglabelwidth{0.97\columnwidth}
\graphicspath{{./figures/}}

\newcommand{\figeps}[3][]{%
 \begin{figure}[tb]
  \begin{center}
   \leavevmode
      \parbox[t]{#1}{%
        \resizebox{#1}{!}{\includegraphics{#2}}
      }
      \vspace{-0.2cm}
   \caption{#3}
   \label{fig:#2}
  \end{center}
 \end{figure}
}

\newcommand{\figepsH}[3][]{%
 \begin{figure}[ht]
  \begin{center}
   \leavevmode
      \parbox[t]{#1}{%
        \resizebox{#1}{!}{\includegraphics{#2}}
      }
      \vspace{-0.2cm}
   \caption{#3}
   \label{fig:#2}
  \end{center}
 \end{figure}
}

\newcommand{\foursubfigeps}[9]{

  \begin{figure}[htb]
    \begin{center}
     \subfigure[#2]{
        \label{fig:#1}
        \parbox[t]{\subfiglabelwidth}{%
            \centering
            \resizebox{\subfigwidth}{!}{\includegraphics{#1}}
        }
     }
     \subfigure[#4]{
        \label{fig:#3}
        \parbox[t]{\subfiglabelwidth}{%
            \centering
            \resizebox{\subfigwidth}{!}{\includegraphics{#3}}
        }
     }
     \subfigure[#6]{
        \label{fig:#5}
        \parbox[t]{\subfiglabelwidth}{%
            \centering
            \resizebox{\subfigwidth}{!}{\includegraphics{#5}}
        }
     }
     \subfigure[#8]{
        \label{fig:#7}
        \parbox[t]{\subfiglabelwidth}{%
            \centering
            \resizebox{\subfigwidth}{!}{\includegraphics{#7}}
        }
     }
    \end{center}
    \caption{#9}
  \end{figure}
}

\newcommand{\SL}[1]{\textcolor{black}{#1}}
\newcommand{\LO}[1]{\textcolor{black}{#1}}

\usepackage{pifont}
\newcommand{\cmark}{\ding{51}}%
\newcommand{\xmark}{-}%

\title{\huge A Survey of Scheduling Algorithms for the Time-Aware Shaper in Time-Sensitive Networking (TSN)}

\author{
  \IEEEauthorblockN{Thomas Stüber, Lukas Osswald, Steffen Lindner, Michael Menth}\\
  \IEEEauthorblockA{
  Chair of Communication Networks, University of Tuebingen, Germany\\
  \{thomas.stueber, lukas.osswald, steffen.lindner, menth\}@uni-tuebingen.de
  }
}

\begin{document}

\maketitle

\begin{abstract}
Time-Sensitive Networking (TSN) is an enhancement of Ethernet which provides various mechanisms for real-time communication. Time-triggered (TT) traffic represents periodic data streams with strict real-time requirements. Amongst others, TSN supports scheduled transmission of TT streams, i.e., the transmission of their frames by end stations is coordinated in such a way that none or very little queuing delay occurs in intermediate nodes. TSN supports multiple priority queues per egress port. The TAS uses so-called gates to explicitly allow and block these queues for transmission on a short periodic timescale. The TAS is utilized to protect scheduled traffic from other traffic to minimize its queuing delay. In this work, we consider scheduling in TSN which comprises the computation of periodic transmission instants at end stations and the periodic opening and closing of queue gates.

In this paper, we first give a brief overview of TSN features and standards. We state the TSN scheduling problem and explain common extensions which also include optimization problems. We review scheduling and optimization methods that have been used in this context. Then, the contribution of currently available research work is surveyed. We extract and compile optimization objectives, solved problem instances, and evaluation results. Research domains are identified, and specific contributions are analyzed. Finally, we discuss potential research directions and open problems.
\end{abstract}

\section{Introduction}

\begin{table*}[ht!]
    \centering
    \begin{tabular}{|M{2.5cm}|c|c|M{6cm}|M{2.7cm}|}
    \hline
        Paper & Date & References & Focus & \# Common referenced papers with this survey \\\hline\hline
        Nasrallah \textit{et al.} \cite{NaTh19} & 2018 & 407 & Overview of TSN, DetNet, and 5G standards & 14 \\\hline
        Minaeva \textit{et al.} \cite{MiHa21} & 2021 & 126 & Scheduling in periodic systems & 6 \\\hline
        Seol \textit{et al.} \cite{SeHy21} & 2021 & 207 & Broad overview of TSN & 29 \\\hline
        Deng \textit{et al.} \cite{DeXe22} & 2022 & 128 & Broad overview of all topics related to AVB and TSN & 17  \\\hline
        This paper & 2022 & 139 & Traffic scheduling in TSN with the TAS & - \\\hline
    \end{tabular}
    \caption{Surveys covering related topics to this paper.}
    \label{tab:surveys}
\end{table*}

\LO{
Modern applications, e.g., Industry 4.0 factory automation and motion control, demand highly deterministic network service.
Exceeding latency and jitter bounds can result in immediate degradation of manufacturing quality or endanger health of machinery and operators.
}
Some of these applications have to exchange data streams with precise timing to keep application-specific deadlines. 
Time-Sensitive Networking (TSN) is an emerging technology which enhances Ethernet networks \SL{with} real-time properties. 
In TSN, talkers send uni- or multicast streams, called streams, to traffic sinks, called listeners. 
The network admits streams and guarantees quality of service (QoS). 
Time-triggered (TT) traffic constitutes periodic data streams with real-time requirements such as bounded latency or jitter.
The transmission times of TT streams at their respective talkers must
be scheduled such that excessive queuing in the network is avoided
and their requirements are met. Although TT traffic has high
priority, it is delayed by low-priority frames in transmission
blocking links for short time. To ensure that links are not
occupied by low-priority traffic when needed for TT traffic, the
standard IEEE Std 802.1Qbv \cite{802.1Qbv} introduces an enhancement for scheduled traffic. 
The Time-Aware Shaper (TAS) can be implemented with this enhancement.
It defines periodic time slices during which queues may send traffic to an output port and delays the respective traffic. In TSN, the TAS is used to protect TT traffic from other
traffic classes. Therefore, TSN requires that appropriate TAS time
slices are scheduled for output queues on all switches, in addition
to the transmission times of all TT streams at their talkers. This
combination guarantees very short delays for TT streams in TSN.

Standardization does not yet cover methods for computing such schedules.
However, the topic has been examined by many publications. 
These research works use different methods for schedule synthesis, evaluation, and objectives for optimization.
We survey the currently available literature for TSN schedule computation. 
The paper focuses on publications published until March of 2023 about TSN schedule planning with the TAS.
Works about stream scheduling related to other technologies than TSN or for other traffic shapers in TSN than the TAS are not covered in this survey.

\subsection{Related Surveys}
To the best of our knowledge, no other review covers scheduling algorithms for TSN as its main topic. 
In fact, there is no survey about scheduling for TT streams for Ethernet networks, regardless of the used standard.
However, there are surveys which intersect with the content of this work.
Table \ref{tab:surveys} compiles the focus and the relationship of these surveys to this paper. 

Nasrallah \textit{et al.} \cite{NaTh19} survey standards for low-latency communication.
Besides DetNet and 5G, they also give a tutorial on the TSN standards.
They reference a small number of papers related to traffic scheduling in TSN.
However, they do not elaborate on the content of these papers and the vast majority of the literature about traffic scheduling is not even referenced. 

Minaeva \textit{et al.} \cite{MiHa21} give a literature summary for scheduling time-triggered real-time systems.
They highlight research works from 1968 to 2020.
As opposed to this work, they not only consider scheduling of streams in networks, but all systems with periodic schedules.
Seminal works for TNS scheduling algorithms in the literature are mentioned, e.g., \cite{St10} and \cite{ScDa17}.
Out of 126 references, only 6 of them intersect with this work.

Deng \textit{et al.} \cite{DeXe22} review a wide range of topics about AVB and TSN from the literature of 2007 -- 2021. 
Besides scheduling approaches, they also give an overview of reliability and security modeling, and delay analysis in the mentioned areas.
As a wider range of topics is covered, only a small part of the survey is concerned with scheduling.
This part is rather superficial, summarizes only 15 works, and is mostly in tabular form.
The algorithms and evaluation results presented in these 15 works are not discussed.
Summing up, only 17 out of the 128 discussed works intersect with this survey.
Seol \textit{et al.} \cite{SeHy21} review TSN as a whole.

The authors cover publications of the years 2014 -- 2020.
An overview of active research directions is given, including computing routings and schedules in TSN.
Not only the literature about scheduling for the TAS is summarized, but also work concerned with other queuing mechanisms, hardware, and simulation frameworks.
Therefore, only a small fraction of the literature about scheduling for TAS-based queuing in TSN is surveyed.
They cover 207 research works, of which 29 are included in this work.
These works are merely mentioned in their review, and the authors do not elaborate on their content. 

\subsection{Contribution}
In contrast to the mentioned surveys of Table \ref{tab:surveys}, we focus on papers about scheduling algorithms and related topics which use the TAS.
This survey claims the following contributions:
\begin{itemize}
    \item We give a tutorial on TSN basics.
    \item We define the TSN scheduling problem for TAS and modifications to it. 
    Additionally, we introduce common solution methods used in the literature
    \item We survey currently available TSN literature about scheduling for the TAS.
    \item We identify research directions, categorize the available literature, and highlight contributions to these topics.
    \item We compare the available algorithms and the presented evaluations to derive open research questions in this area.
\end{itemize}

\subsection{Survey Structure}
This paper is structured as follows. 
In Section \ref{sec:background} we present a brief introduction to TSN with a special focus on the TAS.
Then, we formally define the scheduling problem in TSN and give a tutorial to common solutions methods from literature in Section \ref{sec:formaldef}. 
Section \ref{sec:literature_overview} gives an overview of the state-of-the-art of TSN scheduling and categorizes the presented literature. 
Section \ref{sec:comparison} compares the presented research work with regard to modelling assumptions, optimization objective, problem instances and scalability.
Furthermore, we present the publication history of the surveyed literature in Section \ref{sec:publication}.
We discuss issues and open research questions in Sections \ref{sec:discussion}. 
Finally, we conclude the paper in Section \ref{sec:conclusion}.

\vspace{3cm}

\subsection{List of Frequently Used Acronyms}
The following acronyms are used in this paper.
\begin{tabbing}
\hspace*{4em}\= \hspace*{2em} \= \kill 
\textbf{ASAP}  \> As Soon As Possible \\
\textbf{AVB}  \> Audio Video Bridging \\
\textbf{BE}  \> Best Effort \\
\textbf{CBS}  \>Credit-Based Shaper  \\
\textbf{CP}  \>Constraint Programming \\
\textbf{CQF}  \> Cyclic Queuing and Forwarding \\
\textbf{GA}  \> Genetic Algorithm \\
\textbf{GCL}  \> Gate Control List \\
\textbf{FIFO}  \> First-In-First-Out \\
\textbf{FRER}  \> Frame Replication and Elimination for Reliability \\
\textbf{gPTP}  \> generalized Precision Time Protocol \\
\textbf{GRASP}  \> Greedy Randomized Adaptive Search Procedure \\
\textbf{ILP}  \>Integer Linear Programming \\
\textbf{OMT}  \>Optimization Modulo Theories \\
\textbf{PBO}  \>Pseudo-Boolean Optimization \\
\textbf{PSFP}  \>Per-Stream Filtering and Policing \\
\textbf{QoS}  \> Quality of Service \\
\textbf{SMT}  \>Satisfiability Modulo Theories  \\
\textbf{SRP}  \>Stream Reservation Protocol \\
\textbf{TAS}  \> Time-Aware Shaper \\
\textbf{TSN}  \>Time-Sensitive Networking \\
\textbf{TT}  \> Time-Triggered \\
\textbf{VLAN}  \>Virtual LAN 
\end{tabbing}

\section{Foundations of TSN}
\label{sec:background}
TSN is a set of standards for deterministic data transmission with real-time requirements over Ethernet networks.
In this section, we present a short tutorial about TSN.
First, we present AVB based on which TSN was developed.
Then, we introduce TSN with a special focus on scheduling and the TAS.

\subsection{Audio Video Bridging (AVB)}
Historically, multimedia equipment was interconnected with half-duplex point-to-point links for data transmission.
These links were often dedicated to a single purpose, i.e., the transmission of one specific data stream.
This results in a large number of links which is expensive, hard to maintain, and error prone.
Switched computer networks solved these problems.
The most widely adopted technology for switched local area networks today is Ethernet.
However, professional audio and video applications need bounded latencies and jitter, i.e., real-time guarantees for data streams.
Switching in Ethernet networks was not designed for real-time transmissions.
Therefore, the Audio Video Bridging (AVB) task group of the IEEE was founded to develop a standard to meet the requirements of multimedia applications in switched Ethernet networks.  

AVB is organized in standards for time synchronization, admission control, and traffic shaping.

\subsubsection{Time Synchronisation}
Network devices need a common understanding of time to ensure that all end stations in a network are able to coordinate their actions.
Every AVB-capable device is equipped with a clock.
The standard IEEE 802.1AS \cite{802.1AS} defines a protocol to synchronize the clocks of all devices in an AVB network.
This protocol is based on the \textit{Precise Time Protocol} (PTP) introduced in IEEE 1588 \cite{IEEE1588} and is denoted as \textit{generalized Precise Time Protocol} (gPTP).
The gPTP defines an algorithm to select a so-called \textit{Grandmaster} among the participating nodes of the protocol.
The internal clock of the Grandmaster is used as reference clock.
All other devices synchronize their clocks to the clock of the Grandmaster with time information sent from the Grandmaster.
Intermediate nodes adjust the received time information to compensate propagation delays, processing delays, and different clock speeds before retransmitting them.
The gPTP allows sub-microsecond precision for devices with at most seven hops distance to each other.
This is needed for applications running on different end stations to synchronize their actions.

\subsubsection{Admission Control}
The Stream Reservation Protocol (SRP) introduced in IEEE 802.1Qat \cite{802.1Qat} allows senders of periodic data streams, denoted as \textit{talkers}, to reserve bandwidth in a multi-hop Ethernet network.
A talker which wants to send data advertises a new data stream to its connected bridge.
This advertisement contains information about bandwidth and real-time requirements, the periodicity of the stream, and the destination MAC address.
The destination may be a multicast group.
The bridge forwards the advertisement if the requested resources are available.
Worst-case latencies are calculated at every bridge.
When the request reaches the destination of a data stream, denoted as \textit{listener}, the listener acknowledges that it is ready, and the bandwidth is reserved along the path.

\subsubsection{Traffic Shaping}
Traffic shaping is the generic term for techniques that distribute packet transmissions in time.
The AVB working group defines the so-called \textit{Credit-Based Shaper} (CBS) in IEEE 802.1Qav \cite{802.1Qav}.
It can be leveraged to smooth out bursts such that receiving devices are not overwhelmed.
This reduces buffering and congestion in the network.
The CBS is a leaky bucket traffic shaper with at least two FIFO queues for two traffic classes.
These classes are denoted as class A and class B.
Both queues have a credit measured in bit.
Dispatching and transmitting a frame from a queue is only allowed if the credit of the respective queue is non-negative. 
Credit increases linearly during times no frame is transmitted and decreases linearly during transmissions.
Latency bounds for streams can be guaranteed by using a special configuration for the CBS defined in IEEE 802.1BA \cite{802.1BA}.
These are specific to the requirements of the AVB domain, guaranteeing 2\,ms and 50\,ms for class A and B traffic in networks with at most 7 hops.
However, the average delay of a frame increases to up to $250\,\mu s$ per hop in the worst case when the CBS is used.

\subsection{Time-Sensitive Networking (TSN)}
Ethernet networks are used in a wide range of industrial use cases as Ethernet is cheap and easy to implement.
However, use cases such as industrial automation, in-vehicle communication or avionics have hard real-time requirements and need reliability. 
Data streams not meeting their deadlines may not only be worthless but impose safety risks.
The latency guarantees and average delays offered by AVB fail to comply with the requirements of such use cases.

Time-Sensitive Networking (TSN) is a set of standards enhancing AVB for deterministic and reliable transmission of data over switched Ethernet networks.
TSN is currently developed in the IEEE 802.1 TSN task group and adds new mechanisms for scheduling, traffic shaping, path selection, stream reservation, filtering and policing, and fault-tolerance.
Most of the standards are enhancements of IEEE 802.1Q \cite{802.1Q} which defines bridges and Virtual LANs (VLANs). 
We give a brief tutorial on the standards and mechanisms relevant for the scope of this survey, i.e., traffic scheduling in TSN with the TAS.

\subsubsection{Similarities to AVB}
Similar to AVB, every device in TSN is equipped with a clock.
TSN also uses the gPTP defined in IEEE 802.1AS \cite{802.1AS} to synchronize clocks of all network devices.
The CBS and an enhancement of the SRP are also part of TSN.

\subsubsection{Path Selection}
TSN introduces a new mechanism for path selection in IEEE 802.1Qca \cite{802.1Qca}.
In contrast to traditional Ethernet networks, it is not necessary to use Spanning Tree Protocols or Shortest Path Bridging.
Paths can be computed by an arbitrary algorithm and are only limited to be trees.
Thus, frames can be forwarded on an arbitrary path.
Forwarding information of these so-called Explicit Trees are distributed with the Intermediate System to Intermediate System (IS-IS) protocol and stored in bridges.
The Explicit Tree for the forwarding of a frame is determined by the MAC address of the root bridge of the Explicit Tree and the VLAN ID in the frame's header.

\subsubsection{Priorities}
Every egress port of a TSN bridge is equipped with up to eight egress queues.
These queues are First-In-First-Out (FIFO) queues.
They correspond to the eight VLAN priorities defined in IEEE 802.1Q \cite{802.1Q}.
The VLAN tag in the header of an Ethernet frame determines the egress queue in which the frame waits for transmission.
Every queue is equipped with a so-called Transmission Selection Algorithm (TSA).
The TSA signals whether a frame is ready for transmission to a transmission selection mechanism.
This mechanism selects the next queue from which a frame is dispatched and sent.
TSN uses strict priority as transmission selection, i.e., the next frame is dispatched from the highest priority queue which signals a frame is ready for transmission.

\subsubsection{Frame Preemption}
High-priority traffic can be delayed due to conflicts with lower-priority traffic.
IEEE 802.1Qbu \cite{802.1Qbu} describes a mechanism for frame preemption in TSN which reduces such delays.
Traffic is divided into preemptable frames and so-called express frames.
The transmission of preemptable frames is paused and finished later if an express frame is ready for transmission.
Consequently, a preempted frame is divided into fragments which are reassembled by the receiving node.
The minimum size of a frame fragment is defined to be 64 byte.
However, every fragment of a frame except for the last one has a trailer containing a 4 byte check sequence for error detection. 
Therefore, a frame can only be preempted after at least 60 byte were transmitted and the last 63 byte of a frame cannot be preempted.

\subsubsection{Frame Replication and Elimination for Reliability (FRER)}
Bridging in classical Ethernet networks assumes that no frames are duplicated and therefore no duplicates must be filtered. 
However, safety critical applications may require protection against frame loss and permanent links failures.
IEEE 802.1CB \cite{802.1CB} introduces a mechanism which allows to sent multiple copies of the same frame, possibly over disjoint paths, and to eliminate duplicates.
Thus, only a single copy of the same frame is forwarded or delivered to a higher layer on an end station.

\subsubsection{Traffic Scheduling}
Time-triggered (TT) traffic, also denoted as \textit{scheduled traffic}, consists of periodic data streams with hard real-time requirements such as bounded latency and jitter.
The properties of TT streams such as period, maximum frame size, frames per period, as well as the range of possible transmission offsets from their respective talkers, are known in advance.
The transmission times of these streams at their respective talkers can be controlled and must be coordinated to ensure that all streams meet their real-time requirements.
The computation of their periodic transmission times is denoted as \textit{traffic scheduling}.

\paragraph{Traffic Shaping}
TSN introduces new traffic shapers in addition to the CBS.
An example for another shaper which can be used instead of the CBS is Cyclic Queuing and Forwarding (CQF) defined in IEEE 802.1Qch \cite{802.1Qch}.
Time is divided into slots of predefined length.
The length of a slot is denoted as cycle time.
Bridges buffer all frames received during a slot and transmit them in the subsequent slot.
Thus, stream latencies can easily be calculated from the cycle time and the number of hops. 

IEEE 802.1Qbv \cite{802.1Qbv} defines an enhancement for scheduled traffic.
It can be leveraged to implement the Time-Aware Shaper (TAS).
The TAS allows protecting TT traffic from other traffic such as AVB traffic or best-effort (BE) traffic.
Additionally, the transit of TT streams through a network can be scheduled.
Every egress queue has a so-called \textit{transmission gate} or simply \textit{gate}.
Gates are either open or closed.
Frames can only be dispatched and sent from an egress queue if the respective gate is open.
The closing and opening of a gate is controlled by a so-called Gate Control List (GCL).
A GCL entry consists of a time interval $[T_i, T_{i+1}]$ and a bit-vector.
The bit-vector indicates which gates are opened or closed during the time interval $[T_i, T_{i+1}]$.
Therefore, a GCL entries defines a time slice exclusively available to traffic with a priority corresponding to an open queue. 
These GCLs are executed periodically for an indefinite number of times.
The computation of GCLs and appropriate cycle times, i.e., periods of these GCLs, is denoted as \textit{scheduling} or \textit{GCL synthesis}.
The number of available GCL entries in an egress port is limited and depends on the used bridge.
Figure \ref{fig:switch_model} depicts the architecture of a typical TSN bridge port according to IEEE 802.1Q \cite{802.1Q}, including the components of the TAS.
\figeps[\columnwidth]{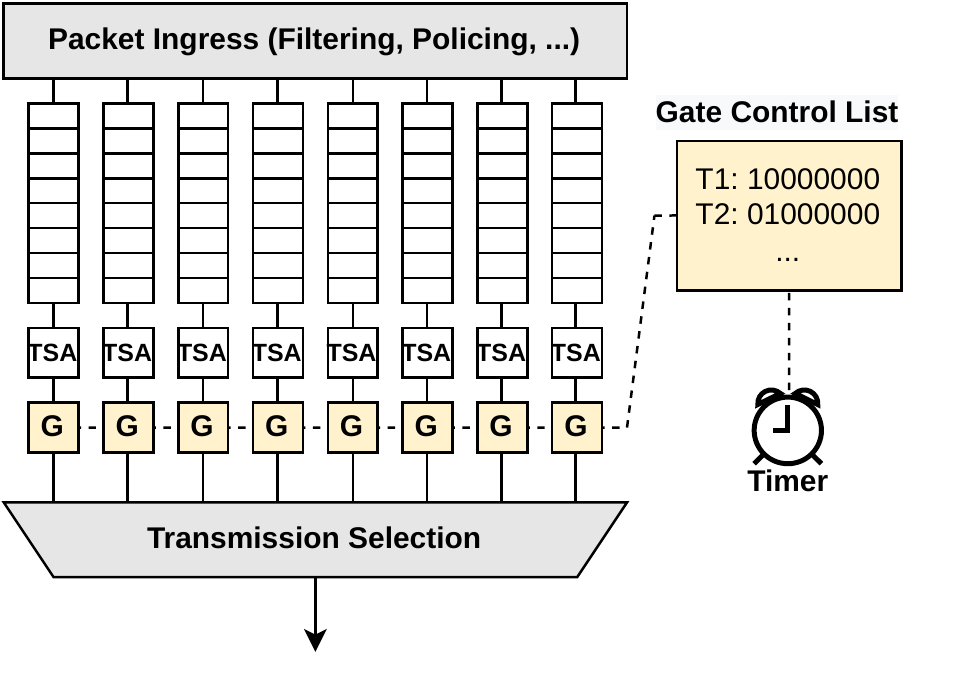}{Components of an egress port implementing the TAS. Every egress ports has eight egress queues guarded by a transmission gate (G). The GCL controls the timed opening and closing of these gates.}
The TAS can be used to protect traffic by scheduling the GCLs accordingly.

\paragraph{Gate Closings and Guard Bands}
If the transmission of a frame is not finished until the end of the time slice the transmission started, a frame in the next time slice may be forced to wait until transmission finishes.
Thus, it would be possible that a frame of a TT stream must wait because of a frame of BE traffic.
This problem is avoided in TSN.
Bridges detect automatically whether a frame transmission would conflict with a gate closing and hold conflicting frames back in this case.
A guard band is a time interval with the length of the transmission of a maximum sized standard Ethernet frame.
The duration of a guard band at the end of a time slice may not be available for transmissions to comply with closed gates.
However, we emphasise that guard bands in TSN are implicit, i.e., they must not be configured explicitly.
Transmissions may even start during a guard band if the transmission finishes before the next gate closing.
\figeps[\columnwidth]{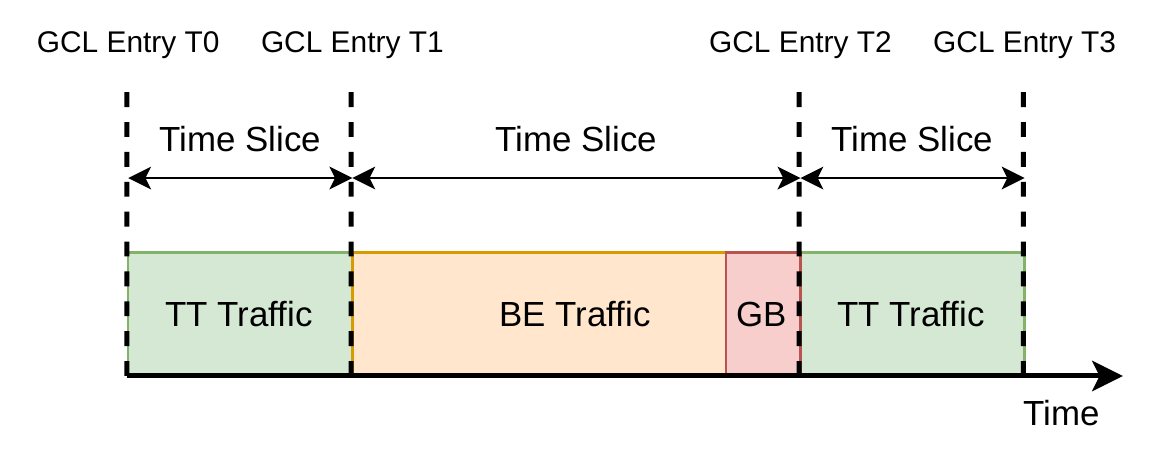}{Time slices, GCL entries, and guard bands. The duration of a guard band may be not available for BE traffic as a frame can only be sent if transmission finishes before the respective gate is closed.}
\fig{guardband.pdf} depicts a guard band which restricts the transmission of BE traffic before the respective gate is closed.
If frame preemption is used, the maximum size of a frame that cannot be preempted is 123 byte.
This is due to the minimum size of a frame fragment, i.e., 60 byte of the frame and an additional 4 byte check sequence.
A frame with 123 byte cannot be preempted until the first 60 byte are transmitted as the resulting first fragment would be too small otherwise.
However, the last 63 byte also cannot be preempted as the resulting last fragment would be too small.
Therefore, guard bands can be reduced to the length of a transmission of 123 byte if frame preemption is used.

\paragraph{Scheduler vs. Traffic Scheduling}
The term \textit{scheduler} is sometimes used as a synonym for \textit{traffic shaper}.
For instance, the CBS and the TAS are schedulers in this terminology.
Unfortunately, the term \textit{scheduler} has also another meaning in the context of this survey.
Many research works denote algorithms to plan GCL entries and frame transmissions in time with the TAS as schedulers.
To avoid confusions, we will only use the second meaning in the remainder of this paper, i.e., a scheduler is a scheduling algorithm for the TAS.
There are research works that use the first meaning in their title or abstract or cover scheduling algorithms for other traffic shapers in TSN, e.g., Cyclic Queuing and Forwarding (CQF). 
Thus, these works give the impression that they are in the scope of this survey, e.g., \cite{PaBe22}, \cite{KiMi20}, \cite{XuSh21}, \cite{ZhXu22}, \cite{CaLi21}, \cite{ZhWu22}, \cite{ZhXu22b}, \cite{WeYa22}, and \cite{WeYa23}.
However, we remark that this survey only covers research works about scheduling algorithms and the scheduling problem for the TAS.
Therefore, we do not discuss works which propose new shapers, i.e., new \textit{schedulers}, or analyse other shapers than the TAS.

\par \medskip

The challenge in planning a TSN network is to compute the schedules that coordinate the transmission times for all streams at their respective talkers and the GCLs of bridges such that the requested real-time requirements of all TT streams are met. 
This problem is formally defined in the next section.

\section{The TSN Scheduling Problem}
\label{sec:formaldef}
First, we introduce a common network model, relevant properties of TT streams, and the definition of schedules.
Second, we state constraints for valid schedules.
Then, we discuss scheduling and optimization in the context of TSN and the computational complexity of these problems.
Furthermore, we present common problem extensions solved in the literature.
Finally, we give an introduction to common solution techniques that have been applied to the scheduling problem.

\subsection{Nomenclature}
A node of a TSN network is either an end station or a TSN-capable bridge. 
End stations are sources and destinations of data streams.
Bridges switch frames based on their header. 
Links are full-duplex Ethernet connections between an end station and a bridge or between two bridges.
TSN bridges are inevitably subject to multiple delays. 
These delays must be considered to ensure deterministic transmissions according to a schedule.
The processing delay of a bridge is the time between a frame arrives at an ingress port, and it is put in an egress queue.
The transmission rate of an egress port is the rate at which data can be transmitted over a link.
The propagation delay of a link is the time needed for electrical signals to traverse the link.
The queuing delay of a frame is the time the frame waits in an egress queue for transmission.
Ethernet uses a preamble before of a frame transmission to signal a new transmission starts, and an inter-frame gap between two frame transmissions to ensure that the receiver can process a new frame.
The maximum size of a frame in TSN is 1542$\,B$, including inter-frame gap and preamble.
A TT stream is a periodically repeated data stream with real-time requirements. 
Every stream has a talker as source and possibly multiple Listeners as destinations.
The earliest and latest transmission offsets describe the time range during which a talker can start transmission relative to the start of a period.
The deadline of a stream is the time at which all frames of the stream must have arrived at all destinations, also relative to the start of a period.
The entire payload of a stream must be delivered before the deadline.
The payload of a stream may be sent with multiple frames.



The hyperperiod $H$ of a set of streams $\mathcal{S}$ is the least common multiple of the periods of the streams.
Let $s \in \mathcal{S}$ be a stream with period $p_s$.
A schedule for all streams in $\mathcal{S}$ contains $\frac{p_s}{H}$ consecutive replications of $s$, each having the hyperperiod as period.
Scheduling algorithms typically consider transmission times, earliest and latest transmission offsets, and deadlines relative to the beginning of the hyperperiod.
\fig{hyperperiod.pdf} depicts an example with two streams A and B.
The period of stream A is three times the period of stream B.
A schedule for both streams thus contains only one period of stream A and three periods of stream B.
A schedule for a set of TT streams in a TSN network consists of the transmission offsets of all streams at their respective talkers, and GCL configurations for all bridges.
Transmission offsets of frames at bridges along their path follow implicitly.
Schedules must be periodic, i.e., repeatable an indefinite number of times.
The hyperperiod of a set of streams is the period of schedules for these streams.

\figeps[\columnwidth]{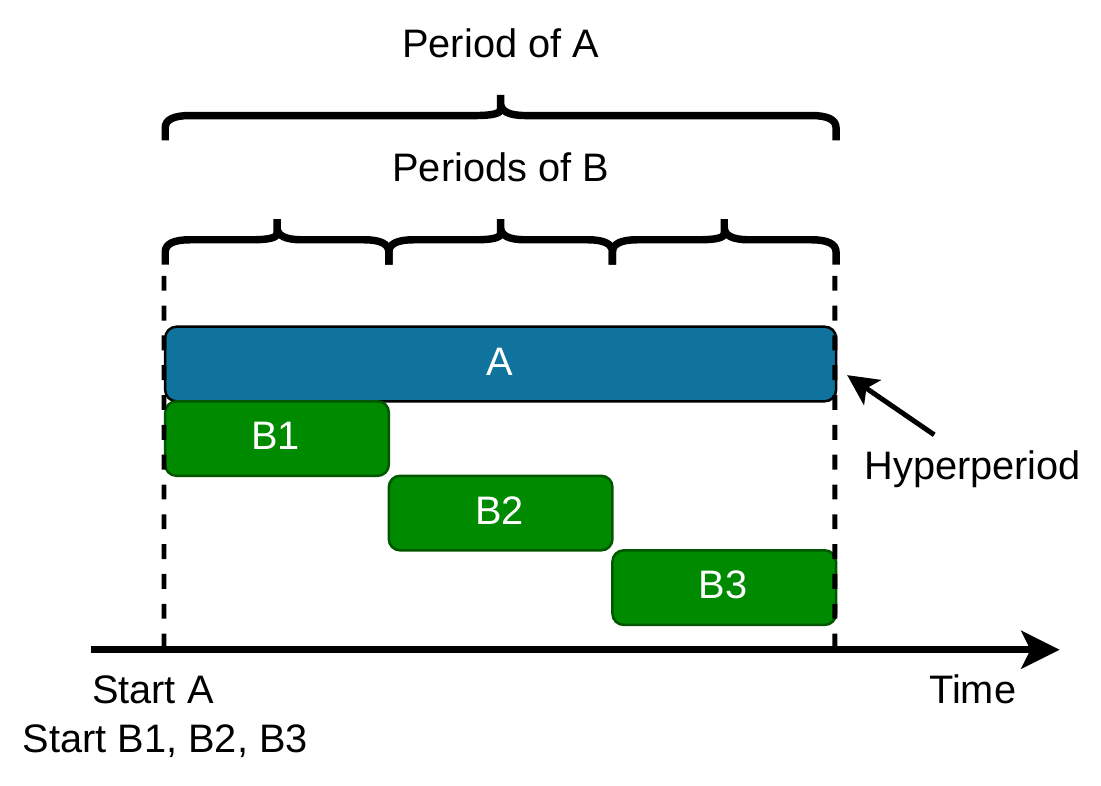}{The period of stream A is three times the period of stream B. For modelling purposes, the hyperperiod is introduced, i.e., all streams are assumed to have that larger period. To cover the full duration of the hyperperiod, B is modelled by three consecutive copies B1, B2, and B3.}

\subsection{Scheduling Constraints}
\label{sec:constraints}
Given a problem instance for the TSN scheduling problem, i.e., a set of TT streams and a network topology.
Every schedule which complies with the real-time requirements of all TT streams is considered a valid solution of the TSN scheduling problem.
Such schedules are denoted as valid schedules in the TSN scheduling literature.
The following constraints restrict the set of all possible schedules to the set of valid schedules.

\subsubsection{Bridge Design} TSN bridges are currently assumed to be store-and-forward bridges. 
Frames cannot be forwarded by a bridge before they have arrived at the egress queue. 
The duration of a transmission depends on the transmission rate of the sending egress port, the size of the frame, and the propagation delay of the used link. 
The processing delay of the bridge must also be considered. 

\subsubsection{Exclusive Link Usage} No two frames can be in transmission over a link in the same direction at the same time.  

\subsubsection{Deadlines} A stream meets its deadline when all of its frames arrive at the stream's destination before its deadline. For multicast streams, this holds for all destinations.

\subsubsection{Routing} The frames of a stream follow some routing. Every instance of the same stream follows the same routing.

\subsubsection{Frame Order} There is no reordering of frames of the same stream. 
Frames that are sent earlier arrive earlier than frames sent later. 
This holds hop-by-hop and end-to-end. 
The source node of a stream sends frames of a stream in-order. 
There are no duplicates, i.e., a frame cannot be replicated by a bridge.

\subsubsection{FIFO Queues} The order of frame arrivals at an egress queue must match the order at which frames are sent. 

\subsubsection{Queue Size} Frames of scheduled traffic must not be dropped for any reason. 

\subsubsection{Gate Control} If a frame waits in an egress queue, it can only be sent when the respective gate is open. 
The gate must stay open until transmission finishes.

\subsubsection{Transmission Selection} If multiple gates of an egress port are open and frames in the respective queues are waiting for transmission, the queue with the highest priority is the next queue to dispatch a frame.

\subsubsection{Additional Features} Various modifications of the problem are presented in the literature. 
Additional constraints may be needed to model these problems. 
For example, multiple queues can be reserved for TT traffic per egress port. 
Queue assignment of streams must be modeled in this case. 
We discuss these problem modifications in Section \ref{sec:problem_modifications}.

\subsection{Finding a Schedule vs. Optimization}
There may be multiple schedules for a given problem instance.
In fact, most problem instances have a large number of schedules as possible solutions.
So far, the definition of the scheduling problem does not differentiate between these solutions.
A common way to compare solutions is to introduce an objective function.
Such a function maps solutions to real numbers.
The solution to an optimization problem is the schedule which minimizes or maximizes the objective function, i.e., has a smaller or larger objective value than any other schedule.
Examples for objectives are minimizing end-to-end delays or jitter of TT streams.
Another possible objective is minimizing the flowspan, i.e., the duration such that all frames have arrived at their respective destinations.

\subsection{Computational Complexity}
The problem of deciding whether there is a valid schedule for a set of TT streams in a TSN network is known to be NP-complete \cite{St10} in general as Bin Packing can be reduced to it. 
This even holds without queuing \cite{DuNa16}. 
NP is a class of decision problems, i.e., contains only problems which can be answered by either \textit{yes} or \textit{no}.
Finding a schedule or finding an optimal schedule are not decision problems.
Therefore, they are not contained in NP.
However, they are computationally at least as hard as the question whether there is a schedule.

\subsection{Problem Extensions and Restrictions}
\label{sec:problem_modifications}
The definition of the basic problem in Section \ref{sec:constraints} only describes the common properties of the problems in the literature reviewed in this survey.
Much research work focuses on special cases or problem extensions with additional constraints.
This section introduces these problem variations in a general way such that they are clear in the remainder of this survey.

\subsubsection{Joint Routing} 
The definition of Section \ref{sec:constraints} assumes that the routing of every stream is a predefined part of the input and fixed. 
Much research work is dedicated to a variation of the scheduling problem with joint routing, which relaxes this assumption.
In contrast to the basic problem, the routing of streams is variable and computed simultaneously with the schedule.
This gives the scheduling algorithm more flexibility, as streams can be routed to omit heavily loaded links and thus conflicting scheduling constraints.
A common approach is that the algorithm gets a set of possible paths as input for every stream, and it selects one per stream as the stream's routing.
Other algorithms select arbitrary paths.
Both approaches are possible due to IEEE 802.1Qca \cite{802.1Qca} as the standard allows arbitrary paths to be configured for every stream.

\subsubsection{Reliability}
Research work dedicated to joint routing and scheduling can take reliability considerations into account.
Such works define a model of possible faults and their probabilities.
Scheduling algorithms can compute schedules which meet the real-time requirements of all streams with high probability for a given fault model.
These schedules are denoted as robust schedules relative to a given fault model.
For instance, scheduling approaches can compute schedules which are robust against single link failures.
This can be achieved by introducing redundant streams with the same payload and routing them through disjoint paths.

\subsubsection{GCL Synthesis}
GCLs for all egress ports must be constructed.
One possible approach is to open the gates for scheduled traffic at the beginning of a hyperperiod and never closing them.
However, this approach comes with the drawback that no other queue can send.
This may be necessary to protect other TT streams with tighter bounds by avoiding congestion in the queues for TT traffic.

Another common approach is to use a postprocessing scheme after scheduling transmission offsets.
GCLs are constructed such that the gate of a queue is opened when a transmission from this queue should start according to the schedule.
The respective gate is closed when the transmission is finished according to the schedule.
This approach allows a scheduler to use gates to delay frames.
However, the number of available GCL entries is limited in real hardware bridges.
Therefore, scheduling transmission offsets and synthesizing GCLs can also be considered in a joint scheduling algorithm instead of a postprocessing.

\subsubsection{Queuing}
Queuing can cause serious problems for schedules of streams with real-time requirements. 
Frames can get lost in non-deterministic events, such as link or end station failure. 
A frame missing in an egress queue may result in another frame being dispatched earlier than expected and scheduled.
As a result, this frame may change the arrival order in some egress queue, ultimately resulting in a stream missing its deadline. 
Such problems can be avoided in two ways.
First, by avoiding queuing at all.
Second, by not allowing frames to wait in the same egress queue at the same time.
In this way, it is not possible that some frame is dispatched earlier than scheduled due to a missing frame in an egress queue.
These restrictions are not imposed by bridges according to IEEE 802.1Q \cite{802.1Q}.
Instead, they are considered during scheduling such that a scheduling algorithm only computes schedules robust against these non-deterministic events.
In the following, we discuss problem extensions and restrictions from the literature.

\paragraph{Unrestricted Queuing}
Allowing frames of different streams to be in the same queue at the same time is denoted as unrestricted queuing.
\fig{unrestricted_queuing.pdf} depicts a schedule by showing frame arrivals and transmissions of a single bridge.
The schedule shows two streams, A and B, with two frames per period.
The queuing state is shown implicitly.
A frame is queued at the same time with all other frames that arrive before the frame is transmitted.
Thus, the frames A1 and B1 are in the egress queue at the same time.
If A1 does not arrive according to the schedule, e.g., due to a permanent link failure, B1 is transmitted earlier than scheduled.
This is the case in the second period depicted in \fig{unrestricted_queuing.pdf}.
Consequently, B1 arrives earlier than scheduled in some other egress queue.
This may result in some other frame experiencing more queuing delay than scheduled, ultimately leading to a missed deadline.

\paragraph{Isolation}
The problems of queuing in case of non-deterministic events can be solved by not allowing frames of different streams to be in the same queue at the same time.
If a frame is missing in a queue and no other frame is scheduled to be queued at the same time, no other frame can be transmitted earlier than scheduled.
This approach is denoted as \textit{frame isolation} in the literature.
It was introduced in \cite{CrOl16}. 
\fig{frame_isolation.pdf} shows a schedule valid with frame isolation.
If A1 does not arrive at the bridge, the schedule of B1 and B2 is unaffected.

\paragraph{No-Wait Scheduling}
In no-wait scheduling, frames are dispatched and sent immediately after arriving at an egress queue.
Queuing is not allowed.
The rational of this constraint is to avoid all consequences of non-deterministic events related to queuing.
\fig{no_wait.pdf} depicts a no-wait schedule for two streams.
All frames are sent immediately after arrival.
For example, B1 is received and transmitted after A1 and before A2 in the same period.

\foursubfigeps{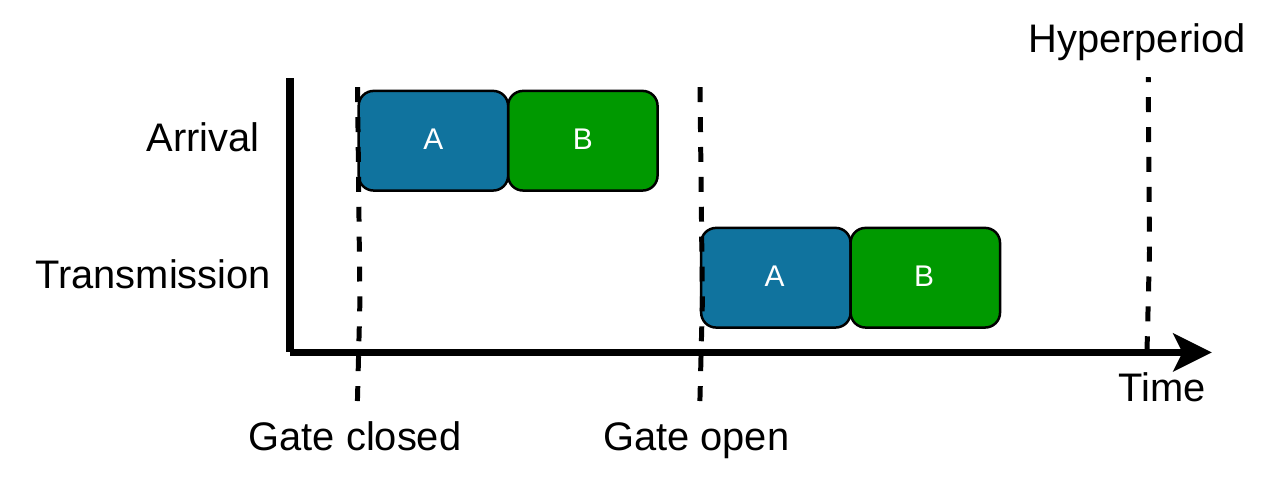}{Unrestricted queuing without fault. Frames A and B wait in the egress queue at the same time.}{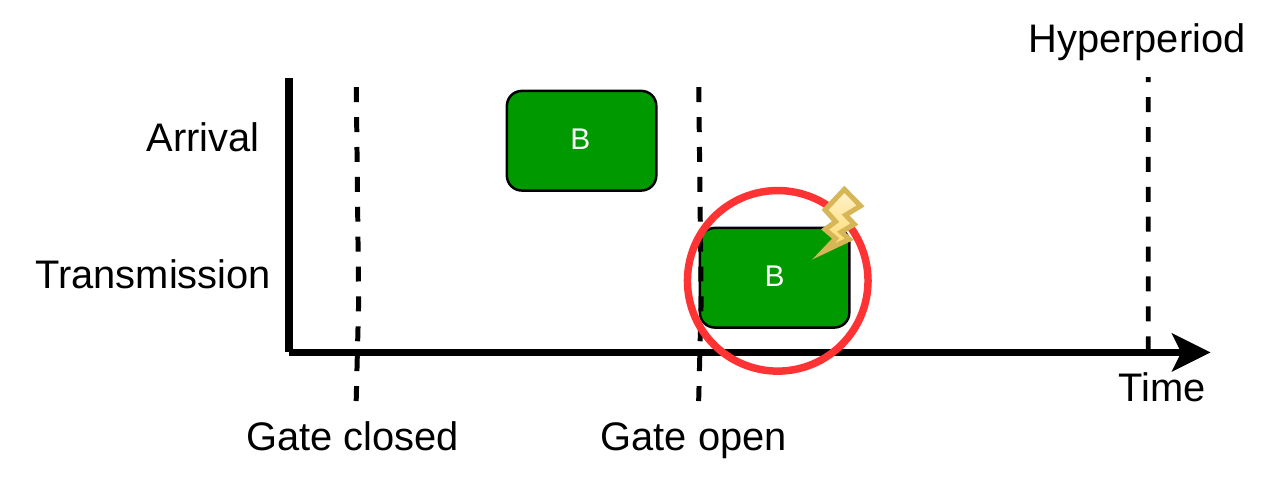}{Unrestricted queuing with fault. Frame B is sent earlier than scheduled if frame A was lost.}{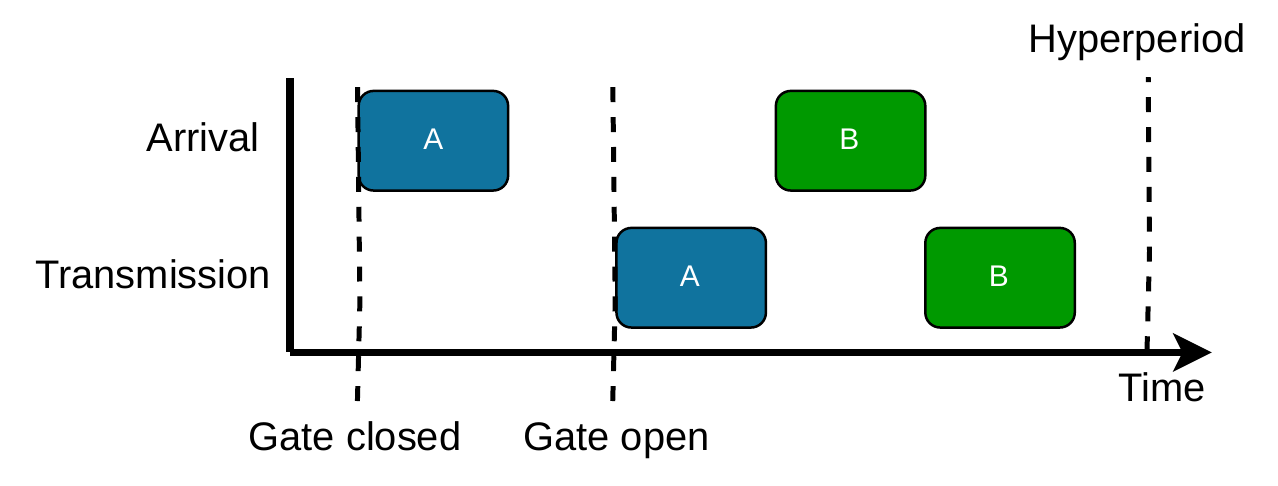}{Frame isolation. Frame A waits at a closed gate for some time before transmission. }{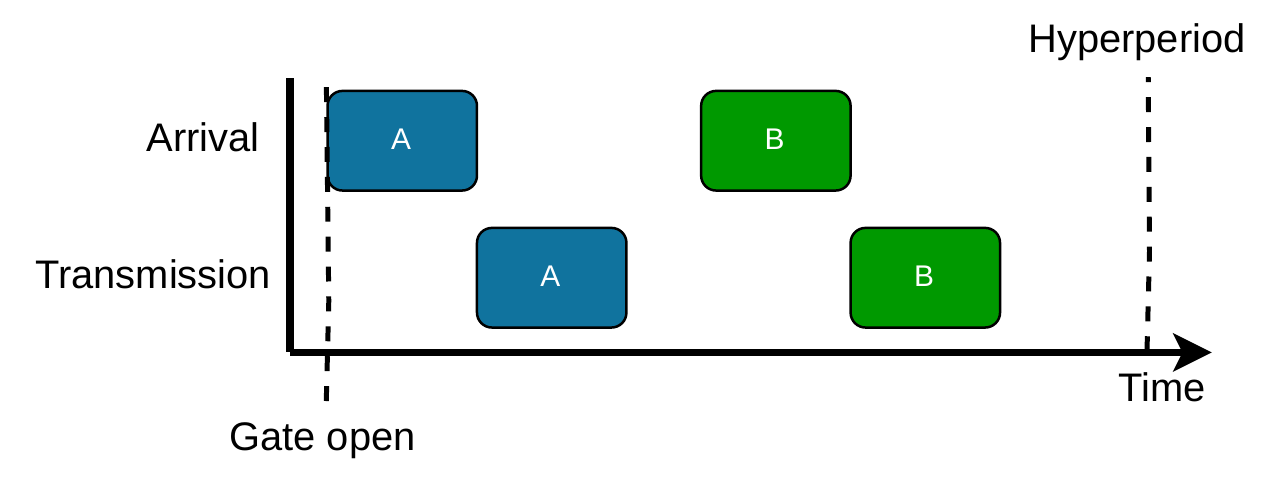}{No-wait scheduling. Frames are transmitted immediately after arrival.}{Queuing restrictions from TSN scheduling literature. Frame arrivals at an ingress port and transmissions at an egress port of the same bridge are shown. Processing delays are omitted to increase comprehensibility.}

\paragraph{Queue Assignment}
Instead of restricting queue usage, the scheduling problem can also be extended by allowing more than one queue per egress port for TT streams.
A scheduling algorithm for such a problem not only schedules transmission offsets and GCL entries, but also assignments of TT streams to egress queues.
This is especially interesting with respect to frame isolation, as frames of multiple streams can simultaneously wait for transmission by the same egress port in different queues.

\subsubsection{Integration of Audio Video Bridging}
TT streams and AVB traffic can coexist in the same network at the same time.
TSN bridges may support to use the CBS and the TAS in parallel. 
TT streams and AVB streams compete for the same links, but use different queues in the egress ports.
Therefore, the scheduling problem can be extended to also include a set of AVB streams as input.
They are scheduled at their respective talkers, and considerations for the behavior of the CBS must be included during scheduling.

\subsubsection{Integration of BE Traffic}
BE streams have no real-time requirements, they are generally
aperiodic and unknown a priori such that they cannot be scheduled.
However, some schedules may be beneficial for BE traffic. For
instance, large bursts of TT traffic within a long hyperperiod
could be avoided to facilitate frequent transmission opportunities
for BE traffic, which may reduce the delay of BE traffic. Another
example is avoiding GCL entries unless they save substantial
capacity for other traffic. For each GCL entry, a guard band is
needed within which packet transmissions cannot start. Therefore,
compact schedules maximize capacity for BE traffic. 

\subsubsection{Dynamic Reconfiguration}
An entirely different problem related to the basic problem is dynamic reconfiguration of existing schedules.
Such reconfigurations are necessary when streams are removed or new streams should be integrated into a schedule.
While removing streams is rather easy, adding new streams to an existing schedule can be complicated for two reasons.
First, the transmission offsets of already scheduled streams may have to be changed.
Second, links and egress queues are occupied by earlier scheduled streams, which places constraints on possible transmission offsets for new streams.
The runtime of a scheduler during reconfiguration must be very low in many scenarios, such as in automotive use cases.
This is due to fast changing real-time requirements and traffic patterns of safety-critical applications.
Recomputing the whole schedule with offline algorithms is computationally infeasible in such cases. 

\subsubsection{Multicast}
Multicast streams have more than one Listener as destinations.
Therefore, the routing of a multicast stream is a tree.
A multicast stream can be modelled by a set of unicast streams.
However, only a single copy of a frame is transmitted per hop in TSN.
Thus, this modelling is not appropriate.
Scheduling algorithms may contain considerations for multicast streams instead of assuming all streams to be unicast.
Joint routing approaches must compute trees instead of paths for every stream.

\subsubsection{Task Scheduling}
Tasks are applications running on end stations.
They are executed periodically.
Their execution depends on data received via TT streams.
Additionally, they can send TT streams after they processed some received data.
Scheduling algorithms for TSN can schedule tasks and TT streams in a joint approach.

\subsection{Optimization Methods}
We classify the scheduling algorithms in the literature in exact and heuristic approaches.
Exact approaches compute a schedule, or an optimal schedule if an objective is given, if one exists, or prove the problem instance infeasible. 
Heuristic approaches do not guarantee to find an optimal schedule.
Instead, they try to find reasonably good solutions within short time. 
In the common case, they cannot deduce whether a problem instance is infeasible, nor is finding a solution guaranteed if one exists.
In this section, we introduce common solution techniques and explain their basics.

\subsubsection{Exact Approaches}
As the Scheduling Problem for TSN is NP-complete, there is probably no polynomial-time algorithm to compute TSN schedules.
Therefore, it is reasonable to rely on the advances of the past decades in mathematical and combinatorial optimization. 
All exact solution approaches in the literature are based on the following four techniques. 

\paragraph{Integer Linear Programming (ILP)} An ILP describes the space of possible solutions to a problem with linear inequalities. 
Every assignment of variables which fulfills all inequalities corresponds to a solution of the problem and vice versa. 
Some variables may be restricted to take only integer values. 
A linear objective function may describe the quality of solutions.
\fig{ilp} depicts an ILP which minimizes an objective function.
\figeps[0.5\columnwidth]{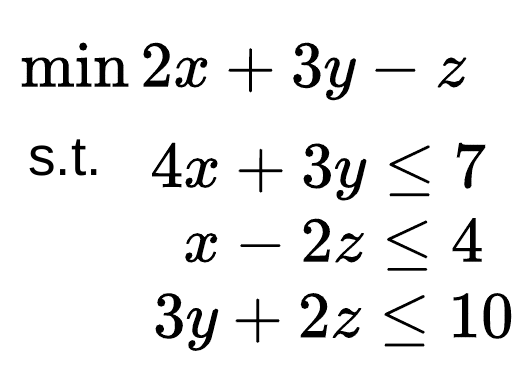}{Example of an ILP model.}
ILP solvers compute a feasible assignment which minimizes the objective function.
Every encountered solution in the solution process corresponds to an upper bound on the objective value of the optimal solution. 
Additionally, the solver can infer lower bounds for the objective value of the optimal solution during the solution process. 
So even when finding the optimal solution is not possible in reasonable time, ILP solvers yield estimations of the maximum gap to the optimum.
Widely used state-of-the-art ILP solvers are CPLEX \cite{cplex} and Gurobi \cite{gurobi}. 

\paragraph{Satisfiability Modulo Theories (SMT)} SMT solvers find solutions to problems described by first-order formulas.
Formulas model a problem with variables and predicates which are connected by logical operators. 
Besides Boolean variables, SMT solvers allow formulating predicates in other logical theories and use them as atomic formulas.
SMT solvers have an interface for theory-specific solvers so that the problem can be modelled with the best suitable theory. 
Examples of theories are the theory of linear arithmetic with integers or the theory of bit vectors.
\fig{smt} depicts a formula with predicates from the theory of linear arithmetic with integers.
The basic structure of a model is a formula from propositional logic, but predicates from integer arithmetic are used as atomic formulas.
\figeps[0.8\columnwidth]{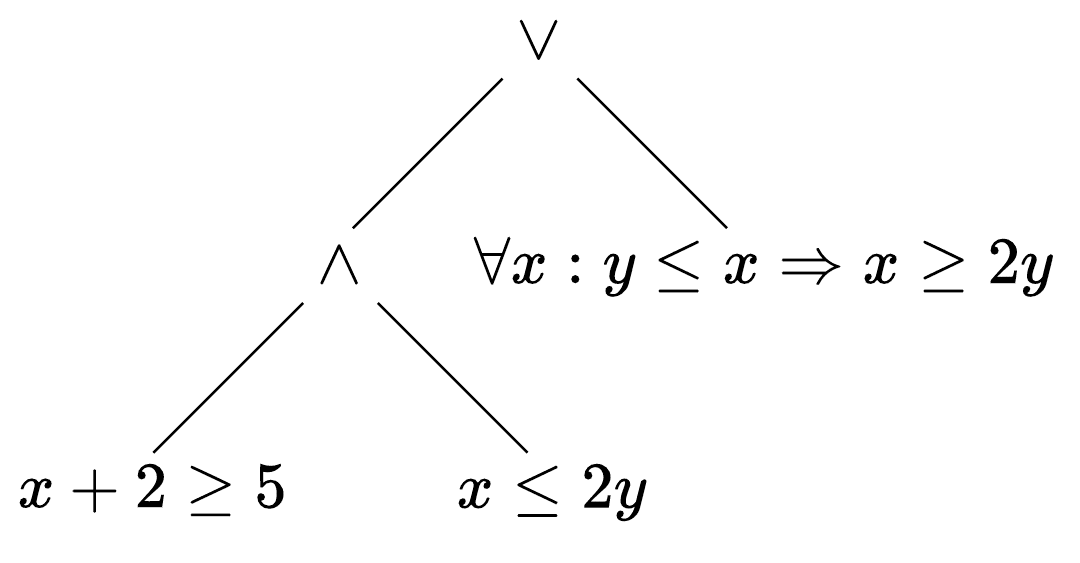}{Example of a formula used in SMT solving. The basic structure is a formula from propositional logic, but predicates from other theories may be used as atomic formulas instead of Boolean variables.}

The solver searches for an assignment of the variables that evaluate the formula to \textit{true}. 
It uses techniques from SAT solving to reason about satisfiability, combined with theory-specific solvers for conjunctions of predicates.
SMT solving is only about finding some satisfying solution.
When the best assignment regarding some objective function is computed, the term OMT is used.
Z3 \cite{z3} is a widely used SMT solver which can also be used for optimization. 

\begin{figure*}[ht!]
    \centering
    \includegraphics[width=0.88\textwidth]{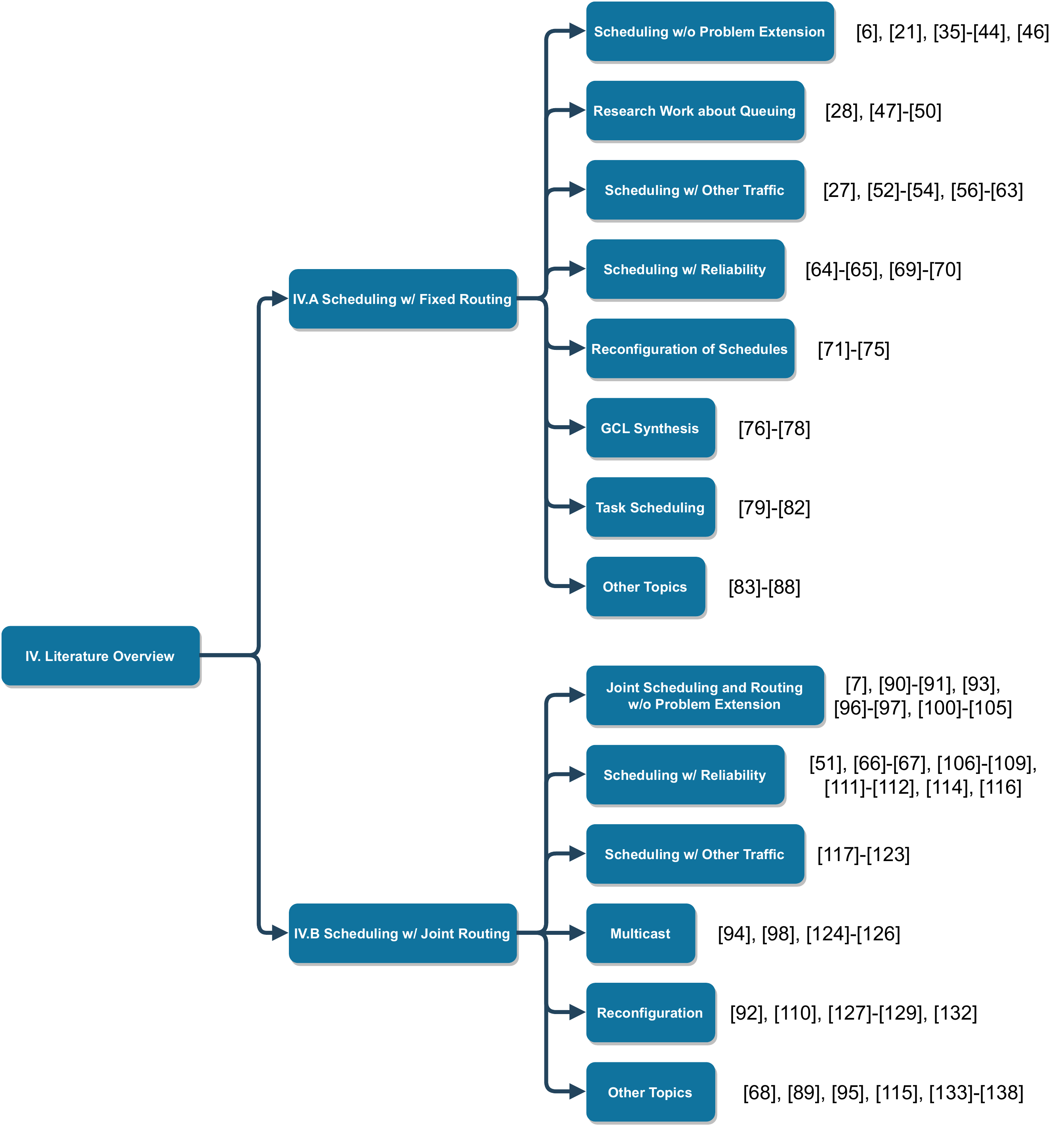}
    \caption{Classification of the surveyed research works in the scope of TSN scheduling.}
    \label{fig:orga}
\end{figure*}

\paragraph{Constraint Programming (CP)} Constraint Programming is a general solution approach to combinatorial problems. 
The set of feasible solutions to a problem is described in a declarative way. 
In this sense, ILP and SMT solving are special cases of CP solving. 
However, CP solvers use backtracking, local search, and constraint propagation techniques to solve CP models as opposed to ILP solvers. 
Another relevant case of CP is the restriction of variable domains to a finite set.
CP-SAT is a widely used CP solver \cite{ortools}.

\paragraph{Pseudo-Boolean Optimization (PBO)} Similar to ILPs the solution space of a problem in PBO is modelled with linear inequalities, but all variables must be binary.
However, instead of using mathematical optimization as in ILP solving, techniques from SAT solving like propagation and conflict refinement are employed.
A linear objective function can be minimized by adding it as a constraint to the model with some bound.
The solver is called multiple times with different bounded objective constraints.
Every infeasible solver run gives a lower bound on the optimal objective values.
Every solution yields an upper bound on the optimal solution. 
The optimal solution is found, with respect to some minimal precision, when the gap between lower and upper bound is smaller than the minimal precision provided by the user.

\subsubsection{Heuristic Approaches}
Because finding optimal solutions for realistic problem instances is infeasible in many cases, heuristic algorithms are used.
Such algorithms are used to find suitable solutions in reasonable time, generally without knowing whether there are better solutions.
Metaheuristic approaches are common algorithms that can be applied for a wide range of problems. 
Alternatively, there are heuristics that use problem-specific knowledge for many problems, and there may be combinations of both.

\paragraph{Greedy Randomized Adaptive Search Procedure (GRASP)} GRASP is a metaheuristic which can be adapted to various problems.
Its building blocks are a greedy-randomized algorithm to construct initial feasible solutions, and a local search algorithm. 
The greedy randomized algorithm incrementally constructs a solution by making random decisions among the set of decisions with the smallest increase in cost until a feasible solution is found.
The local search explores neighboring solutions, i.e., solutions with minimal changes, to the intermediate solution.
It explores the solution space until it finds a local optimum. 
Both steps are repeated a predefined number of times and the best encountered solution is returned. 
To adapt GRASP for a specific problem, a greedy randomized algorithm to generate initial solutions and a local search algorithm must be constructed. 

\paragraph{Tabu Search} Tabu Search is a metaheuristic to systematically explore the solution space. 
It uses an initial solution as start and moves to the best neighboring solution.
The algorithm keeps a tabu list of previously visited solutions or changes to solutions to avoid walking cycles in the solution space.
Only neighboring solutions or changes to solutions which are not contained in the tabu list are considered for the move.
The best encountered solution after a specific number of moves is returned.
To construct a problem-specific heuristic, a heuristic to generate an initial solution and a function returning the possible changes to some given solution must be built.

\paragraph{Simulated Annealing (SA)} is a metaheuristic used to find good approximations of the global optimum of an optimization problem. 
It is inspired by cooling processes in physics.
A global variable for temperature is used.
Temperature decreases slowly to 0 in discrete steps.
In each step, a neighboring solution is randomly selected, and the objective function is evaluated.
The probability of moving to a neighboring solution depends on the current temperature and the objective value of the considered solution. 
A move to a neighboring solution which is worse than the current solution is possible with small probability to escape from local optima.
As temperature decreases, the probability of moving to solutions with worse objective value vanishes.
The best solution encountered after some acceptance criterion holds is returned.
To adapt SA to a specific problem, a heuristic to generate an initial solution and a function returning the possible changes to some given solution must be built. 
Additionally, the way the temperature is decreased and the acceptance criterion must be selected.

\paragraph{Genetic Algorithms (GA)} Genetic algorithms are a metaheuristic approach inspired by evolution processes and natural selection in biology.
Candidate solutions are considered as individuals. 
Chromosomes represent properties of these individuals and are coded into bitstrings. 
At every point in time, there is a pool of individuals, i.e., the population.
New individuals are constructed from two or more existing solutions, i.e., genetic crossover is performed.
Individuals may be altered randomly, i.e., their chromosomes are mutated. 
When transiting to the next generation, some individuals die and are removed from the population.
The probability of dying for an individual depends on its fitness.
The fitness function is the optimization objective of the modeled problem.
The best individual encountered after some number of generations is returned. 
As in biology, high-quality solutions have a higher probability to survive and reproduce, which in terms yields new high-quality solutions.
To construct a problem-specific heuristic, a heuristic to generate initial solutions must be constructed.
Suitable crossover as well as mutation and selection mechanisms have to be used.
Parameters like population size, stopping criterion, and probabilities for selection and mutation must be designed.

\paragraph{List Scheduling (LS)} List scheduling is a metaheuristic to schedule tasks on identical machines. 
The tasks are sorted in a list according to some measure of priority.
In every step, the first task in the list is selected.
If a suitable machine is available, the task is executed on this machine, otherwise the next task in the list is selected.
These steps are repeated until all tasks are executed.
Considering streams as tasks and end stations as machines yields a heuristic for TSN scheduling.
A well-known heuristic from the scheduling literature can be considered to be special case of list scheduling.
As-soon-as-possible (ASAP) scheduling orders streams by priority and schedules them one by one at the earliest possible time along their paths.

\paragraph{Machine Learning} Machine learning is the generic term for a wide range of methods. Tools from linear algebra, statistics, and probability theory are used to construct mathematical models that can make decisions or construct solutions to a problem. The construction of such a model is denoted as \textit{learning} or \textit{training}. Typically, it takes a large amount of time and computational effort to train a model, but answers to request can be obtained really fast afterwards. Examples of machine learning methods are deep learning and reinforcement learning. However, the details of these methods are way beyond the scope of this survey. We refer to \cite{Al20} and \cite{GoBe16} for an introduction.
\section{Literature Survey}
In the following section, we give an overview of the literature about TSN scheduling.
We categorize research work based on whether scheduling with fixed routing or joint routing is considered.
Both sections are further grouped by the main topics of the respective papers.
Comparability of techniques and results of research works in the same group is ensured by this classification.
Figure \ref{fig:orga} depicts this classification.

\label{sec:literature_overview}
\subsection{Scheduling w/ Fixed Routing}
\label{sec:scheduling_fixed_routing}

We give an overview of research works which only deal with the scheduling of TT streams.
In all papers presented in this section, the routing of TT streams is fixed and given as input to the scheduling algorithm.
Such scheduling algorithms cannot change the routing during scheduling in case of conflicting streams.
We group publications in categories based on similar topics, like model assumptions or problem extensions.

\subsubsection{Scheduling w/o Problem Extensions}
We discuss publications solving the unmodified scheduling problem.
\figepsH[0.8\columnwidth]{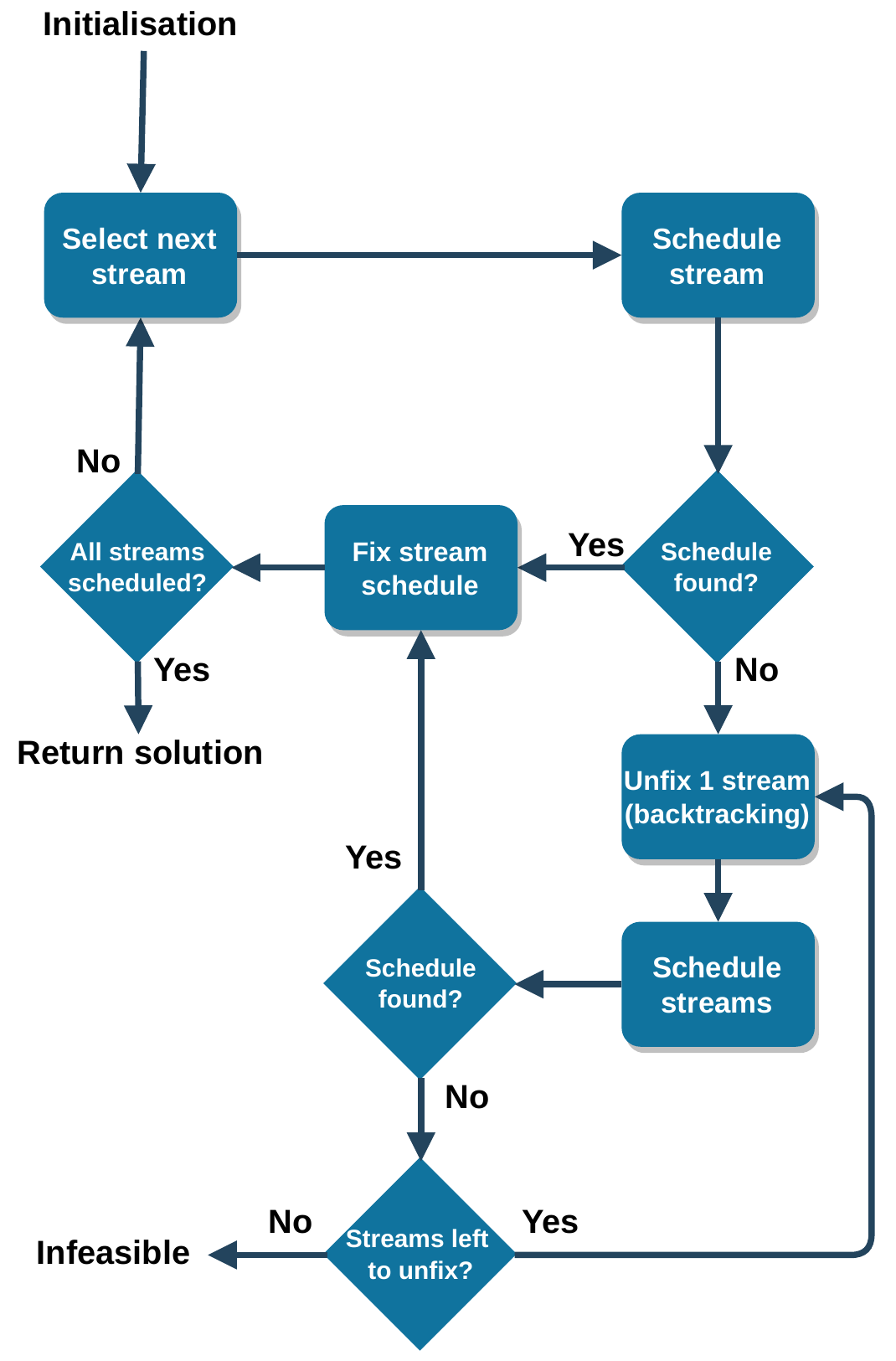}{Incremental approach of Steiner \cite{St10}. Similar ideas were used by other approaches, e.g., in \cite{CrOl16}.}

Early work about scheduling of TT traffic in Ethernet networks was conducted by Steiner \cite{St10}.
Even though this work is not specific to TSN, it influenced many research works covered by this survey.
The author proposes the use of SMT solving for the scheduling of TT streams.
An incremental approach is presented.
\fig{steiner.pdf} depicts this approach.
Streams are scheduled one after another.
Schedules of already scheduled streams are fixed in later iterations. 
Backtracking is used in case of infeasibility, i.e., the schedule of some stream is unfixed and the stream is scheduled again simultaneously with the new stream.
Backtracking is repeated until a schedule is found, or no stream schedules are left to unfix. 
This idea was adopted by many later works for TSN scheduling, e.g., \cite{CrOl16} and \cite{OlCr18}.

Oliver \textit{et al.} \cite{OlCr18} give an SMT model based on mapping streams to transmission time windows of egress queues. 
The number of these transmission windows is fixed per egress port, and their placement and size is computed by the scheduling algorithm.
As a side effect of using a fixed number of transmission windows, the number of gate events and thus guard bands is limited, even though the authors do not explore this matter. 
The authors use isolation to restrict the problems imposed non-deterministic behavior, e.g., frame loss.
Two queues per egress port are dedicated for TT traffic.
They evaluate the solving time of their approach with respect to the number of streams and the number of transmission windows per egress port.
Their results indicate that the solving time increases exponentially with the number of streams.
However, for reasonable numbers of streams and transmission windows, solving time is more sensible to the number of transmission windows.
A comparison to the SMT from \cite{CrOl16} shows that the window-based approach with one window per egress port is faster in finding a schedule.
The average jitter is significantly reduced when the number of transmission windows per egress port is increased.

\figeps[\columnwidth]{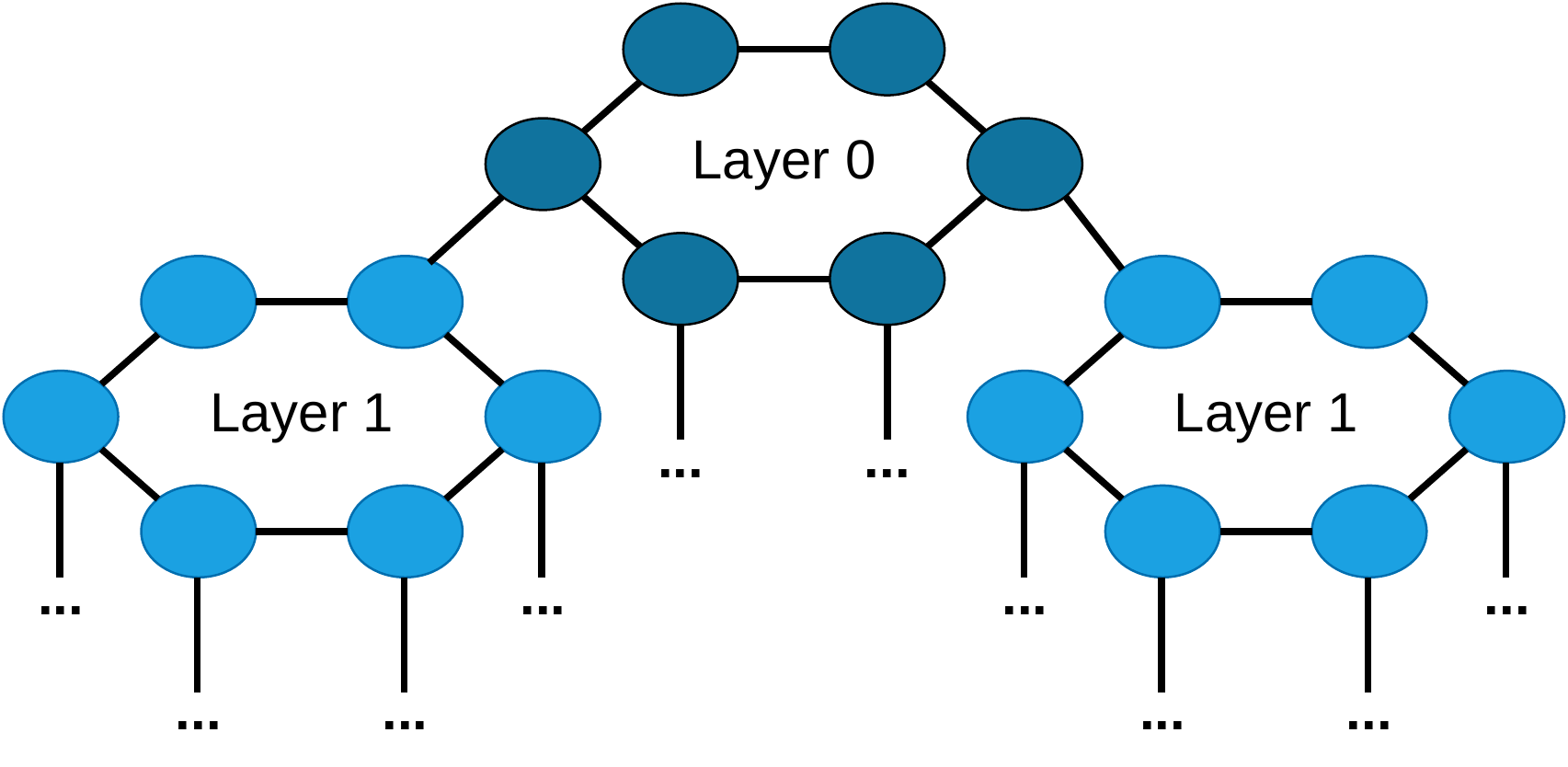}{Multi-layered ring topology used in \cite{HeGl20}.}

Steiner \textit{et al.} \cite{StCr18} suggest the SMT model from \cite{OlCr18} as a starting point for the standardization of TSN scheduling mechanisms.
They demonstrate their model by reporting the same evaluation results as in \cite{OlCr18} with a reduced number of transmission windows. 

Hellmanns \textit{et al.} \cite{HeGl20} extend the Tabu Search algorithm of \cite{DuNa16} for no-wait scheduling.
They construct a 2-stage approach for hierarchical networks which consist of multiple rings on different layers. 
They argue that such topologies are common in factory automation.
\fig{multi_ring.pdf} depicts a model of such a topology.
First, they schedule streams with talker and listener in the same ring.
This step is done individually for every ring.
No queuing is allowed in these schedules.
Then, they simulate the transmission of all other streams as if they were sent at the same time from their respective talkers.
If all streams meet their deadlines in this simulation, the simulated behavior is used as schedule.
They compare this approach with scheduling all streams at once with the Tabu Seach algorithm.
Their evaluations demonstrate that the proposed 2-stage scheduling scales better for problem instances with many streams compared to the original Tabu Search approach.
The latter does not produce results for more than 1000 streams due to memory limitations.
The 2-stage scheduling is two orders of magnitude faster in the special case of multi-layered ring topologies.
The authors report that the number of needed GCL entries is significantly reduced by the 2-stage approach.

Another heuristic for no-wait scheduling is proposed by Zhang \textit{et al.} \cite{ZhXu22}.
They analyze how frame transmissions may conflict and derive the range of possible transmission offsets per frame.
A comparison to the SMT of \cite{OlCr18} and the ILP of \cite{DuNa16} shows clear performance benefits of the proposed heuristic.
The SMT was able to schedule about 300 streams, while the ILP scheduled about 1000 streams, and the heuristic scheduled 1200 streams in the evaluation scenario.

Kim \textit{et al.} \cite{KiCh21} give a heuristic algorithm to compute valid schedules, and a post-processing to reduce end-to-end delays.
Streams are ordered by priority and are scheduled one after another.
The individual frames of a stream are scheduled along the stream's path. 
The hyperperiod is divided into intervals and every frame is assigned to the earliest unoccupied interval.
The presented evaluations indicate that end-to-end delays are reduced by up to one third per stream in the evaluation scenarios.

The authors of \textit{et al.} \cite{KiLe21}\cite{KiLe22} propose a genetic algorithm to schedule TT traffic in automotive scenarios.
Genes encode the scheduling order of frames.
Frames are scheduled as soon as possible according to this order and along the respective stream's path.
The objective function used to compare scheduling orders is the weighted sum of end-to-end delays, jitter, and bandwidth utilization of the corresponding schedule.
As in \cite{DuNa16}, a schedule compression algorithm is employed to reduce the bandwidth occupation of guard bands.
The proposed approach outperformed random schedules regarding all three metrics in almost all evaluation scenarios.
The approach from \cite{KiCh21} is also outperformed with regard to the used objective.

Ansah \textit{et al.} \cite{AnAb19} present a scheduling algorithm in the special case of a line topology where all talkers converge in a single bridge.
Based on this method, they also give an algorithm to compute GCLs in such a topology if the streams are schedulable.

The special case of an in-vehicle network with only a single hop is analyzed in \cite{WaZh22}.
They assume all traffic streams to be send continuously and belonging to different traffic classes and egress queues.
Essentially, they implement round robin traffic shaping with the TAS.
However, we remark that real in-vehicle networks are more complex and thus the gained insights are limited. 

The authors of \cite{MiOh22} compare the suitability of a large set of metaheuristics for TAS scheduling.
They maximize the number of scheduled streams for the same problem instance with various functions of a metaheuristic library.
The authors observed the best results with math based and system based heuristics and interpret this as a hint for future research directions. 

Vlk \textit{et al.} \cite{VlBr22} propose a heuristic algorithm to schedule very large-scale problem instances.
The algorithm shares many similarities with the well known DPLL algorithm from SAT solving \cite{DaLo62}, e.g., probing, backtracking in the case of conflicts, and restarts.
Frames are scheduled one by one. 
If a conflict arises, all decisions are reverted up to the conflicting frame. 
The authors compare the heuristic with SMT-, ILP-, and GRASP-based algorithms.
The schedulability of an approach with respect to some set of problem instances is the fraction of solvable problem instances within some time limit.
The proposed heuristic outperforms all other approaches regarding schedulability and solving time.
In fact, they were able to schedule instances with up to 10812 streams in a tree-like topology with 2000 nodes.
This result outperforms all other approaches in the literature.
Evaluations with a real-world instance from avionics are also presented.

Wang \textit{et al.} propose a deep reinforcement learning approach for no-wait scheduling in \cite{WaYa22}.
They train machine learning models for various network topologies.
The model aims to reduce the maximum arrival time among all frames to reduce the number of guard bands.
For networks with up to 9 bridges and 10 end stations, the authors report solving times of at most 400$\,$s.

\subsubsection{Research Work about Queuing}
We highlight works which allow or deal with the implications of queuing.

Craciunas \textit{et al.} \cite{CrOl16} construct an incremental SMT model to schedule TT streams based on \cite{St10}.
They define flow isolation and frame isolation as properties of a schedule to prevent some sources of non-determinism, e.g., single link failures.
They present models to compute schedules with either flow or frame isolation.
Besides isolation in the time domain, they also employ isolation in the spatial domain by the possibility of assigning different streams to different queues. 
The authors identified the problem of clock synchronization errors and introduce gaps between frame transmission to cope with this problem.
They compare the impact of frame and flow isolation to the solving time of their SMT model.
Their evaluations indicate that flow isolation reduces solving times compared to frame isolation.
However, more problem instances can be scheduled with frame isolation.

Vlk \textit{et al.} \cite{VlHa20} investigate the effect of the isolation constraints from \cite{CrOl16} on schedulability.
When a frame is lost during transmission and does not reach the next egress queue as scheduled, another frame may be dispatched earlier than scheduled from this queue.
This frame in term can cause more non-determinism on its path.
Not allowing frames of different streams to be in the same queue at the same time solves this problem, but reduces the solution space considerably.
A modification for bridges implementing the TAS is proposed to cope with this conflict.
Queues with this modification check whether the next frame is the correct one with respect to the schedule.
If this is not the case, the queue idles until the next frame transmission is scheduled.
A comparison shows clear benefits regarding schedulability. 
The number of streams which are scheduled to arrive before their deadline is also significantly increased compared to isolation models.

The authors of \cite{ReZh20} present a heuristic to schedule streams with queuing.
The heuristic is based on transmission windows similar to \cite{OlCr18}. 
In contrast to earlier works which include queuing in their model \cite{CrOl16}\cite{OlCr18}, they drop isolation constraints.
Network calculus is employed for a worst-case end-to-end latency analysis.
They minimize the occupation percentage of egress ports, i.e., the percentage of the hyperperiod which is reserved for TT traffic.
In this way, long and frequent time intervals for lower-priority traffic are scheduled.
Their evaluations indicate that their approach is superior regarding end-to-end delay and schedulability of streams compared to earlier works from the same authors.

Chaine \textit{et al.} \cite{CaBo22} use queuing for jitter control.
They propose to schedule streams without queuing at all egress ports except for egress ports connected to end stations.
Frames are buffered in these egress ports and are released such that jitter constraints are satisfied.
The authors present a novel isolation approach, denoted as \textit{size based isolation}.
Frames must be buffered in increasing frame size order if they are stored in the same queue.
Two GCL entries are used to close and open the corresponding gate between two frame transmissions.
In this way, frames cannot be transmitted during an earlier time slice than scheduled if another frame is missing in the queue, as earlier time slices are too short.
\fig{size_isolation.pdf} depicts such a scenario.
\figeps[\columnwidth]{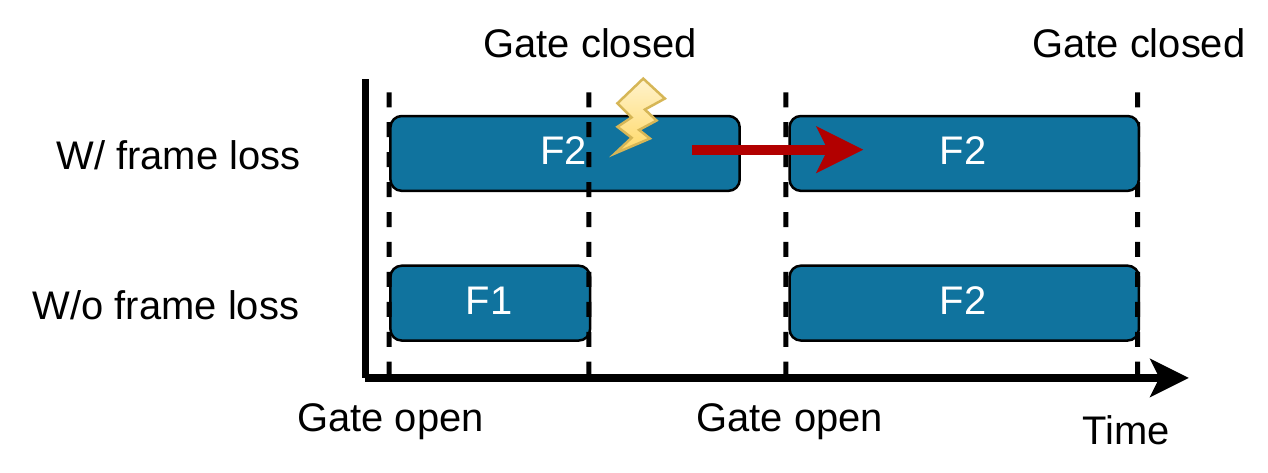}{Size based isolation proposed in \cite{CaBo22}. Frame transmissions from an egress port connected to an end station are shown. Assume F1 and F2 are scheduled to wait some time in the same egress queue. If the first frame F1 has not arrived at the egress queue, F2 cannot be transmitted in the time slice dedicated for F1 as it is too short.}
The authors give an ILP model to compute schedules with their approach.
A comparison of their approach to an unspecified approach for latency minimization demonstrates that their approach reduces scheduling time significantly.
However, this comes with the cost of higher latencies.

Bujosa \textit{et al.} \cite{BuAs22} propose a heuristic scheduling algorithm which handles queue assignment of streams.
Instead of scheduling frames or streams one after another on their entire path, they schedule all transmissions over a single link before scheduling the transmissions over another link.
They present results about the scalability and schedulability of their approach compared to a CP approach from the literature \cite{ReCr22}.
Not surprisingly, scheduling is significantly slower with a CP approach compared to a heuristic.

\subsubsection{Scheduling w/ Other Traffic}
The schedule of TT streams may affect other traffic classes.
AVB and BE traffic cannot be scheduled, but QoS metrics of these classes can be influenced when they are taken into account during the scheduling of TT streams.
We summarize works with such considerations.

Dürr \textit{et al.} \cite{DuNa16} present an ILP and a Tabu Search algorithm to compute no-wait schedules for TT streams. 
They model the problem with job-shop scheduling, a widely used modelling framework in the scheduling literature.
The authors measure the solving times of their Tabu Search and conclude that the network topology and size have no impact on solving times.
They minimize the flowspan to construct a large time slice for BE traffic at the end of the computed schedules.
They propose a compression algorithm as post-processing for schedules which aims to reduce the number of GCL entries needed to deploy a schedule.
The authors note that this increases the available bandwidth for BE traffic as the number of guard bands is reduced.
\fig{compression.pdf} depicts the effect of the schedule compression algorithm.
\figeps[\columnwidth]{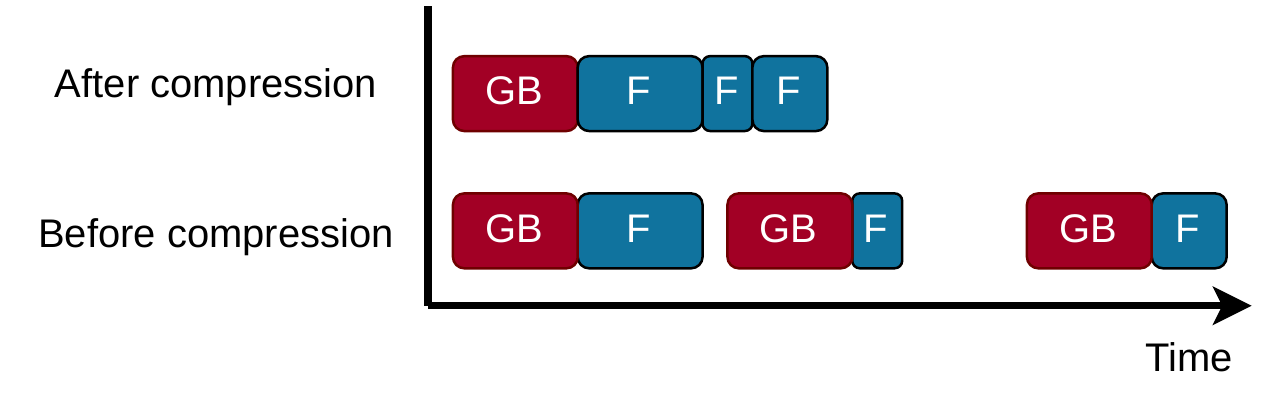}{Effect of the schedule compression algorithm from \cite{DuNa16}. Scheduled frame transmissions (F) and guard bands (GB) over a single link are shown.}
They report that the number of GCL entries can be reduced by 24\% on average.
Parts of the content of this work are also featured in the PhD thesis of Nayak \cite{Na18}.

The authors of \cite{PoRa16} give an ILP to compute schedules for TT streams and additionally present a GRASP heuristic to schedule AVB streams. 
They restrict queuing by enforcing frame isolation.
Their heuristic computes a routing for AVB streams such that they meet their deadlines.
It reduces the search space by only considering a fixed number of shortest paths for every pair of nodes as possible routings. 
The schedule of TT streams, computed by their ILP model, serves as input for the heuristic and cannot be changed.
They compare their AVB routing with the na\"ive approach of always selecting the shortest path.
The comparison demonstrates that more AVB streams can be scheduled with their approach.
A comparison of solving times of their ILP to the SMT from \cite{CrOl16} is conducted. 
They state that their proposed ILP does not scale well for industrial-size instances and further efforts to create a suitable heuristic are needed.

Santos \textit{et al.} \cite{SaCa19} present an extensive SMT-based modelling of the scheduling problem with openly accessible implementation.
Their model contains a range of features known from previous works, e.g., transmission windows, multicast, guard bands, and bandwidth considerations for BE traffic which were not covered by a single approach in the past. 
The starvation of BE traffic is prevented by restricting a user-defined fraction of a hyperperiod exclusively to be used by other traffic which is related to the approach of minimizing the flowspan \cite{DuNa16}.
Additionally, unrestricted queuing is integrated which is uncommon in exact approaches so far.
The authors mention the limitation of only one gate opening per queue per hyperperiod in the presented model which reduces the available bandwidth for other traffic classes.
They evaluate their approach on a realistic sized network and report successful scheduling for up to 10 multicast streams.
The model is also used in the well known simulation framework OMNeT++ \cite{OMNeT}.
The thesis of Santos \cite{Sa20} explains the model in detail. 

Houtan \textit{et al.} \cite{HoAs21} compare schedules computed with various objectives for the same SMT model with respect to the QoS of BE traffic.
They propose minimization and maximization of frame offsets, hoping that grouping frames together increases the QoS. 
Additionally, they also suggest two objectives which maximize the gaps between consecutive frame transmissions over a link.
They integrate frame and flow isolation in their SMT model. 
Unfortunately, their work lacks a description which one was used in the evaluations.
A comparison of the different objective functions indicates that larger gaps between frame transmissions of TT streams increase the QoS of BE streams.
For instance, BE traffic may experience less starvation and average latencies are reduced.
\fig{houtan.pdf} depicts how BE traffic may benefit from maximizing the gaps between TT frame transmissions.
\figeps[\columnwidth]{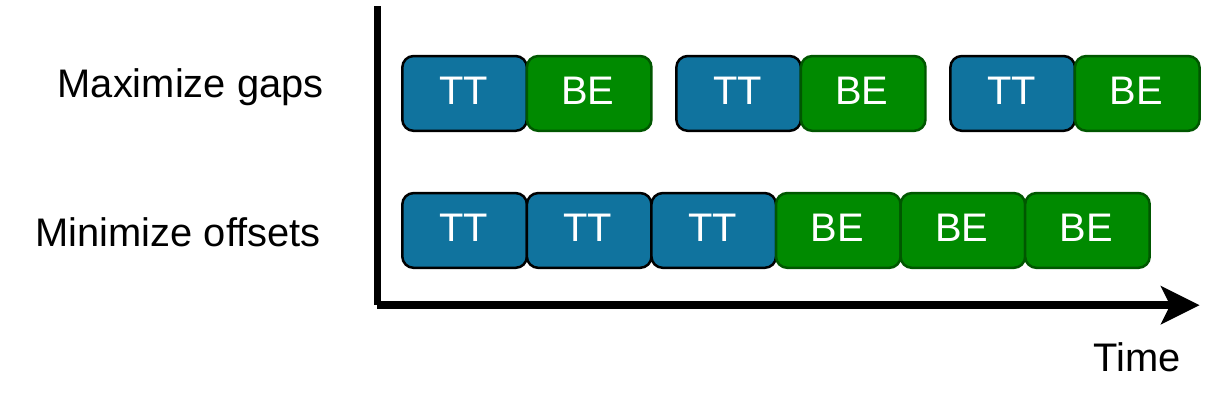}{Effects of different objective functions to BE traffic in \cite{HoAs21}. Frame transmissions of TT and BE traffic over a single link are shown.}
However, we note that the used system model features deadlines for BE traffic, and they measure the number of deadline misses, so a comparison with the other mentioned research works is not possible.
The solving time for their SMT depends heavily on the objective function used.

The model of \cite{CrOl16} is used for an in-vehicle scenario in \cite{ZhLi20}.
The authors present simulation results with typical traffic patterns in such a scenario.
They compare end-to-end delays for schedules using the TAS to schedules using the strict priority mechanism. 
Their results indicate that scheduling with the TAS can ensure real-time requirements of TT streams while the performance of lower-priority traffic is less affected compared to the strict priority mechanism.

Barzegaran \textit{et al.} \cite{BaRe22} give a CP approach to compute transmission windows for TT streams.
In contrast to other window-based approaches \cite{OlCr18}\cite{StCr18}, they assume that not all end stations support TSN.
They use a worst-case delay analysis to eliminate solutions that may violate the real-time requirements of the given problem instance.
They compare their approach to the algorithms presented in  \cite{CrOl16}, \cite{OlCr18}, \cite{ReZh20}, and \cite{PoRa16}.
They outperform these approaches in terms of solving time, but end-to-end delays and bandwidth utilization are significantly worse compared to \cite{CrOl16} and \cite{OlCr18}.
They also perform simulation runs of their schedules with OMNeT++.
The results indicate that their worst-case analysis for end-to-end delays holds but overestimates the simulated delays considerably.

The coexiestence of the TAS and Cyclic Queuing and Forwarding (CQF) in TSN is investigated by Pei \textit{et al.} \cite{PeHu22}.
They propose to use CQF for rate constrained traffic with deadlines, i.e., some egress queues per egress port are shaped by CQF.
Streams of scheduled traffic and rate constrained traffic are scheduled simultaneously.
They are scheduled one after another in least laxity first order, i.e., the next scheduled stream is the stream which deadline expires next.
The same approach is used only for scheduled traffic streams as an alternative for comparison.
The evaluation shows that the joint handling of scheduled traffic and rate constrained streams results in higher schedulability.

Another approach which combines the TAS and CQF is presented by \cite{YaGa22}.
They consider the scenario of multiple traffic classes with different real-time constraints.
Besides of scheduled traffic with low latency and jitter requirements, there are also two other traffic classes with uncritical periodic streams and best effort traffic.
The uncritical periodic streams are assigned by egress queues shaped by the CQF.
The authors present a heuristic to compute a schedule for all traffic classes simultaneously. 
The evaluations show that sorting streams in earliest deadline first order before scheduling is beneficial for the schedulability.
In contrast to that, sorting streams by frame size or period reduces schedulability considerably. 

Wang \textit{et al.} \cite{WaXu22} propose a combined scheduling scheme for TT and AVB streams.
AVB streams are shaped with CQF.
They use guard bands to protect TT frames from AVB frames.
Their heuristic tries to schedule as many AVB streams as possible while load balancing the traffic amount between the time slots of the CQF mechanism.
The authors report that their approach significantly reduces jitter and solving times compared to another approach for CQF.

Huang \textit{et al.} \cite{HuZh22} propose a recursive scheduling heuristic using backtracking.
They use a complex in-vehicle topology to evaluate their approach and also include AVB streams in the evaluation scenario.
We highlight that they give detailed stream parameters which is rare for real-world use cases.
They also include Frame Replication and Elimination for Reliability \cite{802.1CB} to cope with frame loss of safety critical traffic.
Additionally, an SMT model is presented and compared to their heuristic.
The heuristic outperforms the SMT in regard to schedulability, scalability, and end-to-end latencies by far in the evaluation scenario.

\subsubsection{Scheduling w/ Reliability}
Reliable transmission of data streams is one of the design goals of TSN.
Additionally, to hardware features ensuring reliability, schedules can be assembled to mitigate the effects of various faults.
We discuss publications which take such considerations into account.

The clock frequencies of two clocks are not exactly equal for technical reasons.
This results in so-called \textit{clock drift}, i.e., clocks running with different speeds.
\fig{clock_drift.pdf} depicts this problem.
\figeps[0.9\columnwidth]{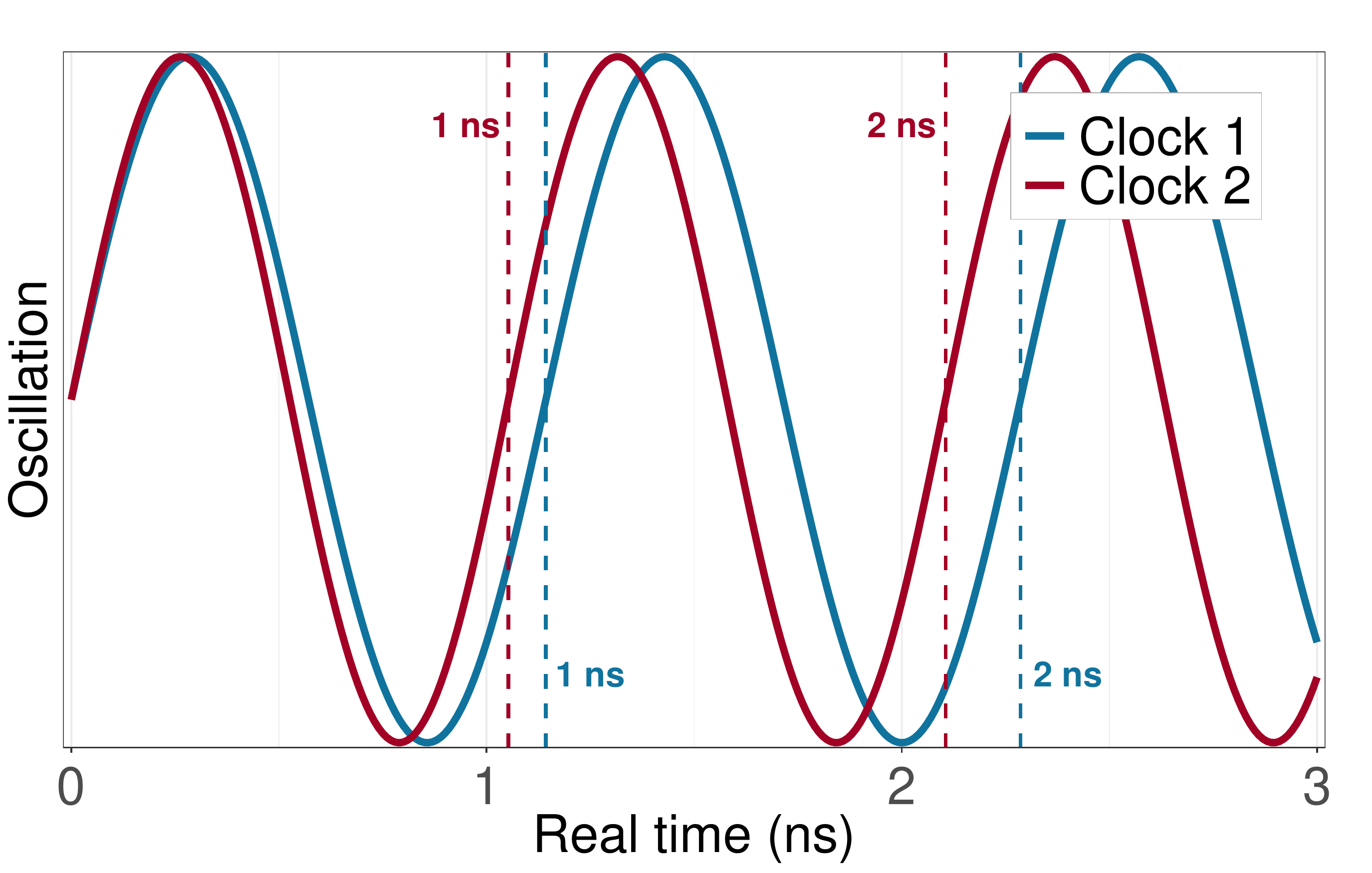}{Timing signals of two clock oscillators. Clock 1 runs slower than clock 2. Therefore, the difference between both clocks increases over time.}
Craciunas \textit{et al.} \cite{CrOl21} extend their model from \cite{CrOl16} to cope with clock drift during scheduling.
They introduce a parameter for the maximum allowed clock drift into all equations which contain transmission offsets or reception times of frames.
Effectively, they merely increase the gap between frame transmissions which is already contained in the model of \cite{CrOl16}. 
Clocks are resynchronized after some predefined out-of-sync detection timeout.
The authors present a design space study which investigates the relationship between the maximum allowed clock drift, the worst-case clock drift rate, the maximum possible diameter of the synchronization spanning tree, and the out-of-sync detection timeout.
Their findings on a number of test networks indicate that shorter out-of-sync detection timeouts are needed for higher clock drift rates.
The maximum possible diameter of the synchronization spanning tree is negatively affected by higher clock drift rates.
Evaluations for a test case regarding schedulability, end-to-end latency, and solving time are conducted.
The results show that end-to-end latency increases for higher allowed clock drifts.
Maximizing the allowed clock drift yields a maximal robust schedule for a given problem instance, but solving time increases by an order of magnitude compared to setting a fixed maximum clock drift in advance.

Feng \textit{et al.} \cite{FeCa21} consider the scheduling problem in the presence of frame loss.
Instead of scheduling redundant streams over disjoint paths, as in \cite{AtBa20}, \cite{RePo20}, and \cite{GaZa17}, reliability is achieved by multiple transmissions of a stream over the same path.
The research work focuses on choosing an appropriate number of repetitions per stream for a trade-off of reliability and network utilization.
In contrast to similar works, considerations for AVB and BE streams are also included in the algorithm, as repeated transmissions of TT streams deplete the available bandwidth and may lead to starvation of other traffic otherwise.
The presented algorithm uses the SMT model from \cite{CrOl16} as a sub-routine. 
Their results show that increasing the fault probability leads to a higher number of retransmissions which in turn results in less available bandwidth for BE traffic.

In later works, Feng \textit{et al.} \cite{FeDe22} studied a similar problem, but also considered ACK and NACK messages and queue assignment of streams.
In contrast to \cite{FeCa21}, every TT stream is sent exactly twice.
Transmission windows for BE streams are computed after the scheduling of TT streams.
The scheduled transmission intervals for the retransmissions can be used to transmit BE traffic when no retransmissions are needed.

Dobrin \textit{et al.} \cite{DoDe19} present a heuristic scheme to schedule streams with reliability considerations.
They consider transmission losses for frames such that only one frame is affected by a fault at a time and the fault is fixed by some predefined number of retransmissions.
Their approach first tightens the deadlines to take some number of retransmissions into account.
Then, they schedule streams in earliest deadline first order.
Additional considerations for rate constrained traffic are also included in their scheme, following the scheduling of the TT streams.
Unfortunately, no evaluations are presented.
The authors note that future works will address more realistic fault models.

\subsubsection{Reconfiguration of Schedules}
Adding and removing streams from an existing schedule is necessary in dynamically changing environments, e.g., automotive use cases.
While removing a stream is straightforward, adding new streams may require more effort.
We summarize research works concerned with this problem extension.

Raagaard \textit{et al.} \cite{RaPo17} propose an algorithm for online scheduling of new TT streams in an existing schedule.
They use a heuristic which schedules streams as early as possible such that schedules comply with isolation.
When a new stream should be added to an existing schedule, they calculate whether there is a starting offset such that the stream can be scheduled without changing the existing schedule.
If this is not possible, the stream is assigned to unused queues of the egress ports along the stream's path.  
They evaluate how many streams can be added to an existing schedule in a specific time.
The authors report that their heuristic is able to schedule about 1300 frames per second in medium-sized test cases. 

Pang \textit{et al.} \cite{PaHu21} compute schedules with an ILP such that updating a schedule does not lead to frame loss or additional update overhead.
In contrast to \cite{DuNa16} and \cite{PoRa16}, their approach is not limited to TSN and streams are scheduled one by one.
Schedules of streams from previous iterations are fixed in later iterations.
When some stream cannot be scheduled, backtracking is used by removing some stream of an earlier iteration from the schedule.
The authors prove that a set of additional constraints of the ILP imply no conflicts during schedule updates.
They evaluate their algorithm with respect to frame loss during updates and update duration on real-world train and automotive networks.
The results confirm that no frames are lost and no time overhead is needed for schedule updates.

Another algorithm for schedule updates is proposed by Wang \textit{et al.} \cite{WaWa22}.
They present a heuristic scheduling algorithm with backtracking similar to \cite{HuZh22}.
Additionally, they present an algorithm for incremental schedule updates which omits frame loss during updates.
A comparison between both algorithms shows that the incremental update algorithm is faster while it has poor schedulability for higher network utilization. 

Gärtner \textit{et al.} \cite{GaRi22} introduce a measure for schedule flexibility denoted as \textit{flexcurve}.
They leverage this measure to add streams incrementally to an existing schedule such that the resulting schedules are beneficial for further reconfigurations.
The authors compare their algorithm to a not specified SMT approach and the algorithm of Santos \textit{et al.} \cite{SaCa19}.
In contrast to the SMT approach, solving time of the proposed reconfiguration algorithm is linear for up to 100 streams in the evaluation scenario.
The approach of Santos \textit{et al.} \cite{SaCa19} results in schedules with lower flexibility and thus is less suitable for dynamic reconfiguration.
A journal extension of this work is presented in \cite{GaRi23}.

\subsubsection{GCL Synthesis}
Most research works use a post-processing to compute GCLs from transmission offsets.
However, this comes with the drawback that GCLs have limited size in bridges and a schedule may not be deployable. 
We present literature which discusses explicit GCL generation.

Jin \textit{et al.} \cite{JiXi20} present an SMT approach to schedule TSN streams with a fixed number of gate openings.
Reducing the number of gate events also reduces the number of guard bands, such that more bandwidth is available for lower-priority traffic.
Their modelling assumptions regarding queuing are even more restrictive than frame isolation as only exactly one frame is allowed to be in a queue at any given time. 
Their approach allows multiple queues for TT traffic per egress port, but assigning streams to queues is not part of the SMT model.
This is done before solving the SMT model by a greedy heuristic which aims to balance the workload of all queues of an egress port.
As their SMT model cannot be solved in reasonable time, they use an incremental scheme to schedule small groups of streams separately. 
Subsets of streams are scheduled one after another such that the schedules of previously scheduled subsets are fixed in later iterations.
The objective when optimizing a subset is to minimize the maximal number of GCL entries for all egress ports.
They also propose a heuristic algorithm which complies with a limited number of GCL entries.
Their evaluations show that the heuristic algorithm is an order of magnitude faster than na\"ive heuristics while reducing the number of GCL entries considerably. 
Instances with up to 10000 streams were scheduled in reasonable time while the SMT approach has not produced a feasible schedule for an instance with 100 streams within 2 days.

Another incremental SMT scheme which aims to reduce the number of GCL entries is given in \cite{LiLi20}.
Their approach divides the hyperperiod into slices, which are scheduled individually. 
GCLs are updated for at the beginning of every slice. 
The authors compare the number of GCL entries needed with schedules computed for an entire hyperperiod.
Their results demonstrate that the number of GCL entries can be reduced while keeping end-to-end delays in reasonable bounds.
However, it is not a surprise that fewer GCL entries are required when updating the GCLs regularly is allowed, as even a single entry per GCL is sufficient with frequent updates.

A rather simple CP model for scheduling on a single link with only four types of constraints is presented in \cite{DaWa21}.
However, they propose a post-processing to reduce the number of GCL entries needed in a schedule.
For a small test case of only three streams, the authors report a reduction of bandwidth loss due to guard bands by 42.8\%.

\subsubsection{Task Scheduling}
Tasks are applications running on end stations.
We highlight publications which consider the scheduling of tasks on end stations, additionally to scheduling data streams between these tasks.

In \cite{FeYa21}, Feng \textit{et al.} compute schedules for streams and tasks sending or receiving streams simultaneously. 
The model includes dependencies between streams and tasks, e.g., an application can only be executed when all frames of some stream were received.
The authors scheduled instances with 11 streams and more than 100 tasks. 
As many other works, the authors note the exponential increase of solving times for larger instances.

The authors of \cite{BaZa20} present a CP model for scheduling of TT traffic which takes characteristics of control applications into account.
Control applications have an execution interval and can only produce output streams for actuators when certain input streams of sensors have arrived.
Although the quality of control application execution covers multiple aspects, the only one taken into account is jitter.
Queuing is allowed in their model, but is restricted to frame isolation.
They compare exact and heuristic search strategies to find solutions to the proposed CP model.
For all presented test cases, both search strategies find the optimal solution with zero jitter, but the heuristic approach is orders of magnitude faster.

These preliminary works were extended in \cite{BaPo21}.
In contrast to \cite{BaZa20}, a more realistic quality measure for streams of control applications is integrated into the CP model.
It constitutes of jitter and end-to-end delays of input and output streams of control applications, and jitter for control application execution.
They compare their model with the model from \cite{CrOl16} which is extended to include stream precedence for input and output streams of control applications.
The authors report that the presented model outperforms the model from \cite{CrOl16} with respect to the proposed quality measure by up to a factor of two on the test cases under consideration.
Additionally, they compute a schedule for a realistic test case of an automotive mobile robot and validate their algorithm on a simulation platform and on real hardware.
The PhD thesis of Barzegaran \cite{Ba21} features this work.

\subsubsection{Other Topics}
This section summarizes research works with unique topics that fit not well into the previous groups.

Jin \textit{et al.} propose an SMT model which also handles an optimized fragmentation of messages in \cite{JiXi19}.
Messages can be transmitted with multiple frames.
How messages are split into frames is an additional degree of freedom in the presented optimization.
Due to performance reasons when solving the model, they also give heuristics for message fragmentation and scheduling.
The presented evaluations demonstrate that schedulability increases considerably by up to 50\% when message fragmentation is also taken into account.
\figeps[\columnwidth]{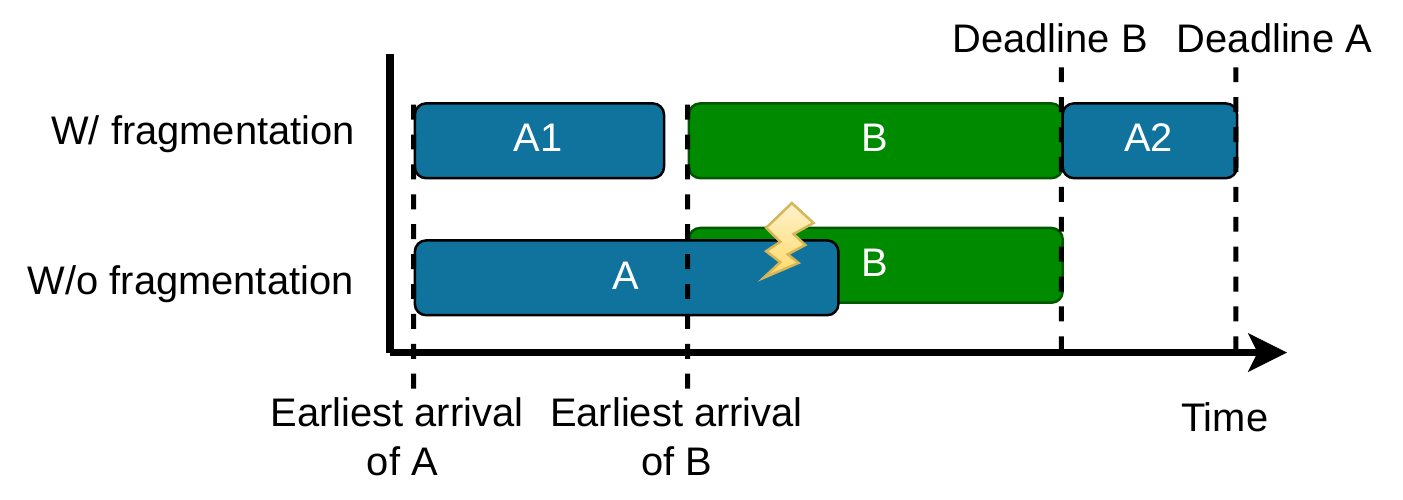}{Transmissions of two frames A and B over a link to a listener. B cannot be scheduled to be transmitted at another time. In this case, scheduling is only possible when the scheduler splits A into two frames.}
\fig{fragmentation.pdf} depicts an example of how schedulability can benefit from message fragmentation.
Additionally, the presented heuristic algorithms can schedule instances an order of magnitude larger than the SMT approach.

A genetic algorithm approach which takes frame preemption into account is presented in \cite{PaSa19}.
Their model contains different kinds of MAC interfaces for preemptable and non-preemptable frames.
Consequently, the presented synthesis problem not only covers the assignment of streams to queues, but also the assignment of queues to interfaces.
Queues are strictly prioritized, i.e., frames contained in a higher-priority queue always preempt frames of a lower-priority queue.
The proposed GA aims to maximize the reliability of a schedule.
Reliability of a stream is defined as the maximum number of allowed retransmissions without missing the deadline, and the reliability of a schedule is the minimum reliability of all streams.
The authors present a comparison of the proposed GA with well-known approaches from automotive traffic scheduling.
The baseline approaches are outperformed with respect to schedulability and reliability.
The authors explain this result with the fact that their algorithm is specifically constructed to use all the available TSN queues and to utilize them in a way suitable for preemption.
\figeps[\columnwidth]{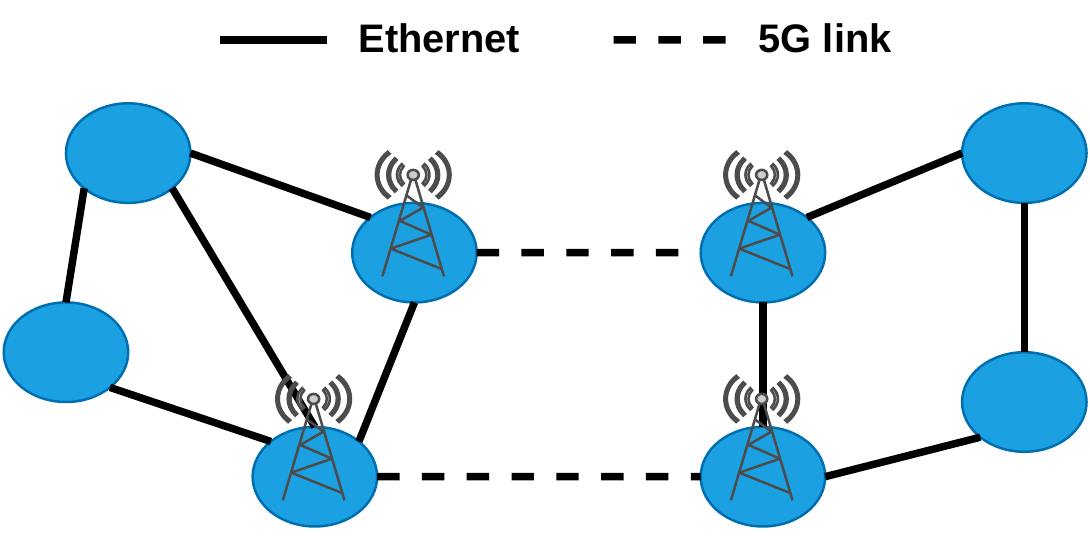}{Example of a converged network with 5G and Ethernet links as considered in \cite{GiGi20}.}

The authors of \cite{GiGi20} give a CP model for TSN in joint converged wired and wireless networks.
Their model integrates Ethernet and 5G links simultaneously.
\fig{5g.pdf} depicts an example for such a converged network.
Frames transmitted over a 5G link must be scheduled to fit into predefined transmission slots.
They aim to minimize unusable resources in both types of links, i.e., time occupied by guard bands for Ethernet links and unused bandwidth resources for 5G links. 
The presented evaluations indicate that minimizing only one kind of unusable resources leads to unsatisfying results for the respective other kind.  

Lin \textit{et al.} \cite{LiLi22} evaluate the impact of the so-called \textit{network cycle} to schedulability.
The network cycle is a design methodology for frame schedules.
All stream periods are assumed to be integer multiples of the network cycle.
Frame transmissions are aligned with network cycles.
The rational of this is to omit conflicts between streams with different periods when streams are scheduled incrementally.
They propose an incremental heuristic which considers the network cycle.
The authors report the highest schedulability when the network cycle is set to the greatest common divisor of all stream periods.

The authors of \cite{FaKu17} developed a graphical modelling tool for TSN scheduling.
They use logic programming to deduce facts about the given problem instance.
These facts are in term used for constraint generation of an SMT model.
If an instance is infeasible, the conflict refinement capabilities of the SMT solver is leveraged to guide the user in changing the network configuration appropriately.
Three test scenarios are presented where the streams causing infeasibility are identified.

Machine learning techniques were introduced to the domain of TSN scheduling in \cite{TuXu21}.
Tu \textit{et al.} present a semi-supervised machine learning model to partition streams in groups before scheduling.
They compare their approach with the partitionings in \cite{AtBa20} and \cite{MaAm18}, and state that they are outperformed regarding schedulability.
However, it is not clear how this statement is backed by the actual computation of schedules with the resulting stream groups.

We highlighted the contributions of research works for the scheduling problem with fixed routings.
We compare and discuss the research works presented in this section together with the research works for the joint routing problem in Section \ref{sec:comparison}.

\subsection{Scheduling w/ Joint Routing}
In this section, we give an overview of research work which inspects the joint routing and scheduling problem.
In contrast to works in \ref{sec:scheduling_fixed_routing}, algorithms proposed by publications in this section compute a routing and a schedule for a given set of streams simultaneously. 
Again, we group the literature based on the main topic of the respective papers.

\subsubsection{Joint Scheduling and Routing w/o Problem Extensions}
This section compiles publications which handle the joint routing and scheduling problem.
Research works are only included when they do not focus on an additional topic highlighted in this survey.

An early ILP model which addresses the problem of joint routing and scheduling is presented in \cite{ScDa17}.
Although it is not exclusively for TSN, the authors state it is applicable for such networks.
Their evaluations show that schedulability increases considerably compared to the same test cases with a fixed routing. 
They compare the solving time of their ILP for joint routing and scheduling with ILPs solely for scheduling.
As expected, the solving time is larger for joint routing and scheduling compared to scheduling with a fixed routing.
Nevertheless, they still recommend joint routing as solution quality increases considerably.

Falk \textit{et al.} \cite{FaDu18} extend the ILP from \cite{DuNa16} to simultaneously compute routing and schedule of TT streams.
They analyze the scalability of the joint routing and scheduling problem using ILPs.
The authors report that solving time is more influenced by the number of streams than the size of the network topology for their ILP. 
The evaluations show that the network topology has an impact on scalability. 
Network topologies with more paths between any pair of nodes tend to yield harder problem instances, as more routings are possible for any stream.

Nie \textit{et al.} \cite{NiLi22} schedule and route streams incrementally.
Streams are grouped by divisibility of their periods, such that streams in the same group can share the same links.
In contrast to similar works, e.g., \cite{AtBa20} and \cite{HuWa21}, they consider only no-wait scheduling.
Time is divided into time slots whose lengths equal the greatest common divisor of the periods of all streams.
Although many evaluations are performed for different network topologies, network sizes, and traffic types, no results not seen in other works were presented.

Xu \textit{et al.} \cite{XuXu22} propose an incremental SMT scheme similar to \cite{AtBa20}\cite{HuWa21}. 
However, they partition streams with machine learning using some of the ideas from \cite{TuXu21}.
The authors compare this partitioning approach with the partitioning algorithms from \cite{AtBa20}, \cite{TuXu21}, \cite{MaAm18}, and \cite{PaTa19}.
The best schedulability was obtained for the proposed partitioning method, second to the methods from \cite{AtBa20} and \cite{TuXu21}.
Schedulability is slightly increased when more streams are scheduled simultaneously, as more conflicting streams are handled in the same iteration.
Additionally, the authors compare the incremental scheme with their global scheduling approach from \cite{XuXu20}.
The incremental scheme outperforms the global approach with respect to schedulability and scalability.
The difference in schedulability between both methods increases for higher link utilization.

The authors of \cite{SmGl17} present a PBO model for joint routing and scheduling.
They compare the solving time of their PB approach with the solving time when routing and scheduling are computed in separate steps by the same model.
Their initial evaluations indicate that solving routing and scheduling in separate steps reduces the overall solving time.
Surprisingly, their evaluations also show that the 2-step approach performs significantly worse for larger instances compared to the joint approach.
They explain this behavior by the capability of SAT solvers to learn from conflicts.
Whenever the solver runs into a conflicting variable assignment, it interferes the cause of the conflict and adds a clause to the model which prevents the conflict explicitly.
The learned clauses are dropped after the routing step in the 2-step approach.
Schedulability increases when routing and scheduling are performed in a single step.
They state that instances with more routing options lead to easier solvable scheduling problems as streams can be distributed over the network.

Arestova \textit{et al.} \cite{ArHi20} construct a genetic algorithm for joint routing and scheduling.
In contrast to other works with genetic algorithms \cite{PaTa19}\cite{PaOb18}, the authors focus on elaborating on the construction for such an approach in detail.
They combine the genetic algorithm with a neighborhood search heuristic to find better solutions efficiently.
They allow queuing with flow isolation constraints from \cite{CrOl16}.
Additionally, a schedule compression algorithm similar to the one in \cite{DuNa16} is presented, which is used to reduce the number of guard bands. 
In a brief evaluation section, they compare their approach with the well-known NEH algorithm \cite{NEH} from job-shop scheduling.
The proposed approach finds feasible schedules faster, while the resulting schedules have comparable flowspans. 
The authors report that scheduling with joint routing takes only slightly more time than scheduling with fixed routing with the genetic algorithm.

Kentis \textit{et al.} \cite{KeBe17} investigate the relation of port utilization and GCL schedule duration.
They employ a simple heuristic for scheduling which is not further explained, and compare the resulting schedule duration with shortest-path routing and the proposed congestion-aware routing.
For most test cases, the duration of the schedule is reduced.
However, they do not motivate why shorter schedule durations are beneficial. 

The authors of \cite{WaCh20} present an algorithm based on ant colony optimization. 
They give the fundamental building blocks of such an algorithm and demonstrate that it can be suitable for TSN scheduling, but also note that further investigation is needed. 
Unfortunately, they do not elaborate on the routing, and the only related evaluation indicates that increasing the number of edges increases the solving time.

Falk \textit{et al.} propose a joint routing and scheduling algorithm in \cite{FaDu20} which is not based on constraints for frame transmissions.
Their approach constructs a conflict graph where each vertex represents a schedule for a single stream. 
Vertices are connected by an edge if and only if the corresponding stream schedules are conflicting.
This reduces the scheduling problem to finding an independent set in a conflict graph, i.e., a set of vertices which are pairwise not connected by an edge.
\figeps[0.8\columnwidth]{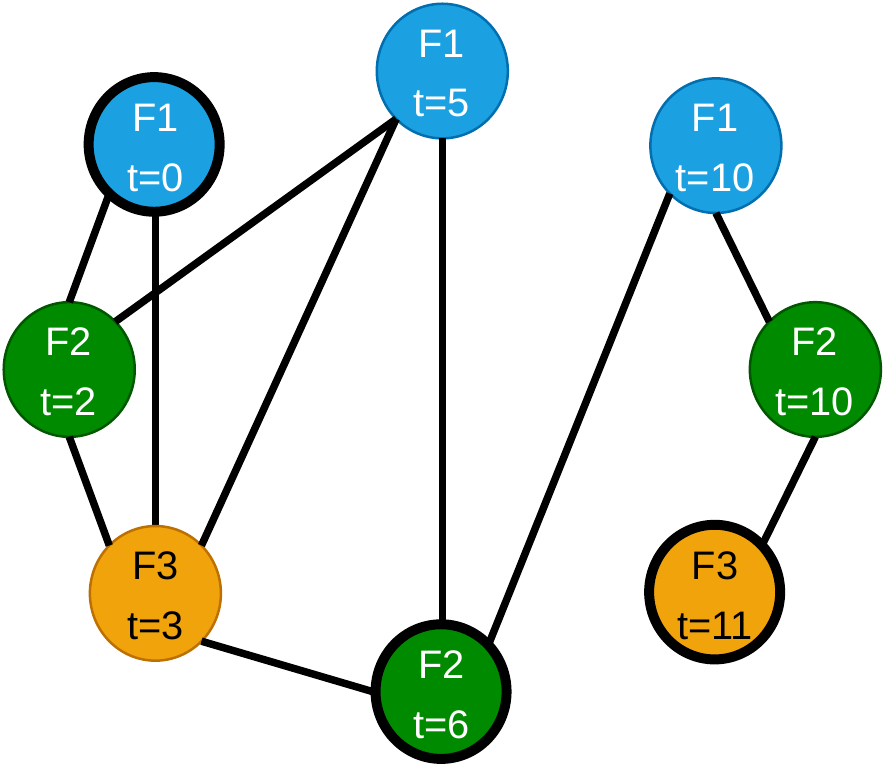}{Example of a conflict graph used in \cite{FaDu20} for a single link topology. Assume the transmission of each frame takes 5 time units. Edges indicate stream schedules which are not compatible. The black circled vertices are an independent set and induce a schedule.}
\fig{independent.pdf} depicts an example for a conflict graph and an independent set.
The authors use incremental heuristics to construct independent sets in such graphs.
Their evaluations demonstrate that the conflict graph approach has advantages regarding runtime and memory consumption compared to ILPs.
They remark that their implementation is just a proof of concept, and more efficient algorithms to find independent sets are known.
They further note that this approach is cheap, as no expensive ILP solver is needed.

\figeps[\columnwidth]{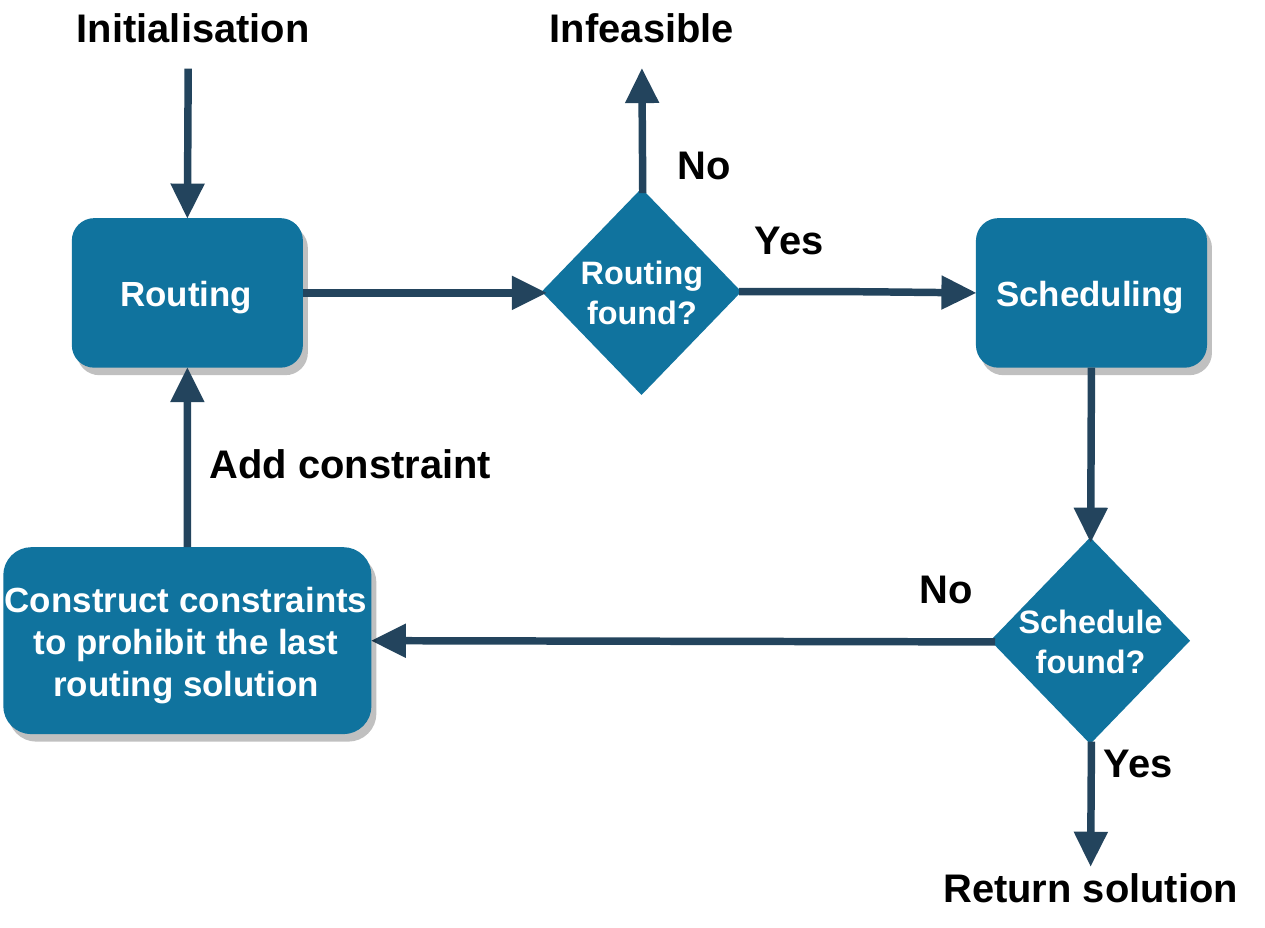}{Flow diagram of the proposed approach in \cite{VlHa21b}.}

An enhanced CP approach for routing and scheduling is presented by Vlk \textit{et al.} \cite{VlHa21b}.
The authors present separate models for routing and scheduling, and use them in a problem decomposition algorithm.
First, they compute a routing for the given streams.
A schedule is computed using this routing.
If no schedule was found, constraints are added to the routing model to prohibit the last routing solution.
These steps are repeated until a schedule is found.
Alternatively, it may be the case that all possible routings for some stream lead to a conflict while scheduling.
In that case, an instance is deemed as infeasible.
\fig{vlk.pdf} depicts a flow diagram of the proposed approach.
Most other research works which performs routing and scheduling in separate steps considers an instance infeasible after only one pass of routing and scheduling.
The model also includes queue assignment of streams as additional degrees of freedom.
The authors substitute their scheduling model with the algorithms from \cite{DuNa16}, \cite{CrOl16}, and \cite{Ca18}, and compared schedulability of the resulting algorithms with their approach.
Their CP model was able to schedule the most instances, while the SMT model scheduled significantly fewer instances than all other algorithms.
Scheduling time was similar for all algorithms except for the SMT model, which needed multiple times longer for most instances.

He \textit{et al.} \cite{HeZu23} present a deep learning based approach for joint routing and scheduling.
They use a graph neural network to handle arbitrary sized network topologies.
They evaluate their approach on various random network topologies and compare it with \cite{ScDa17}, \cite{DuNa16}, \cite{HeGl20}, and \cite{PaOb18}.
All other approaches were outperformed in regard to schedulability and scalability for various numbers of streams and network topology sizes.
They also compare different encodings, policies, and sampling strategies featured in their deep learning approach.
The results give useful insights for future works involving deep learning in TSN scheduling.
Additionally, the authors present measurements of jitter on a real hardware testbed which integrates their deep learning scheduling algorithm.
They report no frame losses and ultra low jitter, which indicates that the constructed schedules are valid.

\subsubsection{Scheduling w/ Reliability}
Reliability is an important topic in the literature of joint routing and scheduling.
The possibility of selecting disjoint paths for redundant stream copies further increases reliability in these research works.

Pozo \textit{et al.} \cite{PoRo18} present an ILP for joint routing which considers reliability constraints.
After a single link failure, all streams over this link must be rescheduled and rerouted.
The authors propose a fast heuristic for this case. 
They also evaluate which properties of a schedule are beneficial for the repairability in case of a failure. 
The results indicate that schedules which maximize the gaps between frame transmissions are much easier to repair than schedules which minimize the flowspan, even for three simultaneous link failures.

Atallah \textit{et al.} \cite{AtBa18} give a heuristic for fault-tolerant joint scheduling, routing, and topology generation.
Their algorithm starts with a full mesh topology.
Streams are routed and scheduled one after another.
A k-shortest-paths algorithm is used to enumerate possible paths for a stream.
If a path with available time slots is found, schedule and routing are fixed for later iterations.
This is repeated multiple times with disjoint paths for redundant copies of a stream to ensure reliability.
Links and bridges are only included in the final topology when they are used by some stream.
The algorithm also selects bridges such that more expensive bridges are only used when necessary.
\fig{atallah.pdf} depicts this approach for topology generation.
The authors compare their algorithm to another approach which realizes redundant paths through multiple copies of the network topology.
The proposed algorithm scheduled all problem instances, while it reduced topology costs considerably. 
\figeps[0.8\columnwidth]{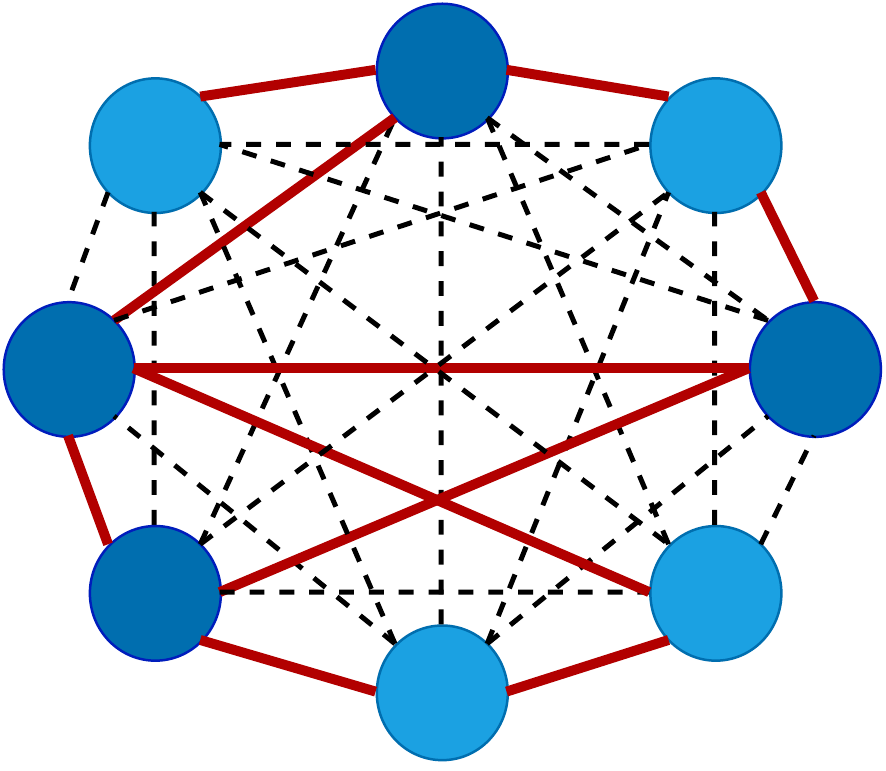}{Topology generation approach of \cite{AtBa18}. A full mesh topology is used during routing and scheduling. Links used in the final routing are indicated in red. Only these links are included in the final topology. The approach covers the selection of bridges from a library depending on the generated topology. More expensive bridges with a higher number of ports are indicated in dark blue.}

The authors of \cite{AtBa20} propose an incremental ILP scheme to comply with requirements for the robustness against single link failures.
First, the authors present a GRASP heuristic for routing which considers reliability constraints.
Streams are replicated and routed over disjoint paths to comply with requirements regarding robustness to single link failures.
The resulting routing is used as input to the incremental ILP approach.
Streams are partitioned into groups by introducing a conflict metric and computing a weighted cut in the conflict graph.
Groups of streams are scheduled one after another.
The computed schedules are fixed when the next group is scheduled.
They model egress ports with only one queue for TT traffic and frame isolation.
A comparison of the  presented ILP with the ILPs from \cite{ScDa17} and \cite{SmGl17} demonstrates that it outperforms both approaches regarding solving times.

Instead of robustness against link failures, Zhou \textit{et al.} \cite{ZhSa21} consider reliability against frame loss, i.e., frames that are lost spontaneously during transmission without a permanent link failure.
They integrate constraints regarding the loss probability along the routed path of a stream in their SMT model.
However, these probabilities are only approximated, as the used theory solver cannot handle exponentials. 
An incremental scheme similar to \cite{AtBa20} is used.
Subsets of streams are scheduled and routed one after another until all streams are scheduled.
Stream schedules are fixed in subsequent iterations.
They use redundant copies of streams to further reduce the probability of frame loss, as there may be no single path with the required reliability. 
In contrast to other works, e.g., \cite{AtBa20}, \cite{RePo20}, and \cite{GaZa17}, they do not enforce paths to be link disjoint.
Their evaluations show that schedulability with a given level of reliability against frame loss increases with a higher number of redundant copies per stream.

Syed \textit{et al.} \cite{SyAy22} present an ILP and a heuristic for scheduling and path selection in in-vehicle networks.
They leverage Frame Replication and Elimination for Reliability (FRER) \cite{802.1CB} to ensure robustness against frame loss of safety-critical streams.
The ILP model is similar to the ILP used in prior works by the same authors \cite{SyAy20}.
They report solving times of about a day with the ILP while the heuristic solved the same instances in a few minutes.

Following their works in \cite{SyAy22}, Syed \textit{et al.} \cite{SyAy23} developed an alternative to FRER.
The authors propose a network coding scheme to mitigate temporary and permanent link failures.
Two disjoint paths are used to transmit two frames.
A third path disjoint from the other two is used to transmit the XORed data of these two frames.
The loss of one frames can be tolerated as the lost frame can be reconstructed from the other two.
Therefore, the redundant transmission of $n$ frames results in $\frac{3}{2n}$ frame transmission with this scheme.
This is a significant reduction compared to the $2n$ frames needed with FRER.

Another incremental scheme for scheduling and routing in safety-critical automotive applications is presented by Zhou \textit{et al.} \cite{ZhSa21b}.
They consider possibly undetected systematic faults of bridges, i.e., implementation bugs or divergence from specifications, instead of randomly arising errors like frame loss. 
Messages and bridges have an Automotive Safety Integrity Level (ASIL) assigned \cite{Iso26262} which defines reliability constraints of the respective component.
A higher ASIL corresponds to higher reliability.
Additionally, to computing routing and scheduling, their algorithm also selects which bridges to use from a library. 
Messages can be decomposed into redundant message copies with lower ASIL to be sent over bridges with lower ASIL.
A comparison of the presented algorithm with and without message decomposition shows that the total costs for bridges can be reduced by up to 23.55\% in the evaluation scenarios.
The authors note that higher ASILs increase the required number of message copies, which leads to more congestion in the network and thus higher end-to-end delays. 
Evaluations on a real-life automotive test case are presented. 
Synthesis time increases considerably with ASIL decomposition.
However, the selected bridges cost only a third compared to using only bridges with the highest ASIL.
The PhD thesis of Zhou \cite{Zh22} compiles the content of \cite{ZhSa21}, \cite{ZhSa21b}, and \cite{ZhSa22}.

The authors of \cite{RePo20} present separate CP models to compute schedules and routings as work-in-progress.
These models take security and reliability considerations into account. 
The routing computed with the first model is used as fixed input for the scheduling model, similar to \cite{AtBa20}.
Redundant streams with disjoint paths are added if needed to comply with security and reliability constraints. 

The ideas from \cite{RePo20} are extended in \cite{ReCr22}.
Besides the CP model, a simulated annealing algorithm combined with a list scheduler is given.
Additionally, a post-processing for latency reduction of scheduled streams is applied after scheduling.
Their evaluations without security and reliability constraints indicate that the SA approach is able to find a feasible schedule in reasonable time, even for huge problem instances.
However, they also introduce a measure for schedule and routing costs.
This measure contains stream latencies, penalties for streams which were not scheduled, and penalties for overlapping paths.
The results demonstrates that schedules computed by SA have up to three times higher costs compared to schedules computed by the CP approach.
Introducing the security and reliability constraints to the same problem instances increased the costs of schedules computed by SA by up to a factor of $\sim 3.5$.
Nevertheless, the authors state that the SA approach can be useful in comparing costs and reliability capabilities of topologies, or to reconfigure the network in case of link failures.

Li \textit{et al.} \cite{LiCh21} propose a heuristic for joint routing and scheduling with reliability constraints.
A greedy algorithm is used to select paths for redundant copies of streams such that link utilization is balanced. 
Frames are scheduled as soon as possible.
Both algorithms are combined in an iterative local search scheme.
Already scheduled and routed streams are randomly removed from time to time, and the remaining streams are rescheduled and rerouted in random order. 
While the authors report a reduction in frame losses by their approach, end-to-end delays are significantly larger compared to schedules with routings computed by Dijkstra's algorithm.

\subsubsection{Scheduling w/ Other Traffic}
This section presents research works concerned with scheduling in the presence of other traffic classes.

Gavrilu{\c{t}} \textit{et al.} \cite{GaZh18} construct a GRASP heuristic to schedule TT streams while taking AVB traffic into account.
In contrast to \cite{PoRa16}, their approach handles TT and AVB streams simultaneously. 
The routing of AVB streams is given and cannot be changed.
In the first evaluation, the authors use a shortest paths algorithm to compute the routing.
The results show that using the GRASP heuristic with an objective that considers tardiness of AVB streams leads to AVB streams meeting their deadlines.
In contrast to that, the GRASP heuristic with other objectives not considering AVB streams results in schedules with late AVB streams.
Even better results are obtained when a routing with load balancing is used before scheduling, which also decreases overall runtime. 
They report that AVB traffic does not benefit from minimizing the number of queues for TT streams.
However, this is rather obvious, as their system model assumes that AVB streams already have an assignment to AVB queues.
Solving time of the GRASP  heuristic increases considerably when AVB streams are taken into account.
A short preview of these results was previously published in \cite{GaPo18} and the implementation details of the heuristic are elaborated in \cite{RaPo17b}.

Berisa \textit{et al.} \cite{BeZh22} propose a heuristic for the joint routing and scheduling of TT streams in the presence of AVB streams.
They make use of frame preemption to increase the schedulability of AVB streams.
To that end, they present a worst-case response-time analysis of AVB streams with preemption.
The heuristic is based on prior works by Gavrilu{\c{t}} \textit{et al.} \cite{GaZh18}.
Their evaluations demonstrate that the schedulability of AVB streams can be increased by allowing frame preemption while the runtime of the heuristic also increases significantly in this case.

Alnajin \textit{et al.} \cite{AlSa19} give a QoS-aware routing algorithm for TSN streams with respect to various metrics.
They present four scheduling heuristics combined with these routings.
They compare these algorithms regarding the number of guard bands needed in the resulting schedules.
Their evaluations show that their heuristics can reduce the number of guard bands significantly.
They note that reducing the number of guard bands is beneficial for BE traffic.

Li \textit{et al.} \cite{LiJi22} present a heuristic for joint routing and scheduling to eliminate non-deterministic queuing delay in networks with mixed time-critical traffic.
Their scheduling algorithm divides the bandwidth resources into time slots and assigns streams to these slots such that transmission conflicts cannot arise. 
Similar to \cite{OlCr18}, the maximum length of the resulting GCLs is bounded instead of being computed by a post-processing from transmission offsets. 
The solving time with the heuristic is compared to \cite{AtBa20}, \cite{FaDu18}, and \cite{ArHi20}.
Solving time is the only inspected metric as the presented algorithm, and the three compared approaches have no objective functions.
In the presented scenarios, all three methods were outperformed by multiple orders of magnitude. 
The authors report schedules for 4000 streams with just 12 GCL entries per egress port on average.  

Yang \textit{et al.} \cite{YaWe22} use deep reinforcement learning for the joint routing and scheduling problem.
Additionally, they take AVB and BE streams into account.
The authors elaborate on the details of their machine learning model and present evaluations about the learning phase.
They introduce three baseline approaches also based on machine learning for comparison.
The proposed model results in slightly lower average latencies for all traffic classes in the evaluation scenario.

\subsubsection{Multicast}
This section highlights research works specifically concerned with multicast streams in a joint routing and scheduling approach.

The joint routing and scheduling model from \cite{ScDa17} is extended for multicast support in \cite{ScTi20}.
The authors state that while this extension is trivial for pure scheduling models, joint routing and scheduling with multicast is more complicated as additional constraints for the routing are needed.
Various pre-processing steps are presented to reduce the solution space and thus solving time.
The authors report that the time to find a feasible solution was reduced by up to 82.4\% while the overall solving time was reduced by up to 47.6\%. 

Another approach for joint routing and scheduling with multicast streams is presented by Li \textit{et al.} \cite{LiZh21}.
Similar to \cite{ScTi20}, the authors use pre-processing to simplify solving of the model.
The streams are divided into groups by spectral clustering based on their properties.
Similar to \cite{AtBa20} and \cite{RePo20}, these groups are routed and scheduled one after another such that previously computed schedules and routes are fixed.
The authors report that random clustering result in slightly longer flowspans. 
As in similar incremental approaches, reduced overall solving times and increased schedulability are reported.

Yu \textit{et al.} \cite{YuGu20} propose an incremental approach with ILPs.
In contrast to \cite{ScTi20}, they route and schedule multicast streams one by one.
Multiple queues per egress port and queue assignment are also integrated in their model.
Additionally, they propose a pre-processing scheme which aims to simplify the topology.
The pre-processing merges cliques in the topology to a single link.
If routing and schedule can be computed, both are modified for the original graph.
Otherwise the conflicting links are expanded and routing and scheduling are repeated.
Compared to \cite{St10} with a Steiner tree as fixed routing, the proposed approach can schedule significantly more instances.

A biology-inspired algorithm is given by Pahlevan \textit{et al.} \cite{PaOb18}.
They construct a genetic algorithm for joint routing and scheduling which also comprises features like multicast streams and dependencies between streams.
The authors state multicast capabilities as one of their main contributions, but consider multicast streams simply as multiple unicast streams.
In contrast to \cite{CrOl16}, \cite{OlCr18}, and \cite{GaPo18}, only a single queue per egress port is dedicated to TT traffic.
While their evaluations indicate that solving time increases compared to scheduling with a fixed routing, they show that schedulability increases by joint routing and scheduling. 

In later works, Pahlevan \textit{et al.} \cite{PaTa19} present a heuristic list scheduling algorithm for the same purpose.
They model queuing and multicast streams in the same way as in \cite{PaOb18}.
Their evaluations again demonstrate that joint routing and scheduling increases schedulability.

\subsubsection{Reconfiguration}
Reconfiguration of streams can benefit from modifying not only a schedule, but also the respective routing.
Newly added streams can be routed over paths with low utilization.
This section compiles the literature about reconfiguration in joint routing and scheduling apporaches.

Research work from Syed \textit{et al.} \cite{SyAy20} deals with joint routing and scheduling in in-vehicle networks. 
They propose an ILP model for static streams which takes load balancing into account.
They compute schedules and routings for static streams such that as many streams as possible can later be added dynamically.

Following their work in \cite{SyAy20}, Syed \textit{et al.} present multiple heuristics for the dynamic scheduling of new streams in an existing schedule \cite{SyAy21}. 
Their heuristics are based on modelling the scheduling problem as a vector bin packing problem.
They evaluate the time needed for adding new streams in an automotive use case, as reconfiguration in such scenarios has strict timing requirements. 

The authors of \cite{SyAy21b} construct multiple heuristics for dynamic scheduling and routing with reliability constraints. 
All heuristics are based on the same idea as in \cite{SyAy21}.
Every stream is replicated twice when added to an existing schedule to ensure reliability.
The best heuristic is able to schedule 500 streams in about 410$\,$ms.
The authors state this is a reasonable response time in automotive use cases. 

Yu \textit{et al.} \cite{YuWa20} developed a heuristic for online rescheduling in scenarios with virtual machines as communication endpoints.
Virtual machines in a cloud computing environment may be migrated from one physical device to another such that schedules and routings must be updated.
Additionally, all streams are multicast streams, which complicates rescheduling after a VM migration.
Therefore, the multicast tree for a stream is computed such that the maximal distance from any possible device where a VM could run to any destination is minimized. 
The authors state that this will reduce conflicts when a VM is migrated, as the new paths are short.
Given a schedule and a stream that is migrated, a greedy heuristic computes the new schedule based on the precomputed multicast tree.
The authors compare their proposed routing heuristic with an optimal routing obtained by an ILP.
Solving times are significantly reduced, while the routing objective grows only slightly compared to the optimal routing.
Schedulability in case of a migration is considerably increased in comparison to the same scheduling heuristic used with a routing computed by the KMB algorithm \cite{KMB}.

Li \textit{et al.} \cite{LiXi22} consider the reconfiguration of routing, scheduling, and mapping of applications to end stations in case of permanent end station failure.
They extend the ILP of \cite{NaDu18} to schedule applications to end stations for global reconfiguration.
As the resulting ILP instances are hard to solve, they propose a heuristic routing and mapping algorithm as alternative.
The results of this heuristic are fixed in the ILP model, such that only a schedule is computed.
The heuristic approach is able to reconfigure almost all instances, while the ILP times out for most of them in their evaluations.
While both algorithms have exponential runtime in the number of streams, the heuristic is two orders of magnitude faster for the considered instances.

An incremental approach which schedules streams one by one is presented in \cite{HuWa21}. 
In contrast to \cite{YuGu20}, the computed schedules are constrained to no-wait scheduling, i.e., no queuing delays are allowed. 
The authors compare the proposed approach with \cite{ScDa17} and \cite{AtBa20} with respect to schedulability and show that schedulability is slightly increased.
The proposed pre-processing for the routing approach gives only minor improvements regarding schedulability.
The authors report that 97.5\% of the streams in an instance with 2000 streams were scheduled in less than 10 seconds per stream.
They state this is fast enough for online scenarios.

\subsubsection{Other Topics}
This sections summarizes research works with unique topics that do not fit well into the previous groups.

The authors of \cite{ZhSa22} propose a heuristic model to schedule streams in the presence of frame preemption similar to \cite{PaSa19}.
Additionally, they also include route computation in their algorithm.
They present an SMT model for this purpose and use it in an incremental approach, similar to \cite{MaAm18}\cite{SmGl17}\cite{MaAm21}.
The presented results show that scheduling time not only increases with the network size and the number of streams, but also with the maximum number of allowed preemptions and retransmissions.
However, allowing more preemptions increases schedulability only to some instance-specific threshold.

Gavrilu{\c{t}} \textit{et al.} \cite{GaZa17} give multiple algorithms to simultaneously compute scheduling and routing of TT streams.
In contrast to \cite{AtBa20}, \cite{RePo20}, and \cite{FaDu18}, these algorithms additionally generate the network topology with minimized costs.  
They present a problem-specific heuristic, a GRASP heuristic, and a CP approach, and compare them to each other regarding solving time and solution quality.
Their optimization objective captures worst-case end-to-end delays as well as topology costs, i.e., costs for links and bridges which are selected from a library.
Redundant copies of streams are included for reliability considerations.
Their evaluations focus on a comparison of the three presented algorithms.
As expected, the CP approach does not scale well.
The GRASP heuristic finds better solutions in minutes compared to the CP approach in two days.

An SMT model which includes scheduling, routing, and queue assignment of streams simultaneously is presented by \cite{XuXu20}.
The authors state that saving bandwidth by not using the same GCL cycle for all egress ports is also novel to their approach.
However, this is not true as other works even schedule GCL closing events, e.g. \cite{StCr18}, \cite{JiXi20}.
They propose to minimize the number of bridges used by scheduled traffic in order to maximize utilization.
In comparison to the list scheduler of \cite{PaTa19}, schedulability is increased while the solving time approaches the timeout after 40\,h for fairly small instances.

Zhou \textit{et al.} \cite{ZhWa21} construct a heuristic which allows different routes for frames of the same stream to enable load balancing.
The required mechanism is implemented by an SDN architecture.  
The scheduling procedure is a mix of evolutionary algorithm and greedy algorithm.
Multiple variations of the heuristic are compared in the evaluations.
In the presented scenarios, scheduling time increased linearly with the number of streams.

Another incremental scheme for scheduling and routing is presented by Mahfouzi \textit{et al.} \cite{MaAm18}\cite{MaAm21}.
The authors investigate the stability of control applications, i.e., latency and jitter of messages sent by these applications.
Instead of grouping streams by some conflict measure, they divide the network period in slices and group the streams by the time slice in which their transmission can start.
They allow routing only over some fixed number of precomputed shortest paths per source-destination pair, similar to \cite{GaZh18} and the AVB routing in \cite{PoRa16}.
Their model allows unrestricted queuing without further discussion of this topic.
Their evaluations indicate that the number of allowed paths has a huge impact to solving time.
However, they note that three paths per pair of nodes may be sufficient, as schedulability is over 90\% in this case.
They conclude that the search space can be considerably reduced with negligible impact on schedulability. 

Yang \textit{et al.} \cite{YaGo22} present a network architecture for industrial use cases based on TSN hardware and software-defined networking (SDN).
They focus on so-called chain flows.
Chain flows consist of multiple stream which are joined at one or more nodes.
For instance, an industrial controller may join streams from multiple sensors and forward a single stream to an actuator. 
The authors propose a tabu search heuristic and an ILP for the scheduling of chain flows.
They report benefits in scalability and schedulability in comparison to handling every stream of a chain flow individually.

Chain flows are further investigated by Gong \textit{et al.} \cite{GoYa23}.
They propose a heuristic time-tabling algorithm combined with a tabu search for schedule reconfiguration when the network topology is changed.
In contrast to the magazine article \cite{YaGo22}, they elaborate on the details of these algorithms.
However, the reported results are consistent with the results in \cite{YaGo22}.

Hellmanns \textit{et al.} \cite{HeHa21} focus their work not on the ability to compute schedules, but analyze how input pre-processings and solver configuration influence the scalability of solving a joint routing and scheduling ILP model.
They categorize optimizations by whether they are input pre-processing, e.g., topology reduction, model generation related, e.g., tighter variable bounds, or solver configurations, e.g. the use of value hints for variables. 
They give an ILP without any optimizations as baseline for their evaluations.
Different combinations of the proposed optimizations are tested on the same set of problem instances and compared with respect to scalability and schedulability.
Their evaluations indicate that solving time can greatly benefit from input pre-processings, but the effects of model generation optimizations and solver configurations are negligible.
Some of the optimizations even increase solving time.
However, queuing is not supported by their base model. 
Thus, the observations only hold for the no-wait case without queuing delay.
It is not clear whether these results can be transferred to other problem extensions from the literature.

Bhattacharjee \textit{et al.} \cite{BhAl22} propose two algorithms for the placement of talker applications in a network.
Additionally, both approaches also solve the joint routing and scheduling problem.
The first algorithm is an ILP for the placement of talker applications that is combined with the GRASP heuristic of \cite{GaZh18}.
The second algorithm is a simulated annealing (SA) heuristic which computes the placement of talker applications, the schedule, and the routing for a given problem instance.
The evaluations show that both algorithms behave approximately similar with respect to load balancing and solving times.
However, the authors report considerably reduces stream latencies with the SA heuristic.
\section{Comparison and Discussion}
\label{sec:comparison}
We compare the presented research work from Section \ref{sec:literature_overview}.
First, we compile modelling assumptions and problem extensions.
Second, we present common scheduling objectives.
Then, we investigate problem instances used for evaluations.
Finally, we summarize results regarding the scalability of the presented approaches.

\subsection{Modelling Assumptions and Problem Extensions}

\begin{table*}[ht]
    \centering
    \begin{tabular}{|M{2.8cm}|M{1cm}|M{1cm}|M{1cm}|M{1.6cm}|M{2cm}|M{1.2cm}|M{1.3cm}|M{1.5cm}|}
    \hline
        Research work & AVB & BE & Queuing & Fixed GCL length & Reconfiguration & Reliability & Multicast & Task Scheduling \\\hline\hline
    Ansah \textit{et al.} \cite{AnAb19} & \xmark & \xmark & \cmark & \xmark & \xmark & \xmark & \xmark & \xmark \\\hline
Barzegaran \textit{et al.} \cite{BaPo21} & \xmark & \xmark & (\cmark) & \xmark & \xmark & \xmark & \cmark & \cmark \\\hline
Barzegaran \textit{et al.} \cite{BaRe22}& \xmark & (\cmark) & \cmark & \cmark & \xmark & \xmark & \xmark & \xmark \\\hline
Barzegaran \textit{et al.} \cite{BaZa20} & \xmark & \xmark & (\cmark) & \xmark & \xmark & \xmark & \xmark & \cmark \\\hline
Bujosa \textit{et al.} \cite{BuAs22} & \xmark & \xmark & \cmark & \xmark & \xmark & \xmark & \cmark & \xmark \\\hline
Chaine \textit{et al.} \cite{CaBo22} & \xmark & \xmark & (\cmark) & \xmark & \xmark & \xmark & \xmark & \xmark \\\hline
Craciunas \textit{et al.} \cite{CrOl16}  & \xmark& \xmark  & (\cmark) & \xmark & \xmark & \xmark & \xmark & \xmark \\\hline
Craciunas \textit{et al.} \cite{CrOl21} & \xmark & \xmark & (\cmark) & \xmark & \xmark & \cmark & \xmark & \xmark \\\hline
Dai \textit{et al.} \cite{DaWa21} & \xmark & (\cmark) & \xmark & \xmark & \xmark & \xmark & \xmark & \xmark \\\hline
Dobrin \textit{et al.} \cite{DoDe19} & \xmark & \xmark & \cmark & \xmark & \xmark & \cmark & \xmark & \xmark \\\hline
Duerr \textit{et al.} \cite{DuNa16}  & \xmark & \cmark & \xmark & \xmark & \xmark & \xmark & \xmark & \xmark \\\hline
Farzaneh \textit{et al.} \cite{FaKu17}  & \xmark & \xmark  & \xmark & \xmark & \xmark & \xmark  & \xmark & \xmark \\\hline
Feng \textit{et al.} \cite{FeCa21} & \cmark & \cmark & (\cmark) & \xmark & \xmark & \cmark & \xmark & \xmark \\\hline
Feng \textit{et al.} \cite{FeDe22} & \xmark & \cmark & \cmark & \xmark & \xmark & \cmark & \xmark & \xmark \\\hline
Feng \textit{et al.} \cite{FeYa21} & \xmark & \xmark & (\cmark) & \xmark & \xmark & \xmark & \cmark & \cmark \\\hline
Gärtner \textit{et al.} \cite{GaRi22}\cite{GaRi23} & \xmark & \xmark & (\cmark) & \xmark & \cmark & \xmark & \xmark & \xmark \\\hline
Gavrilu{\c{t}} \textit{et al.} \cite{GaPo18}  & \cmark & \xmark  & (\cmark) & \xmark & \xmark & \xmark & \xmark & \xmark\\\hline
Ginthör \textit{et al.} \cite{GiGi20} & \xmark & \cmark & \cmark & \cmark & \xmark & \xmark & \xmark & \xmark \\\hline
Hellmanns \textit{et al.} \cite{HeGl20}  & \xmark &  (\cmark) & (\cmark) & \xmark & \xmark & \xmark  & \xmark & \xmark \\\hline
Houtan \textit{et al.} \cite{HoAs21}  & \xmark & \cmark & (\cmark) & \xmark & \xmark & \xmark & \xmark & \xmark \\\hline
Huang \textit{et al.} \cite{HuZh22}  & \cmark &  \xmark & (\cmark) & \xmark & \xmark & \cmark  & \xmark & \xmark \\\hline
Jin \textit{et al.} \cite{JiXi19} & \xmark & \xmark & \xmark & \xmark & \xmark & \xmark & \xmark & \xmark \\\hline
Jin \textit{et al.} \cite{JiXi20}  & \xmark & (\cmark) & (\cmark) & \cmark & \xmark & \xmark & \xmark & \xmark \\\hline
Kim \textit{et al.} \cite{KiCh21} & \xmark & (\cmark) & \cmark & \xmark & \xmark & \xmark & \xmark & \xmark \\\hline
Kim \textit{et al.} \cite{KiLe21}\cite{KiLe22} & \xmark & (\cmark) & \cmark & \xmark & \xmark & \xmark & \xmark & \xmark \\\hline
Lin \textit{et al.} \cite{LiLi22} & \xmark & \xmark & \xmark & \xmark & \cmark & \xmark & \xmark & \xmark \\\hline
Min \textit{et al.} \cite{MiOh22} & \xmark & \xmark & \xmark & \xmark & \xmark & \xmark & \xmark & \xmark \\\hline
Oliver \textit{et al.} \cite{OlCr18}  & \xmark & (\cmark)  & (\cmark) & \cmark & \xmark &  \xmark & \cmark & \xmark \\\hline
Pang \textit{et al.} \cite{PaHu21} & \xmark & \xmark & \cmark & \xmark & \cmark & \xmark & \cmark & \xmark \\\hline
Park \textit{et al.} \cite{PaSa19} & \xmark & \xmark & \cmark & \xmark & \xmark & \cmark & \xmark & \xmark \\\hline
Pei \textit{et al.} \cite{PeHu22} & \xmark & \cmark & \cmark & \xmark & \xmark & \xmark & \xmark & \xmark \\\hline
Pop \textit{et al.} \cite{PoRa16}  & \cmark & \xmark & (\cmark) & \xmark & \xmark & \xmark & Only AVB & \xmark \\\hline
Raagaard \textit{et al.} \cite{RaPo17}  & \xmark & \xmark  & (\cmark) & \xmark & \cmark & \xmark & \xmark & \xmark \\\hline
Reusch \textit{et al.} \cite{ReZh20}  & \xmark & (\cmark) & \cmark & \cmark & \xmark & \xmark  & \xmark & \xmark \\\hline
Santos \textit{et al.} \cite{SaCa19} & \xmark & \cmark & \cmark & \cmark & \xmark & \xmark & \cmark & \xmark \\\hline
Steiner \cite{St10}  & \xmark & \xmark & \cmark & \xmark & \xmark & \xmark & \cmark & \xmark \\\hline
Steiner \textit{et al.} \cite{StCr18}  & \xmark & (\cmark)  & (\cmark) & \cmark & \xmark & \xmark & \cmark & \xmark \\\hline
Vlk \textit{et al.} \cite{VlBr22} & \xmark & \xmark & (\cmark) & \xmark & \xmark & \xmark & \xmark & \xmark \\\hline
Vlk \textit{et al.} \cite{VlHa20} & \xmark & \xmark & \cmark & \xmark & \xmark & \xmark & \xmark & \xmark \\\hline
Wang \textit{et al.} \cite{WaYa22} & \xmark & (\cmark) & \xmark & \xmark & \xmark & \xmark & \xmark & \xmark \\\hline
Wang \textit{et al.} \cite{WaZh22}  & \xmark & \xmark  & \cmark & \cmark & \xmark & \xmark  & \xmark & \xmark\\\hline
Wang \textit{et al.} \cite{WaWa22} & \xmark & \xmark  & \cmark & \xmark & \cmark & \xmark  & \xmark & \xmark\\\hline
Wang \textit{et al.} \cite{WaXu22} & \cmark & \cmark & \xmark& \xmark& \xmark& \xmark& \xmark& \xmark \\\hline
Yao \textit{et al.} \cite{YaGa22} & \xmark & \cmark  & \cmark & \cmark & \xmark & \xmark  & \xmark & \xmark\\\hline
Zhang \textit{et al.} \cite{ZhXu22} & \xmark & \xmark & \xmark & \xmark & \xmark & \xmark & \cmark & \xmark \\\hline
Zhou \textit{et al.} \cite{ZhLi20} & \xmark & \xmark & (\cmark) & \xmark & \xmark & \xmark & \xmark & \xmark \\\hline
    \end{tabular}
    \caption{Overview of considered problem extensions and restrictions in the literature of scheduling approaches with a fixed routing.  }
    \label{tab:features}
\end{table*}

\begin{table*}[ht]
    \centering
    \begin{tabular}{|M{2.8cm}|M{1cm}|M{1cm}|M{1cm}|M{1.6cm}|M{2cm}|M{1.2cm}|M{1.3cm}|M{1.5cm}|}
    \hline
        Research work & AVB & BE & Queuing & Fixed GCL length & Reconfiguration & Reliability & Multicast & Task Scheduling \\\hline\hline
    Alnajim \textit{et al.} \cite{AlSa19} & \xmark & \cmark  & \cmark & \xmark & \xmark & \xmark & \xmark & \xmark \\\hline
Arestova \textit{et al.} \cite{ArHi20} & \xmark & \xmark & (\cmark) & \xmark & \xmark & \xmark & \xmark & \xmark \\\hline
Atallah \textit{et al.} \cite{AtBa18} & \xmark & \xmark  & \xmark & \xmark & \xmark & \cmark & \xmark & \xmark \\\hline
Atallah \textit{et al.} \cite{AtBa20} & \xmark & \xmark  & (\cmark) & \xmark & \xmark & \cmark & \xmark & \xmark \\\hline
Berisa \textit{et al.} \cite{BeZh22} & \cmark & \xmark & \cmark & \xmark & \xmark & \xmark & \xmark & \xmark \\\hline
Bhattacharjee \textit{et al.} \cite{BhAl22} & \xmark & \xmark & (\cmark) & \xmark & \xmark & \xmark & \xmark & \xmark \\\hline
Falk \textit{et al.} \cite{FaDu18}  & \xmark & \xmark  & \xmark & \xmark & \xmark & \xmark & \xmark & \xmark \\\hline
Falk \textit{et al.} \cite{FaDu20}  & \xmark &  \xmark & \xmark & \xmark & \xmark & \xmark  & \xmark & \xmark \\\hline
Gavrilu{\c{t}} \textit{et al.} \cite{GaZa17}  & \xmark & \xmark & \xmark & \xmark & \xmark & \cmark & \cmark & \xmark \\\hline
Gavrilu{\c{t}} \textit{et al.} \cite{GaZh18}  & \cmark & \xmark & (\cmark) & \xmark & \xmark &  \xmark & \xmark & \xmark \\\hline
Gong \textit{et al.} \cite{GoYa23} &  \xmark & \xmark  & \cmark & \xmark & \xmark & \xmark  & \xmark & \xmark\\\hline
He \textit{et al.} \cite{HeZu23} &  \xmark & \xmark & \cmark & \xmark & \xmark &  \xmark & \xmark & \xmark \\\hline
Huang \textit{et al.} \cite{HuWa21} & \xmark & \xmark & \xmark & \xmark & \cmark & \xmark & \xmark & \xmark \\\hline
Kentis \textit{et al.} \cite{KeBe17}  & \xmark & \xmark & \cmark & \xmark & \xmark & \xmark & \xmark & \xmark \\\hline
Li \textit{et al.} \cite{LiCh21} & \xmark & \xmark & \cmark & \xmark & \xmark & \cmark & \cmark & \xmark \\\hline
Li \textit{et al.} \cite{LiJi22} & \xmark & \cmark & \cmark & \cmark & \xmark & \xmark & \xmark & \xmark \\\hline
Li \textit{et al.} \cite{LiLi20} & \xmark & & \xmark  & \xmark & \xmark & \xmark & \xmark & \xmark \\\hline
Li \textit{et al.} \cite{LiXi22} & \xmark & \xmark & \xmark & \xmark & \cmark & \xmark & \xmark & \xmark \\\hline
Li \textit{et al.} \cite{LiZh21} & \xmark & \xmark & (\cmark) & \xmark & \xmark & \xmark & \cmark & \xmark \\\hline
Mahfouzi \textit{et al.} \cite{MaAm18}\cite{MaAm21} & \xmark & \xmark & \cmark & \xmark & \xmark & \xmark & \xmark & \cmark \\\hline
Nie \textit{et al.} \cite{NiLi22} & \xmark & \xmark & \xmark & \xmark & \xmark & \xmark & \xmark & \xmark \\\hline
Pahlevan \textit{et al.} \cite{PaOb18}  & \xmark &  (\cmark)  & \xmark & \xmark & \xmark & \xmark & \xmark & \cmark \\\hline
Pahlevan \textit{et al.} \cite{PaTa19}  & \xmark &  (\cmark)  & \xmark & \xmark & \xmark & \xmark & \xmark & \cmark \\\hline
Pozo \textit{et al.} \cite{PoRo18}  & \xmark & \xmark  & \xmark & \xmark & \xmark &  \cmark & \cmark & \xmark \\\hline
Reusch \textit{et al.} \cite{ReCr22} & \xmark & \xmark & (\cmark) &\xmark & \xmark & \cmark & \cmark & \cmark \\\hline
Reusch \textit{et al.} \cite{RePo20} & \xmark & \xmark  & \xmark & \xmark & \xmark & \cmark & \xmark & \cmark \\\hline
Schweissguth \textit{et al.} \cite{ScDa17}  & \xmark & \xmark & \xmark & \xmark & \xmark & \xmark & \xmark & \xmark \\\hline
Schweissguth \textit{et al.} \cite{ScTi20} & \xmark & \xmark & \xmark & \xmark & \xmark & \xmark & \cmark & \xmark \\\hline
Smirnov \textit{et al.} \cite{SmGl17}  & \xmark & \cmark & \xmark & \xmark & \xmark & \xmark & \cmark & \xmark \\\hline
Syed \textit{et al.} \cite{SyAy20} & \xmark & \xmark  & \xmark & \xmark & \xmark & \xmark & \xmark & \xmark\\\hline
Syed \textit{et al.} \cite{SyAy21b}  & \xmark & \xmark & \xmark & \xmark & \cmark & \cmark & \xmark & \xmark\\\hline
Syed \textit{et al.} \cite{SyAy21}  & \xmark & \xmark  & \xmark & \xmark & \cmark & \xmark & \xmark & \xmark\\\hline
Syed \textit{et al.} \cite{SyAy22}  & \xmark & \xmark  & \xmark & \xmark & \xmark & \cmark & \xmark & \xmark\\\hline
Syed \textit{et al.} \cite{SyAy23}  & \xmark & \xmark  & \xmark & \xmark & \xmark & \cmark & \xmark & \xmark\\\hline
Vlk \textit{et al.} \cite{VlHa21b} & \xmark & \xmark & (\cmark) & \xmark & \xmark & \xmark & \xmark & \xmark \\\hline
Wang \textit{et al.} \cite{WaCh20}  & \xmark & \xmark  & \xmark & \xmark & \xmark & \xmark  & \xmark & \xmark\\\hline
Xu \textit{et al.} \cite{XuXu20} & \xmark & \xmark  &(\cmark) & \cmark & \xmark & \xmark & \xmark & \xmark \\\hline
Xu \textit{et al.} \cite{XuXu22} & \xmark & \xmark  &(\cmark) & \xmark & \xmark & \xmark & \xmark & \xmark \\\hline
Yang \textit{et al.} \cite{YaGo22} &  \xmark & \xmark  & \cmark & \xmark & \xmark & \xmark  & \xmark & \xmark\\\hline
Yang \textit{et al.} \cite{YaWe22} & \cmark & \cmark  &\cmark & \xmark & \xmark & \xmark & \xmark & \xmark \\\hline
Yu \textit{et al.} \cite{YuGu20} & \xmark & \xmark  & (\cmark) & \cmark & \xmark & \xmark & \cmark & \xmark \\\hline
Yu \textit{et al.} \cite{YuWa20} & \xmark & \xmark & (\cmark) & \xmark & \cmark & \xmark & \cmark & \xmark \\\hline
Zhou \textit{et al.} \cite{ZhSa21b} & \xmark & \xmark & \xmark & \xmark & \xmark & \cmark & \xmark & \xmark \\\hline
Zhou \textit{et al.} \cite{ZhSa21} & \xmark & \xmark  & \xmark & \xmark & \xmark & \cmark & \xmark & \xmark \\\hline
Zhou \textit{et al.} \cite{ZhSa22} & \xmark & \xmark  & \cmark & \xmark & \xmark & \xmark & \xmark & \xmark \\\hline
Zhou \textit{et al.} \cite{ZhWa21} & \xmark & \xmark & (\cmark) & \xmark & \xmark & \xmark & \xmark & \xmark \\\hline
 \end{tabular}
    \caption{Overview of considered problem extensions and restrictions in the literature of joint routing approaches. }
    \label{tab:features_routing}
\end{table*}

Table \ref{tab:features} compiles important modelling assumptions and problem extensions in the surveyed research works with fixed routings.
Table \ref{tab:features_routing} shows the same information for research works about the joint routing problem.
In the following section, we compile the contributions to each of these topics.

\subsubsection{Other Traffic}
Only five works examine TT and AVB streams simultaneously.
Pop \textit{et al.} \cite{PoRa16} present a GRASP heuristic for the handling of AVB streams.
The heuristic gets a schedule of TT streams as input and cannot change it.
The authors of \cite{GaPo18} present a short preview of AVB-aware scheduling, which was later extended in \cite{GaZh18}.
Feng \textit{et al.} \cite{FeYa21} consider the bandwidth available to AVB and BE traffic in their approach as they consider repeated frame loss which may result in starvation.
Berisa \textit{et al.} \cite{BeZh22} use frame preemption and a worst-case end-to-end delay analysis to increase the schedulability of AVB streams.
Huang \textit{et al.} \cite{HuZh22} give parameters for AVB streams in an in-vehicle network and include them in their evaluation scenario.
Wang \textit{et al.} \cite{WaXu22} consider the joint handling of AVB and TT streams.
AVB streams are shaped by CQF.
Their objective aims to reduce the influence of non-periodic BE traffic to AVB streams.
Other research works handle BE traffic by minimizing the flowspan which yields a large time slot at the end of a schedule exclusively for other traffic, e.g., \cite{DuNa16}, \cite{HeGl20}, and \cite{HoAs21}.
Pei \textit{et al.} \cite{PeHu22} evaluate their approach in the presence of BE traffic and rate constrained traffic.
Yao \textit{et al.} \cite{YaGa22} consider the joint scheduling of periodic stream without real-time requirements in their approach.
The tables indicate works that do not mention BE traffic, but use objective functions that are beneficial to other traffic or that limit the number of guard bands with (\cmark).
For instance, Oliver \textit{et al.} \cite{OlCr18} limit the number of guard bands indirectly by introducing a fixed number of transmission time windows per egress port.

\subsubsection{Queuing}
Queuing is a controversial topic in the TSN scheduling literature as non-determinism, e.g., frame loss, may cause serious problems.
Some research works do not allow queuing at all, e.g., \cite{DuNa16}, \cite{AtBa20}, and \cite{FaDu18}.
The majority of the algorithms in the literature uses frame isolation introduces by \cite{CrOl16}.
These works are indicated by (\cmark) in the Tables \ref{tab:features} and \ref{tab:features_routing}.
Vlk \textit{et al.} \cite{VlHa20} discuss the effects of isolation constraints.
They report results indicating that isolation constraints reduce schedulability significantly. 
However, they propose a different solution to deal with non-determinism, effectively modifying the TAS.
Thus, their results are not applicable to current TSN implementations.
There are some research works which allow unrestricted queuing, e.g., \cite{SaCa19} and \cite{DoDe19}.
Most of these works do not elaborate on the consequences of unrestricted queuing.
In contrast to that, Reusch \textit{et al.} \cite{ReZh20}, Barzegaran \textit{et al.} \cite{BaRe22}, and Berisa \textit{et al.} \cite{BeZh22} introduced countermeasures for the mentioned consequences.
The authors include a worst-case end-to-end delay analysis in their algorithms such that even in the case of non-determinism, deadlines are met.

\subsubsection{Fixed GCL Length}
Most algorithms presented in the literature handle the generation of GCLs indirectly.
They schedule transmission offsets of frames at end stations and intermediate bridges.
The GCLs are generated by a post-processing after scheduling.
This step is only mentioned, and the respective authors do not elaborate on it.
Examples for such works are \cite{CrOl16}, \cite{VlBr22}, and \cite{ReCr22}.
However, computing GCLs by a post-processing comes with two drawbacks.
First, GCLs have limited size in bridges.
Thus, GCLs obtained by a post-processing may not be deployable.
Second, the scheduling algorithm cannot include considerations for guard bands.
There are some exceptions to this in the literature. 
Jin \textit{et al.} \cite{JiXi20} present a heuristic to compute schedules with a limited number of GCL entries per egress port.
Santos \textit{et al.} \cite{SaCa19} give a detailed SMT model for TSN scheduling which includes the explicit representation of GCLs.
Yao \textit{et al.} \cite{YaGa22} limit the number of GCL entries and the maximum queue size in their heuristic, i.e., schedules which do not meet these constraints are not considered valid solutions. 
Some works limit the number of GCL entries indirectly by introducing transmission windows for egress ports.
Streams are mapped to these transmission windows and their number is fixed before scheduling.
Examples of such works are \cite{OlCr18}, \cite{StCr18}, and \cite{ReZh20}.
All schedules for no-wait scheduling can be deployed with a fixed number of GCL entries.
As no queuing delay is allowed, frames cannot be scheduled to wait at closed gates.
Such a schedule can be deployed by opening all gates for TT traffic at the start of a hyperperiod and never closing them.

\subsubsection{Reconfiguration}
In some scenarios it may be infeasible to compute new schedules every time a stream should be integrated into or removed from an existing schedule.
For instance, automotive scenarios may include ad-hoc connections between cars and infrastructure.
Computing new schedules every time a new stream is added may take too much time, even with heuristic algorithms.
Syed \textit{et al.} \cite{SyAy21} \cite{SyAy21b} consider reconfiguration in such automotive scenarios.
Additionally, they also present preliminary work about computing schedules suitable for later reconfiguration in \cite{SyAy20}.
Raagaard \textit{et al.} \cite{RaPo17} present a heuristic to add streams to an existing schedule.
When the heuristic fails, they assign the new stream to other egress queues which were unused before.
Pang \textit{et al.} \cite{PaHu21} present work about deploying an updated schedule to a network already executing another schedule.
Their approach allows updating the schedule without frame loss or new streams interfering with the old schedule.
Another use case for reconfiguration is the reallocation of tasks sending and receiving TT streams.
Yu \textit{et al.} \cite{YuWa20} use virtual machines as end stations in their model.
These virtual machines may be migrated from one physical device to another, which requires updating schedules and routings. 
A similar example for reconfiguration is presented in \cite{LiXi22}.
The authors propose an approach for updating a schedule in case of a permanent end station failure.
Schedules and routings must be updated in this case as in \cite{YuWa20}.
Gärtner \textit{et al.} \cite{GaRi22} introduce a measure for schedule flexibility which also considers deadlines.
They use this measure to update schedules in a beneficial way for future updates.
Lin \textit{et al.} \cite{LiLi22} show in their evaluations that aligning frame transmissions to the greatest common divisor results in schedules that are suitable for adding streams later.

\subsubsection{Reliability}

\begin{table}[hb]
    \centering
    \begin{tabular}{|c|p{4cm}|}
    \hline
        Fault Model & Research work \\\hline\hline
        Permanent link failure & Atallah \textit{et al.} \cite{AtBa18}, Atallah \textit{et al.} \cite{AtBa20}, Gavrilu{\c{t}} \textit{et al.} \cite{GaZa17}, Pozo \textit{et al.} \cite{PoRo18}, Reusch \textit{et al.} \cite{ReCr22}, Reusch \textit{et al.} \cite{RePo20}, Syed \textit{et al.} \cite{SyAy21b}, Syed \textit{et al.} \cite{SyAy23} \\\hline
        Frame loss & Atallah \textit{et al.} \cite{AtBa18}, Atallah \textit{et al.} \cite{AtBa20}, Dobrin \textit{et al.} \cite{DoDe19}, Feng \textit{et al.} \cite{FeCa21} \cite{FeDe22}, Gavrilu{\c{t}} \textit{et al.} \cite{GaZa17}, Huang \textit{et al.} \cite{HuZh22}, Li \textit{et al.} \cite{LiCh21}, Park \textit{et al.} \cite{PaSa19}, Reusch \textit{et al.} \cite{ReCr22}, Reusch \textit{et al.} \cite{RePo20}, Syed \textit{et al.} \cite{SyAy21b}, Syed \textit{et al.} \cite{SyAy22}, Syed \textit{et al.} \cite{SyAy23}, Zhou \textit{et al.} \cite{ZhSa21} \\\hline
        Clock drift & Craciunas \textit{et al.} \cite{CrOl21} \\\hline
        Hardware bugs & Zhou \textit{et al.} \cite{ZhSa21b} \\\hline
    \end{tabular}
    \caption{Fault models in research works dedicated to reliability.}
    \label{tab:reliability}
\end{table}

Table \ref{tab:reliability} compiles fault models used in the literature.
The listed research works construct schedules robust in the respective fault model.
Computing schedules robust against frame loss is the most common kind of reliability in the TSN scheduling literature.
Park \textit{et al.} \cite{PaHu21} handle frame loss by allowing the retransmission of frames.
Schedules for such a scenario must schedule enough time between frame transmissions such that retransmissions do not interfere with other frames.
Another way to deal with frame loss is proposed by Feng \textit{et al.} \cite{FeCa21}\cite{FeDe22}. 
In contrast to \cite{PaHu21}, they do not use retransmissions, but they schedule redundant copies of the same stream over the same path.
Zhou \textit{et al.} \cite{ZhSa21} approximate the probability of frame loss in a joint routing and scheduling model. 
Redundant copies of streams are routed over not necessarily disjoint paths to reduce the probability of frame loss. 
Robustness against permanent single link failures are also covered in several works.
There are two approaches in the TSN scheduling literature to handles such failures.
First, redundant copies of streams are scheduled and routed over link-disjoint paths before a link failure arises.
Examples for such works are \cite{AtBa18}, \cite{AtBa20}, \cite{GaZa17}, and \cite{SyAy21b}.
Huang \textit{et al.} \cite{HuZh22} and Syed \textit{et al.} \cite{SyAy22} use Frame Replication and Elimination for Reliability \cite{802.1CB} to implement this approach.
Second, streams can be rescheduled and rerouted after a link failure occurred. 
Pozo \textit{et al.} \cite{PoRo18} present a heuristic for fast rescheduling and rerouting in this case.
We remark that all works about computing schedules robust against permanent link failures are also robust against frame loss.
Both countermeasures against link failures are also effective against frame loss.
Another kind of reliability is considered in \cite{CrOl21}.
The authors compute schedules robust against clock drift, i.e., clocks of different devices running not with the same speed. 
They introduce gaps between frame transmissions such that the maximum possible clock drift does not affect other frame transmissions.
Unknown hardware bugs or deviations from TSN standards are treated by \cite{ZhSa21b}.
The proposed algorithm selects expensive bridges with higher certification for paths of streams with higher safety requirements.
In contrast to all other mentioned works with reliability, Syed \textit{et al.} \cite{SyAy23} use an encoding scheme to reconstruct lost frames.
The XORed data of two frames is transmitted over a path disjoint to the paths of both frames.

\subsubsection{Multicast}
Every algorithm in literature can be used for multicast streams, as a multicast stream can be substituted by a set of unicast streams.
However, the number of streams negatively affects the solving time for a problem instance. 
Tables \ref{tab:features} and \ref{tab:features_routing} indicate multicast support only for works which include some considerations for the efficient integration of multicast streams without introducing a set of new streams.
Most such works handle multicast streams by scheduling only a single frame per link, regardless of the number of consecutive links in the multicast tree of the respective stream.
Examples for such works are \cite{OlCr18}, \cite{StCr18}, \cite{SaCa19}, \cite{FeYa21}, \cite{BaPo21}.
An analysis of the joint routing and scheduling problem with multicast streams is presented in \cite{ScTi20}.
Yu \textit{et al.} \cite{YuWa20} compute routings and schedules for multicast streams such that migrating a virtual machine sending or receiving TT streams can be done easily.
Some works allow multicast streams, but do not elaborate on the implementation details, e.g., \cite{ZhXu22}.

\subsubsection{Task Scheduling}
Only a few research works are concerned with the joint scheduling of streams and tasks.
Some of them have integrated dependencies between streams and tasks, i.e., tasks can only be scheduled after some stream has arrived.
Such works are presented in \cite{FeYa21}, \cite{BaZa20}, \cite{PaOb18} and \cite{BaPo21}.
Other works focus on the scheduling of tasks which produce TT streams while considering safety and security considerations.
Preliminary results for this scenario are presented in \cite{RePo20} and extended in \cite{ReCr22}.

\begin{table}[ht]
    \centering
    \begin{tabular}{|M{2.5cm}|M{5.0cm}|}
    \hline
        Objective Category & Research work \\\hline\hline
        No objective & Alnajim \textit{et al.} \cite{AlSa19}, Ansah \textit{et al.} \cite{AnAb19}, Atallah \textit{et al.} \cite{AtBa18}, Atallah \textit{et al.} \cite{AtBa20}, Bujosa \textit{et al.} \cite{BuAs22}, Dai \textit{et al.} \cite{DaWa21}, Dobrin \textit{et al.} \cite{DoDe19}, Falk \textit{et al.} \cite{FaDu20}, Farzaneh \textit{et al.} \cite{FaKu17}, He \textit{et al.} \cite{HeZu23}, Kim \textit{et al.} \cite{KiCh21}, Li \textit{et al.} \cite{LiJi22}, Li \textit{et al.} \cite{LiLi20}, Mahfouzi \textit{et al.} \cite{MaAm18}, Mahfouzi \textit{et al.} \cite{MaAm21}, Raagaard \textit{et al.} \cite{RaPo17} Falk \textit{et al.} \cite{FaDu18}, Santos \textit{et al.} \cite{SaCa19}, Steiner \cite{St10}, Steiner \textit{et al.} \cite{StCr18}, Syed \textit{et al.} \cite{SyAy21} \cite{SyAy21b} \cite{SyAy23}, Vlk \textit{et al.} \cite{VlBr22}, Wang \textit{et al.} \cite{WaWa22}, Wang \textit{et al.} \cite{WaZh22}, Xu \textit{et al.} \cite{XuXu22}, Yao \textit{et al.} \cite{YaGa22}, Zhou \textit{et al.} \cite{ZhLi20}, Zhou \textit{et al.} \cite{ZhSa21}, Zhou \textit{et al.} \cite{ZhSa22} \\\hline
        Latency and Jitter & Arestova \textit{et al.} \cite{ArHi20}, Barzegaran \textit{et al.} \cite{BaPo21}, Barzegaran \textit{et al.} \cite{BaZa20}, \cite{PaTa19}, Dürr \textit{et al.} \cite{DuNa16}, Feng \textit{et al.} \cite{FeCa21}, Gärtner \textit{et al.} \cite{GaRi22}\cite{GaRi23}, Hellmanns \textit{et al.} \cite{HeGl20}, Houtan \textit{et al.} \cite{HoAs21}, Huang \textit{et al.} \cite{HuWa21}, Huang \textit{et al.} \cite{HuZh22}, Kim \textit{et al.} \cite{KiLe21}\cite{KiLe22}, Nie \textit{et al.} \cite{NiLi22}, Oliver \textit{et al.} \cite{OlCr18}, Pahlevan \textit{et al.} \cite{PaOb18}, Pang \textit{et al.} \cite{PaHu21}, Pei \textit{et al.} \cite{PeHu22}, Schweissguth \textit{et al.} \cite{ScDa17} \cite{ScTi20}, Vlk \textit{et al.} \cite{VlHa20}, Wang \textit{et al.} \cite{WaYa22}, Yang \textit{et al.} \cite{YaWe22}, Zhang \textit{et al.} \cite{ZhXu22}, Zhou \textit{et al.} \cite{ZhWa21} \\\hline
        Queuing  & Craciunas \textit{et al.} \cite{CrOl16}, Feng \textit{et al.} \cite{FeDe22}, Gavrilu{\c{t}} \textit{et al.} \cite{GaPo18}, Pop \textit{et al.} \cite{PoRa16}, Vlk \textit{et al.} \cite{VlHa20}, Vlk \textit{et al.} \cite{VlHa21b} \\\hline
        Other Traffic & Barzegaran \textit{et al.} \cite{BaRe22}, Berisa \textit{et al.} \cite{BeZh22}, Gavrilu{\c{t}} \textit{et al.} \cite{GaPo18} \cite{GaZh18}, Houtan \textit{et al.} \cite{HoAs21}, Reusch \textit{et al.} \cite{ReZh20}, Smirnov \textit{et al.} \cite{SmGl17}, Wang \textit{et al.} \cite{WaXu22} \\\hline
        Routing & Li \textit{et al.} \cite{LiCh21}, Li \textit{et al.} \cite{LiXi22}, Li \textit{et al.} \cite{LiZh21}, Schweissguth \textit{et al.} \cite{ScTi20}, Wang \textit{et al.} \cite{WaCh20}, Yu \textit{et al.} \cite{YuGu20}, Yu \textit{et al.} \cite{YuWa20} \\\hline
        Topology Synthesis & Gavrilu{\c{t}} \textit{et al.} \cite{GaZa17}, Reusch \textit{et al.} \cite{ReCr22}, Xu \textit{et al.} \cite{XuXu20}, Zhou \textit{et al.} \cite{ZhSa21b}\\\hline
        Reliability & Craciunas \textit{et al.} \cite{CrOl21}, Park \textit{et al.} \cite{PaSa19}, Pozo \textit{et al.} \cite{PoRo18} \\\hline
        GCL Synthesis & Kentis \textit{et al.} \cite{KeBe17}, Jin \textit{et al.} \cite{JiXi20} \\\hline
        Other &  Bhattacharjee \textit{et al.} \cite{BhAl22}, Chaine \textit{et al.} \cite{CaBo22}, Feng \textit{et al.} \cite{FeYa21}, Ginthör \textit{et al.} \cite{GiGi20}, Gong \textit{et al.} \cite{GoYa23}, Jin \textit{et al.} \cite{JiXi19}, Lin \textit{et al.} \cite{LiLi22}, Min \textit{et al.} \cite{MiOh22}, Syed \textit{et al.} \cite{SyAy20}, Syed \textit{et al.} \cite{SyAy22}, Yang \textit{et al.} \cite{YaGo22}, Yang \textit{et al.} \cite{YaGo22} \\\hline
    \end{tabular}
    \caption{Categorization of research works based on optimization objectives. }
    \label{tab:objectives}
\end{table}

\subsection{Scheduling Objective}
\label{sec:objectives}
Objective functions are used to measure the quality of solutions and to compare them.
We discuss common objectives from the literature and classify research works by their objective.
Table \ref{tab:objectives} shows which research work features which kind of objective.

\subsubsection{No Objective}
Many research works have no scheduling objective and only try to find some schedule which fulfills all constraints, e.g., \cite{RaPo17}, \cite{FaDu18}, \cite{HeZu23} and \cite{FaDu20}.
We note that many SMT approaches feature no objective \cite{St10}\cite{StCr18}\cite{SaCa19}\cite{FaKu17}\cite{MaAm18}\cite{MaAm21}.
In contrast to ILP solving, SMT solvers were not originally designed for optimization.
Therefore, many SMT approaches focus on finding a feasible schedule.

\subsubsection{Latency and Jitter}
TSN and the TAS were designed for traffic with hard real-time requirements.
Therefore, latency and jitter of streams are interesting properties of schedules.
Objective functions including them are the most common kind of objectives in TSN schedule optimization.
Oliver \textit{et al.} \cite{OlCr18} and Barzegaran \textit{et al.} \cite{BaZa20} minimize the per-stream jitter.
Minimizing the flowspan, i.e., the time all TT stream arrive at their destination, is a common objective.
Examples of approaches using this objective include \cite{DuNa16}, \cite{HeGl20}, \cite{WaYa22}, \cite{HoAs21}, \cite{HuZh22}, \cite{PaTa19}, \cite{ArHi20}, and \cite{PaOb18}.
A related but different objective is the minimization of end-to-end delays of TT streams \cite{ScDa17}\cite{VlHa20}\cite{PeHu22}\cite{PaHu21}\cite{GaRi22}\cite{YaWe22}\cite{ScTi20}.
Kim \textit{et al.} \cite{KiLe21}\cite{KiLe22} minimize multiple objectives weighted by constant factors.
They take end-to-end delays, jitter, and bandwidth occupation into account.
Barzegaran \textit{et al.} \cite{BaPo21} use a combination of jitter and end-to-end latency as measure of schedule quality.
Nie \textit{et al.} \cite{NiLi22} minimize end-to-end latency and transmission offsets simultaneously.
We remark that these objectives are not competing, as opposed to most multi-criterion problems. 
Minimization of transmission offsets is also pursued by \cite{HuWa21} and \cite{ZhXu22} which is related but not equal to flowspan or end-to-end latency minimization.
Zhou \textit{et al.} \cite{ZhWa21} use a combination of jitter, end-to-end delays, number of scheduled streams, and link utilization.

\subsubsection{Queuing}
Research works which apply isolation constraints for queuing often use more than one queue per egress port for scheduled traffic.
In that way, they are able to schedule more streams as isolation only concern streams in the same egress queue.
The assignment of streams to egress queues per egress port is a degree of freedom in the respective scheduling problems.
Therefore, they try to minimize the number of queues reserved for TT streams per egress port, as the remaining queues are available for other traffic.
Examples for such works include \cite{CrOl16}, \cite{VlHa20}, \cite{PoRa16}, \cite{FeDe22}, \cite{VlHa21b}, and \cite{GaPo18}.

\subsubsection{Other Traffic}
The schedule of TT streams has an influence on the Quality of Service for other traffic classes.
Current approaches for scheduling in TSN focus on AVB traffic and BE traffic.
For the joint scheduling of TT and AVB streams, Gavrilu{\c{t}} \textit{et al.} \cite{GaZh18}\cite{GaPo18} minimize the tardiness of AVB streams as their deadlines are considered to be not strict.
Another objective related to AVB streams is used in \cite{BeZh22}.
The presented heuristic has the objective to schedule as many AVB streams as possible.
Yang \textit{et al.} \cite{YaWe22} minimize the the weighted sum of stream latencies of scheduled traffic, AVB, and BE streams.
Wand \textit{et al.} \cite{WaXu22} use CQF to shape AVB streams in their approach.
The objective function aims to load balance the AVB frame transmissions between the time slots of the CQF mechanism. 
This reduces the probability that non-periodic BE traffic overloads such a slot.
The authors of \cite{ReZh20} minimize the occupation percentage of egress ports, i.e., the percentage of a hyperperiod with no active transmission window for TT traffic.
The rational of this is that low occupation corresponds to long and frequent time intervals available to other traffic.
A similar objective is used in \cite{BaRe22} as the authors minimize the average bandwidth occupied by transmission windows for scheduled traffic.
Smirnov \textit{et al.} \cite{SmGl17} use a multi-criterion objective for joint routing and scheduling.
They reduce the influence of scheduled traffic to other traffic, and simultaneously minimize the number of GCL entries needed to deploy a schedule.
The work in \cite{HoAs21} focuses on comparing the influence of different objective functions to the QoS of BE traffic.
They propose minimization and maximization of frame offsets, hoping that grouping frames together increases the QoS.
Additionally, they also suggest two objectives which maximize the gaps between consecutive frame transmissions on a link.
They assume that starvation of other traffic classes is reduced in this way.

\subsubsection{Routing}
Research works about joint routing and scheduling often consider the quality of the routing in their objective.
All of them have in common that the length of the paths is minimized.
This is reasonable as longer paths correspond to higher link utilizations, end-to-end latencies, and harder scheduling instances.
Schweissguth \textit{et al.} \cite{ScTi20} propose a multi-objective optimization for joint routing and scheduling.
First, routing and schedule with minimized path lengths are computed.
The obtained path lengths are used as maximum path lengths per stream in a second run.
The second run minimizes end-to-end latencies.
The joint routing approach of \cite{WaCh20} minimizes the number of links in the routing.
Li \textit{et al.} \cite{LiXi22} simultaneously minimize the path lengths and a measure for scheduling conflicts of streams routed over the same link. 
Yu \textit{et al.} \cite{YuGu20} schedule and route streams one after another. 
They minimize a weighted sum of the number of links used for the currently scheduled stream, and the bandwidth utilization.
Li \textit{et al.} \cite{LiZh21} simultaneously minimize path lengths in the routing, and the flowspan. 
Yu \textit{et al.} \cite{YuWa20} consider the migration of sources of TT streams. 
They minimize the maximum distance from all possible source nodes of a stream to all destination nodes in a multicast tree.
Li \textit{et al.} \cite{LiCh21} maximize the number of streams which are scheduled and routed, and also try to minimize the maximum link load as a secondary objective.

\subsubsection{Topology Synthesis}
In addition to joint routing and scheduling, some works also construct the network topology.
TSN bridges are expensive, and thus such objectives always include costs for bridges.
Gavrilu{\c{t}} \textit{et al.} \cite{GaZa17} minimize multiple objectives weighted by constant factors.
The first objective is the tardiness of TT streams to guide their GRASP heuristic to solutions with no deadline misses.
The second objective is topology costs.
A similar objective is used in \cite{ReCr22}.
The weighted sum of routing and schedule costs is minimized.
Routing costs constitute of overlap penalties for redundant paths and path lengths.
Schedule costs constitute of punishments for not schedulable streams and stream latencies.
Another approach which minimizes topology costs is proposed in \cite{ZhSa21b}.
Selecting bridges from a library is part of the presented problem, which imposes costs for bridges and additional costs when multiple vendors are used.
Xu \textit{et al.} \cite{XuXu20} minimize the number of bridges needed to schedule and route all streams such that the utilization is maximized.

\subsubsection{Reliability}
Reliability requirements can be ensured by constraining the set of feasible solutions.
However, some works choose to maximize reliability for their respective fault model.
Pozo \textit{et al.} \cite{PoRo18} maximize the idle times of links and frames, as such schedules are easier to repair upon link failure.
Craciunas \textit{et al.} \cite{CrOl21} maximize the allowed out-of-sync clock drift to cope with synchronization problems and maximize robustness against clock drift.
Park et. al. \cite{PaSa19} maximize the number of times a frame can be retransmitted without missing its deadline, as they include preemption in their model. 

\subsubsection{GCL Synthesis}
TSN bridges do not have an unlimited number of GCL entries per egress port.
The minimization of GCL entries is considered by \cite{JiXi20}. 
The reason for this is that the authors propose an incremental approach and the overall number of needed GCL entries is not known in advance.
Reducing the number of gate events also reduces the number of guard bands which is beneficial for BE traffic.
Kentis \textit{et al.} \cite{KeBe17} minimize the GCL schedule duration.
However, it is not clear why schedule duration matters, as the limiting factor in TSN hardware is the number of GCL entries.

\subsubsection{Others}
Some research works use a problem specific objective not comparable to other works.
We present them for the sake of completeness.
The authors of \cite{JiXi19} minimize the number of frames as they propose a joint approach for scheduling and message fragmentation.
Syed \textit{et al.} \cite{SyAy20} and \cite{SyAy22} use a modelling specific objective which is related to load balancing of ports in an in-vehicle architecture with one central processing unit.
Ginthör \textit{et al.} \cite{GiGi20} minimize the wasted bandwidth for different link layer technologies, i.e., Ethernet and 5G links.
Feng \textit{et al.} \cite{FeYa21} minimize the response time of tasks which may be dependent on streams as they consider the joint scheduling of streams and tasks.
Chaine \textit{et al.} \cite{Ca18} maximize the length of transmission time windows of streams at their respective talkers such that latency and jitter requirements are met.
This is the only work in TSN scheduling which employs a quadratic objective function.
Bhattacharjee \textit{et al.} \cite{BhAl22} employ a multi-criterion objective.
Their first objective is to minimize the maximum load across all servers as the considered problem includes the placement of talker applications.
The second objective is to minimize the average hop count of all streams.
Lin \textit{et al.} \cite{LiLi22} present a heuristic for incremental scheduling that aims to maximize the probability that more streams can be added later.
This is required in industrial use cases as turning off machines to deploy a new schedule may be expensive.
Yang \textit{et al.} \cite{YaGo22} state that they maximize the number of scheduled streams while minimizing the link occupancy rate. 
However, this rate is not defined in the published magazine article.
Gong \textit{et al.} \cite{GoYa23} also maximize the occupancy rate while minimizing the maximum link utilization.
They define the occupancy rate as the fraction of bandwidth reserved for scheduled traffic actually used for transmissions.
Min \textit{et al.} \cite{MiOh22} compare metaheuristics by maximizing the number of scheduled stream for the same problem instance.

\begin{table}[htbp]
\scriptsize
    \centering
    \begin{tabular}{|M{2.3cm}|M{1.3cm}|M{0.7cm}|M{0.8cm}|M{1.2cm}|}
    \hline
     Research work & Topology & ES & Bridges & TT Streams \\\hline\hline
    Barzegaran \textit{et al.} \cite{BaPo21} & Various & 6 -- 20 & 2 -- 20 & 8 -- 27  \\\hline
Barzegaran \textit{et al.} \cite{BaRe22} & Various & 3 -- 31 & 2 -- 15 & 7 -- 137 \\\hline
Barzegaran \textit{et al.} \cite{BaZa20} & Various & 5 -- 20 & 2 -- 20 & 8 -- 27  \\\hline
Bujosa \textit{et al.} \cite{BuAs22} & Line & 3 -- 9 & 1 -- 3 & N/A \\\hline
Chaine \textit{et al.} \cite{CaBo22} & Line, spacecraft & 2 -- 31 & 4 -- 15 & 15 -- 304 \\\hline
Craciunas \textit{et al.} \cite{CrOl16}  & N/A & 3--7 & 1--5 & 5--1000 \\\hline
Craciunas \textit{et al.} \cite{CrOl21} & Tree & 16 & 7 & 96 \\\hline
Dai \textit{et al.} \cite{DaWa21} & Link & N/A & N/A & 3  \\\hline
Dürr \textit{et al.} \cite{DuNa16}  & ER, RRG, BA & 24--100 & 5--20 & 30--1500  \\\hline
Farzaneh \textit{et al.} \cite{FaKu17}  & Snowflake & 12 & 2 & 14--100 \\\hline
Feng \textit{et al.} \cite{FeCa21} & N/A & 2 -- 8 & 2 -- 5 & 2 -- 16 \\\hline
Feng \textit{et al.} \cite{FeDe22} & N/A & 5 -- 8 & 3 -- 5 & N/A \\\hline
Feng \textit{et al.} \cite{FeYa21} & N/A & 50 & 10 & 11 \\\hline
Gärtner \textit{et al.} \cite{GaRi22}\cite{GaRi23} & Line, machine & 2 -- 54 & 2 -- 15 & 1 -- 104 \\\hline
Gavrilu{\c{t}} \textit{et al.} \cite{GaPo18}  & Star, Snowflake & 3--32 & 1--18 & 4--35 \\\hline
Ginthör \textit{et al.} \cite{GiGi20} & N/A & N/A & N/A & 10  \\\hline
Hellmanns \textit{et al.} \cite{HeGl20}  & Ring of rings & \multicolumn{2}{c|}{100--2500} & 10--2000 \\\hline
Hellmanns \textit{et al.} \cite{HeHa21} & Ring of rings & N/A & N/A & 50--500  \\\hline
Houtan \textit{et al.} \cite{HoAs21}  & Snowflake & 6 & 2 & 10 \\\hline
Huang \textit{et al.} \cite{HuZh22} & Automotive & 33 & 14 & 20 -- 480 \\\hline
Jin \textit{et al.} \cite{JiXi19} & N/A & 5--30 & 5--30 & 10--60 \\\hline
Jin \textit{et al.} \cite{JiXi20}  & N/A & N/A & 6--20 & 10--10000 \\\hline
Kim \textit{et al.} \cite{KiCh21} & Various & 7 & 5 & 6 \\\hline
Kim \textit{et al.} \cite{KiLe21}\cite{KiLe22} & Automotive & 17 & 4 & 27 \\\hline
Li \textit{et al.} \cite{LiLi20} & Various & 30 -- 100 & 7 -- 20 & 3 -- 100  \\\hline
Lin \textit{et al.} \cite{LiLi22} & Tree, mesh, ring & 4 -- 16 & 1 -- 4 & $\sim 48$ -- $223$ \\\hline
Min \textit{et al.} \cite{MiOh22} & Real-world & \multicolumn{2}{c|}{14} & $300$ \\\hline
Oliver \textit{et al.} \cite{OlCr18}  & Line & 50 & 10 & 10--50 \\\hline
Pang \textit{et al.} \cite{PaHu21} & In-train, spacecraft & 31 -- 54 & 13 -- 31 & N/A \\\hline
Park \textit{et al.} \cite{PaSa19} & Automotive & 5 -- 20 & 3 -- 7 & 100 -- 500 \\\hline
Pei \textit{et al.} \cite{PeHu22} & Spacecraft & 31 & 15 & 20 \\\hline
Pop \textit{et al.} \cite{PoRa16}  & N/A & 3--5 & 1--2 & 3--5 \\\hline
Raagaard \textit{et al.} \cite{RaPo17}  & N/A & \multicolumn{2}{c|}{4--402} & 15--290 \\\hline
Reusch \textit{et al.} \cite{RePo20} & Various & 4--32 & 1--16 & N/A \\\hline
Reusch \textit{et al.} \cite{ReZh20}  & Snowflake & 3--256 & 1--146 & 14--316 \\\hline
Santos \textit{et al.} \cite{SaCa19} & N/A & 50 & 10 & 1 -- 10 \\\hline
Steiner \cite{St10}  & Star, tree, snowflake & N/A & N/A & 100 -- 1000  \\\hline
Steiner \textit{et al.} \cite{StCr18}  & N/A & 50 & 10 & 10--50  \\\hline
Vlk \textit{et al.} \cite{VlBr22} & Ring of lines, ring of trees, spacecraft & $\sim$ 20 -- 1700 & $\sim 20$ -- 300 & 1500 -- 12000 \\\hline
Vlk \textit{et al.} \cite{VlHa20} & Mesh, ring, tree & 4 -- 16 & 1 -- 4 & 6 -- 450 \\\hline
Wang \textit{et al.} \cite{WaYa22} & N/A & 4 -- 10 & 3 -- 9 & 10 -- 100 \\\hline
Wang \textit{et al.} \cite{WaZh22} & Automotive & 4 & 1 & 3 \\\hline
Wang \textit{et al.} \cite{WaWa22} & Mesh & 15 -- 16 & 6 -- 8 & N/A \\\hline
Wang \textit{et al.} \cite{WaXu22} & Line, ring, tree & N/A & 5 -- 200 & N/A \\\hline
Yao \textit{et al.} \cite{YaGa22} & N/A & N/A & N/A & 10 -- 20 \\\hline
Zhou \textit{et al.} \cite{ZhLi20} & Automotive & 5 & 16 & N/A \\\hline
Zhang \textit{et al.} \cite{ZhXu22} & Tree, line, tree & \multicolumn{2}{c|}{6 -- 8} & 200 -- 1200 \\\hline
    \end{tabular}
    \caption{Overview of investigated problem instances in the literature of the scheduling problem with fixed routing.}
    \label{tab:instances}
\end{table}

\begin{table}[htbp]
\scriptsize
    \centering
    \begin{tabular}{|M{2.3cm}|M{1.3cm}|M{0.7cm}|M{0.8cm}|M{1.2cm}|}
    \hline
     Research work & Topology & ES & Bridges & TT Streams \\\hline\hline
    Alnajim \textit{et al.} \cite{AlSa19} & N/A & 30--150 & 10--50 & 100--1500\\\hline
Arestova \textit{et al.} \cite{ArHi20} & N/A & 50 & 10 & 10 -- 100 \\\hline
Atallah \textit{et al.} \cite{AtBa18} & N/A & 6 -- 24 & N/A & 30 -- 600   \\\hline
Atallah \textit{et al.} \cite{AtBa20} & N/A & N/A & 3--21 & 20--60   \\\hline
Berisa \textit{et al.} \cite{BeZh22} & Ring, full mesh, spacecraft & 13 -- 31 & 4 -- 15 & 222 \\\hline
Bhattacharjee \textit{et al.} \cite{BhAl22} & Ring, BA, ER & 46 -- 1024 & 24 -- 512 & 90 -- 1845 \\\hline
Falk \textit{et al.} \cite{FaDu18}  & Line, ring, BA, ER & \multicolumn{2}{c|}{5--36} & 2--30\\\hline
Falk \textit{et al.} \cite{FaDu20} & Ring w/ $k$ neighbors & \multicolumn{2}{c|}{50--400} & 50--150 \\\hline
Gavrilu{\c{t}} \textit{et al.} \cite{GaZa17}  & N/A & 4--20 & N/A & 4--38  \\\hline
Gavrilu{\c{t}} \textit{et al.} \cite{GaZh18}  & Various & 3--256 & 2--146 & 4--427 \\\hline
Gong \textit{et al.} \cite{GoYa23} & N/A & N/A & N/A & 15 -- 180 \\\hline
He \textit{et al.} \cite{HeZu23} & ER, RRG, BA & \multicolumn{2}{c|}{20} & 25 -- 200 \\\hline
Huang \textit{et al.} \cite{HuWa21} & Spacecraft, ER & N/A & N/A & 500 -- 2000 \\\hline 
Kentis \textit{et al.} \cite{KeBe17}  & Ring, mesh & 12 & 12  & 20--52 \\\hline
Li \textit{et al.} \cite{LiCh21} & ER, RRG, BA & 10 & 10 & 10 -- 100 \\\hline
Li \textit{et al.} \cite{LiJi22} & Mesh, automotive & 15 -- 105 & 3 -- 21 & 2 -- 4000 \\\hline
Li \textit{et al.} \cite{LiXi22} & Real-world & 31 & 13 & 100 -- 300 \\\hline
Li \textit{et al.} \cite{LiZh21} & N/A & 39 & 16 & 10 -- 40 \\\hline
Mahfouzi \textit{et al.} \cite{MaAm18}\cite{MaAm21}  & ER, Automotive & 20 & 10--45 & 19--106 \\\hline
Nie \textit{et al.} \cite{NiLi22} & Ring, mesh & 11 -- 14 & 4 -- 15 & 10 -- 40 \\\hline
Pahlevan \textit{et al.} \cite{PaOb18}  & Grid, ring & 27--45 & 9 & 30--40  \\\hline
Pahlevan \textit{et al.} \cite{PaTa19}  & Grid, ring & 50 & 10 & 60--100 \\\hline
Pozo \textit{et al.} \cite{PoRo18}  & Synthetic & 6--8 & 3--8 & 10--50 \\\hline
Reusch \textit{et al.} \cite{ReCr22} & Various & 4 -- 128 & 2--64 & 2 -- 144 \\\hline
Reusch \textit{et al.} \cite{RePo20} & Various & 4--32 & 1--16 & N/A \\\hline
Schweissguth \textit{et al.} \cite{ScDa17}  & Ring, mesh & 13 & N/A  & 60 \\\hline
Schweissguth \textit{et al.} \cite{ScTi20} & Ring, mesh & 12 & 12 & 25 -- 40 \\\hline
Smirnov \textit{et al.} \cite{SmGl17}  & N/A & N/A & N/A & 5 -- 75 \\\hline
Syed \textit{et al.} \cite{SyAy20} & Automotive & N/A & N/A & 20--90 \\\hline
Syed \textit{et al.} \cite{SyAy21}  & Automotive & N/A & N/A & 100--500 \\\hline
Syed \textit{et al.} \cite{SyAy21b}  & Automotive & N/A & N/A & 100--500 \\\hline
Syed \textit{et al.} \cite{SyAy22}  & Automotive & N/A & N/A & 20--90 \\\hline
Syed \textit{et al.} \cite{SyAy23}  & Automotive & N/A & N/A & 112 \\\hline
Vlk \textit{et al.} \cite{VlHa21b} & Ring, mesh & 12 -- 48 & 12 -- 48 & 10 -- 300\\\hline
Wang \textit{et al.} \cite{WaCh20}  & N/A & 25 & 5 & 10--100 \\\hline
Xu \textit{et al.} \cite{XuXu20} & Mesh & 5 & 1 -- 23 & 4 -- 12  \\\hline
Xu \textit{et al.} \cite{XuXu22} & Mesh, ring & N/A & 7--15 & 40 -- 80 \\\hline
Yang \textit{et al.} \cite{YaGo22} & N/A & 2 & N/A & 45 -- 180 \\\hline
Yang \textit{et al.} \cite{YaWe22} & Star, tree, BA & N/A & 8 -- 18 & 150 -- 900 \\\hline
Yu \textit{et al.} \cite{YuGu20} & Mesh & 72 & 24 & 1 -- 500  \\\hline
Yu \textit{et al.} \cite{YuWa20} & BA & 72 & 24 & 10 -- 350 \\\hline
Zhou \textit{et al.} \cite{ZhSa21b} & ER, automotive & N/A & 6--10 & 50 -- 240 \\\hline
Zhou \textit{et al.} \cite{ZhSa21} & ER, automotive & 16 & 6--10 & 50--380\\\hline
Zhou \textit{et al.} \cite{ZhSa22} & ER, real-world & 16 & 4--64 & 30 -- 220 \\\hline
Zhou \textit{et al.} \cite{ZhWa21} & Mesh & $\geq 4$ & 3 -- 15 & 50 -- 650 \\\hline
    \end{tabular}
    \caption{Overview of investigated problem instances in the literature of the joint routing problem.}
    \label{tab:instances-routing}
\end{table}

\subsection{Problem Instances}
Before we describe evaluation results, we describe the problem instances used for evaluations in the literature.
We present an overview of used network topologies, network sizes, and numbers of streams.
Tables \ref{tab:instances} and \ref{tab:instances-routing} compile this information about the problem instances used for evaluations with fixed routing and joint routing, respectively.
Unfortunately, some research works do not elaborate on the used topologies, which makes assessing and comparing the results to other works harder. 
Most research works only use synthetic test cases. 
Ring topologies are commonly used in evaluations, e.g., in \cite{ScDa17}, \cite{HeGl20}, \cite{WaXu22}, \cite{FaDu18}, \cite{NiLi22}, \cite{PaTa19}, \cite{XuXu20}, \cite{PaOb18}, \cite{FaDu20}, \cite{VlHa21b}, and \cite{ScTi20}.
Hellmanns \textit{et al.} \cite{HeGl20} argue that rings are a common topology in real-world industrial facilities.
Other systematic topologies used include line \cite{OlCr18}\cite{WaXu22}\cite{FaDu18}, grid \cite{PaTa19}\cite{PaOb18}, and snowflake-like \cite{ReZh20}\cite{HoAs21}\cite{FaKu17}\cite{GaPo18} networks. 
Various randomly generated topologies are also used in evaluations.
Erdős–Rényi graphs (ER) are the most common ones \cite{DuNa16}\cite{FaDu18}\cite{HeZu23}\cite{ZhSa21}\cite{ZhSa21b}\cite{ZhSa22}\cite{LiCh21}, but Barabási-Albert graphs (BA) \cite{DuNa16}\cite{FaDu18}\cite{HeZu23}\cite{LiCh21}\cite{YuGu20} and random regular graphs (RRG) \cite{DuNa16}\cite{HeZu23}\cite{LiCh21} are also used.
A few research works features evaluations with real-world topologies. 
Syed \textit{et al.} \cite{SyAy22}\cite{SyAy20}\cite{SyAy23}\cite{SyAy21}\cite{SyAy21b} use a real-world automotive architecture for their evaluations. 
A large automotive architecture including stream parameters is discussed in \cite{HuZh22}.
Other automotive architectures are used by Kim \textit{et al.} \cite{KiLe21}\cite{KiLe22}, Li \textit{et al.} \cite{LiJi22}, Mahfouzi \textit{et al.} \cite{MaAm18}\cite{MaAm21}, and Wang \textit{et al.} \cite{WaZh22}.
Zhou \textit{et al.} \cite{ZhSa21}\cite{ZhSa21b} conclude their evaluations by investigating a real-world example from General Motors. 
The authors of \cite{GaZa17} also use a real-world problem instance from General Motors.
However, this instance is only a set of streams without topology. 
Min \textit{et al.} \cite{MiOh22} use the network topology of the National Science Foundation of the United States of America.
Barzegaran \textit{et al.} \cite{BaRe22} presents evaluations with real-world test cases from General Motors and a real-world spacecraft.
Pang \textit{et al.} \cite{PaHu21} evaluate an algorithm for schedule updates in a real-world in-train network and a spacecraft.
Similarly, the authors of \cite{GaRi22} evaluate their algorithm for schedule reconfigurations with the topology of a not specified machine.
Vlk \textit{et al.} \cite{VlBr22}, Chaine \textit{et al.} \cite{CaBo22}, Huang \textit{et al.} \cite{HuWa21}, and Berisa \textit{et al.} \cite{BeZh22}, perform evaluations with a real-world spacecraft topology.

All research works concerned with synthetic test cases use randomly generated streams.
Sources and destinations of these streams are selected uniformly from the sets of talkers and listeners in the respective topology.
The number of streams varies considerably between different research works.  
It ranges from 2 streams in the smallest instance of Falk \textit{et al.} \cite{FaDu18} to up to 10812 streams in the largest instance of Vlk \textit{et al.} \cite{VlBr22}.
All works, which describe the placement of deadlines, place them at the end of the respective stream's period. 
No research work allows deadlines to be after the end of the hyperperiod a frame was sent.
Stream periods range from $32\,\mu s$ in \cite{JiXi20} to $500\,ms$ in \cite{CrOl16}.
All research works assume transmission rates of either 100\,Mb/s or 1\,Gb/s per egress port.

\subsection{Scalability}
Scheduling in TSN is known to be NP-complete.
Therefore, solving times and sizes of feasible problem instances matter.
Almost all research works about TSN scheduling include or even focus on evaluating the scalability of the respective proposed approach.
These evaluations measure the solving times for selected problem instances.
Tables \ref{tab:runtime-fixed-routing-exact} and \ref{tab:runtime-fixed-routing-heuristic} compile the reported runtimes needed to solve the largest problem instance for which a schedule was found in the respective research work.
Tables \ref{tab:runtime-joint-routing-exact} and \ref{tab:runtime-joint-routing-heuristic} report the same results for research works which feature the joint computation of schedules and routings.
We divided results in separate tables for exact and heuristic algorithms for better comparability. 
In cases where it was not clear which problem instance can be considered as the largest one, we used the number of streams as tie-breaker.
This is justified by several research works surveyed in this paper, e.g., in \cite{FaDu18}.
The tables are meant to show general tendencies and improvements, not to suggest one approach over the other.
Caution is needed when interpreting the tables. 
It shows the reported times after which an algorithm terminated, not the time until a first valid schedule was obtained, as almost all papers do not report this time. 
This is a systematic disadvantage of exact approaches as they only terminate when the optimal solution is found or some timeout is reached, while heuristic algorithms may terminate much earlier with suboptimal solutions.
Some research works deal with more parameters than the size of the network and the number of streams, e.g., Oliver \textit{et al.} \cite{OlCr18} present evaluations about the influence of the number of transmission windows per egress port to scalability.
Other works handle problem extensions, e.g., AVB or task scheduling. 
Approximations are given when results are not stated in the text and had to be estimated by the presented figures.
Ranges are given when multiple instances are considered to be the largest.
We identified two tendencies with respect to solving times.

First, heuristic approaches can handle larger instances than approaches with exact solution methods.
While the number of network nodes is approximately in the same range, heuristic algorithms can schedule problem instances with more streams compared to exact approaches.
Typical numbers of streams in exact approaches are less than 100, e.g., in \cite{ScDa17}, \cite{CrOl21}, and \cite{FaDu18}.
However, there are some notable exceptions.
Craciunas \textit{et al.} \cite{CrOl16} present an incremental scheduling algorithm with backtracking, which scheduled instances with 1000 streams in their evaluations.
Later works present incremental approaches which were able to schedule as many as 2000 streams \cite{HuWa21}.
Oliver \textit{et al.} \cite{OlCr18} assigned streams to transmission windows of egress ports and report solved instances with 750 streams.
Heuristic approaches were able to schedule instances with more than 10000 streams, e.g., \cite{VlBr22} and \cite{JiXi20}.

Second, exact approaches which solve the joint routing and scheduling problem can only handle instances with smaller numbers of nodes compared to approaches solely for scheduling. 
Typical networks in evaluations of joint routing algorithms contain less than 50 nodes \cite{AtBa20}\cite{RePo20}\cite{GaZa17}.
This is due to the solution space growing heavily with an increased number of possible paths per stream.
However, there are approaches able to compute routings and schedules for problem instances with up to 96 nodes \cite{VlHa21b}\cite{YuGu20}.
Most networks in the literature of scheduling with a fixed routing contain less than 96 nodes.
The range of the number of streams is approximately the same for approaches with fixed routing and joint routing.

\begin{table}[ht!]
\scriptsize
    \centering
    \begin{tabular}{|M{2.3cm}|M{1.0cm}|M{0.75cm}|M{0.75cm}|M{1.3cm}|}
    \hline
    Research work & Solution approach & Nodes & TT Streams & Runtime (s) \\\hline\hline
Barzegaran \textit{et al.} \cite{BaRe22} & CP & 120 & 500 & N/A \\\hline
Barzegaran \textit{et al.} \cite{BaZa20} & CP & 40 & 27 & 2563  \\\hline
Chaine \textit{et al.} \cite{CaBo22} & ILP & 46 & 304 & $\sim 240$ \\\hline
Craciunas \textit{et al.} \cite{CrOl16}  & SMT & 12 & 1000 & $< 18000$ \\\hline
Craciunas \textit{et al.} \cite{CrOl21} & SMT & 23 & 96 & $\sim 0.343$ -- $0.437$ \\\hline
Dai \textit{et al.} \cite{DaWa21} & CP & N/A & 3 & N/A  \\\hline
Farzaneh \textit{et al.} \cite{FaKu17}  & SMT & 14 & 100 & $< 240$ \\\hline
Feng \textit{et al.} \cite{FeCa21} & SMT & 13 & 16 & N/A \\\hline
Feng \textit{et al.} \cite{FeDe22} & SMT & 13 & N/A & $\sim 900$ \\\hline
Feng \textit{et al.} \cite{FeYa21} & SMT & 60 & 11 & 384 -- 694 \\\hline
Ginthör \textit{et al.} \cite{GiGi20} & CP & N/A & 10 & N/A  \\\hline
Houtan \textit{et al.} \cite{HoAs21}  & SMT & 8 & 10 & $0.37$--$1153.52$ \\\hline
Jin \textit{et al.} \cite{JiXi19}  & SMT & 4 & 4 & $< 600$ \\\hline
Jin \textit{et al.} \cite{JiXi20}  & SMT & 6 & 50 & $660$--$1080$ \\\hline
Li \textit{et al.} \cite{LiLi20} & SMT & 120 & 100 & $\sim 320$ -- $450$\\\hline
Nie \textit{et al.} \cite{NiLi22} & ILP & 26 & 40 &  $\sim$ 0 -- 10 \\\hline
Oliver \textit{et al.} \cite{OlCr18}  & SMT & 20 & 750 & $\sim 600$ \\\hline
Pang \textit{et al.} \cite{PaHu21} & ILP & 86 & N/A & $\sim$ 620 -- 900 \\\hline
Pop \textit{et al.} \cite{PoRa16}  & ILP & 7 & 5 & $80.19$ \\\hline
Santos \textit{et al.} \cite{SaCa19} & SMT & 60 & 10 & $< 288000 $  \\\hline
Steiner \cite{St10}  & SMT & N/A & 1000 & $\sim 180$ -- $1140$ \\\hline
Steiner \textit{et al.} \cite{StCr18}  & SMT & 60 & 50 & $\sim 0.01$ -- $100$ \\\hline
Vlk \textit{et al.} \cite{VlHa20} & ILP & 20 & 414 & $\leq 600$ \\\hline
Zhou \textit{et al.} \cite{ZhLi20} & SMT & 21 & N/A & N/A \\\hline
    \end{tabular}
    \caption{Overview of solving times of the respective largest reported problem instance for which a schedule was found. Only research works with exact approach and fixed routing are included for comparability.}
    \label{tab:runtime-fixed-routing-exact}
\end{table}  

\begin{table}[ht!]
\scriptsize
    \centering
    \begin{tabular}{|M{2.3cm}|M{1.0cm}|M{0.75cm}|M{0.75cm}|M{1.3cm}|}
    \hline
    Research work & Solution approach & Nodes & TT Streams & Runtime (s) \\\hline\hline
Atallah \textit{et al.} \cite{AtBa18} & Heuristic & 24 & 600 & $\sim 8$ \\\hline
Barzegaran \textit{et al.} \cite{BaPo21} & CP + heuristic & 40 & 27 & 3553 -- 9153  \\\hline
Barzegaran \textit{et al.} \cite{BaZa20} & CP + heuristic & 40 & 27 & 161  \\\hline
Bujosa \textit{et al.} \cite{BuAs22} & Heuristic & 12 & N/A & $< 0.01$ \\\hline
Dürr \textit{et al.} \cite{DuNa16}  & Tabu search & 120 & 1500 & 11520 \\\hline
Gärtner \textit{et al.} \cite{GaRi22}\cite{GaRi23} & Heuristic & 69 & 104 & $\sim 0.1$ \\\hline
Gavrilu{\c{t}} \textit{et al.} \cite{GaPo18}  & GRASP & 50 & 35 & $612.8$ \\\hline
Hellmanns \textit{et al.} \cite{HeGl20} & Tabu search & 2500 & 2000 & $\sim 4400$ \\\hline
Huang \textit{et al.} \cite{HuZh22} & Heuristic & 47 & 480 & $\sim 2700$ \\\hline
Jin \textit{et al.} \cite{JiXi20} & Heuristic & 20 & 10000 & $\sim 5100$ \\\hline
Kim \textit{et al.} \cite{KiCh21} & Heuristic & 12 & 6 & N/A \\\hline
Kim \textit{et al.} \cite{KiLe21}\cite{KiLe22} & GA & 21 & 27 & 12300 \\\hline
Lin \textit{et al.} \cite{LiLi22} & Heuristic & 20 & $\sim 223$ & N/A \\\hline
Park \textit{et al.} \cite{PaSa19} & GA & $<$ 27 & 500 & N/A \\\hline
Pei \textit{et al.} \cite{PeHu22} & Heuristic & 46 & 20 & N/A \\\hline
Raagaard \textit{et al.} \cite{RaPo17}  & Heuristic & 402 & 290 & 20--54 \\\hline
Reusch \textit{et al.} \cite{ReZh20} & Heuristic & 402 & 316 & $10.52$ \\\hline
Vlk \textit{et al.} \cite{VlBr22} & Heuristic & 2000 & 10812 & $\sim 1000$ \\\hline
Wang \textit{et al.} \cite{WaYa22} & Machine learning & 19 & 100 & $\sim 400$ \\\hline
Wang \textit{et al.} \cite{WaZh22} & Heuristic & 5 & 3 & $< 1$ \\\hline
Wang \textit{et al.} \cite{WaWa22} & Heuristic & 23 & N/A & N/A \\\hline
Wang \textit{et al.} \cite{WaXu22} & Heuristic & 16 & 200 & $\sim 1000$ \\\hline
Yao \textit{et al.} \cite{YaGa22} & Heuristic & N/A & 20 & $\sim 2$ \\\hline
Zhang \textit{et al.} \cite{ZhXu22} & Heuristic & 8 & 1200 & $\sim 11$ \\\hline
\end{tabular}
    \caption{Overview of solving times of the respective largest reported problem instance for which a schedule was found. Only research works with heuristic approach and fixed routing are included for comparability.}
    \label{tab:runtime-fixed-routing-heuristic}
\end{table}

\begin{table}[ht!]
\scriptsize
    \centering
    \begin{tabular}{|M{2.3cm}|M{1.0cm}|M{0.75cm}|M{0.75cm}|M{1.3cm}|}
    \hline
    Research work & Solution Approach & Nodes & TT Streams & Runtime (s) \\\hline\hline
    Atallah \textit{et al.} \cite{AtBa20} & ILP & 14 & 60 & $\sim 100$ \\\hline
Falk \textit{et al.} \cite{FaDu18} & ILP & 8 & 30 & $\sim 170$ -- $1580$ \\\hline
Gavrilu{\c{t}} \textit{et al.} \cite{GaZa17} & CP & 20 & 38 & $172800$ \\\hline
Hellmanns \textit{et al.} \cite{HeHa21} & ILP & N/A & N/A & 50--500  \\\hline
Huang \textit{et al.} \cite{HuWa21} & ILP & 44 & 2000 & 1620 \\\hline
Li \textit{et al.} \cite{LiZh21}  & ILP & 55 & 40 & $\sim 80$ \\\hline
Mahfouzi \textit{et al.} \cite{MaAm18}\cite{MaAm21}  & SMT & 65 & 45 & $\sim 21$ \\\hline
Pozo \textit{et al.} \cite{PoRo18}  & ILP & 16 & 50  & 2--85 \\\hline
Reusch \textit{et al.} \cite{ReCr22}  & CP & 48 & 33 & $1500$ \\\hline
Reusch \textit{et al.} \cite{RePo20}  & CP & 48 & N/A & $\sim 1800$ \\\hline
Schweissguth \textit{et al.} \cite{ScDa17}  & ILP & 24 & 52  & $1284.53$ \\\hline
Schweissguth \textit{et al.} \cite{ScTi20} & ILP & 24 & $\sim 46$ -- 60 & N/A \\\hline
Smirnov \textit{et al.} \cite{SmGl17}  & PBO & N/A & 75 & $\sim$ 42 -- 78 \\\hline
Syed \textit{et al.} \cite{SyAy20} & ILP & N/A & 90 & $\sim 21000$ \\\hline
Vlk \textit{et al.} \cite{VlHa21b} & CP & 96 & 300 & $\sim$ 100  \\\hline
Xu \textit{et al.} \cite{XuXu20}  & SMT & 24 & 12 & $\sim 144000$ \\\hline
Xu \textit{et al.} \cite{XuXu22}  & SMT & 12 & 80 & $\sim$ 850 \\\hline
Yu \textit{et al.} \cite{YuGu20}  & ILP & 96 & 450 & N/A \\\hline
Zhou \textit{et al.} \cite{ZhSa21b}  & SMT & 10 & 240 & $\sim 3500$ -- $4000$ \\\hline
Zhou \textit{et al.} \cite{ZhSa21}  & SMT & 10 & 380 & $\sim 2700$ \\\hline
Zhou \textit{et al.} \cite{ZhSa22}  & SMT & 24 & 220 & $\sim$ 55 -- 440 \\\hline
    \end{tabular}
    \caption{Overview of solving times of the respective largest reported problem instance for which a schedule was found. Only research works with exact approach and joint routing are included for comparability.}
    \label{tab:runtime-joint-routing-exact}
\end{table}

    \begin{table}[ht!]
    \scriptsize
    \centering
    \begin{tabular}{|M{2.3cm}|M{1.0cm}|M{0.75cm}|M{0.75cm}|M{1.3cm}|}
    \hline
    Research work & Solution Approach & Nodes & TT Streams & Runtime (s) \\\hline\hline
Alnajim \textit{et al.} \cite{AlSa19} & Heuristic & 200 & 1500 & $2718$ \\\hline
Arestova \textit{et al.} \cite{ArHi20} & Genetic algorithm & 60 & 100 & $\sim$ 470 \\\hline
Berisa \textit{et al.} \cite{BeZh22} & Heuristic & 46 & 222 & 602 -- 7090 \\\hline
Bhattacharjee \textit{et al.} \cite{BhAl22} & SA & 1536 & 1845 & $\sim 1800$ \\\hline
Falk \textit{et al.} \cite{FaDu20} & Heuristic & 400 & 400 & $\sim 6000$ -- $11000$ \\\hline
Gavrilu{\c{t}} \textit{et al.} \cite{GaZh18}  & GRASP & 402 & 427 & 534.6 \\\hline
\multirow{2}{*}{Gavrilu{\c{t}} \textit{et al.} \cite{GaZa17}}  & GRASP & 20 & 38 & $558$ \\\cline{2-5}
     & Heuristic & 20 & 38 & $130$ \\\hline
Gong \textit{et al.} \cite{GoYa23} & Tabu search & N/A & 180 & $\sim 11160$ \\\hline
He \textit{et al.} \cite{HeZu23} & Machine learning & 20 & 200 & 7 \\\hline
Kentis \textit{et al.} \cite{KeBe17}  & Heuristic & $\geq 13$ & 60 & N/A \\\hline
Li \textit{et al.} \cite{LiCh21} & Heuristic & 20 & 100 & N/A \\\hline
Li \textit{et al.} \cite{LiJi22} & Heuristic & 126 & 4000 & 0.437 -- 0.88 \\\hline
Li \textit{et al.} \cite{LiXi22} & ILP + heuristic & 44 & 300 & $\sim 0.3$ \\\hline
\multirow{3}{*}{Pahlevan \textit{et al.} \cite{PaOb18}} & List scheduler & 45 & 40 & $0.014$ \\\cline{2-5}
& Genetic algorithm & 45 & 40 & $56.75$ \\\hline
\multirow{2}{*}{Pahlevan \textit{et al.} \cite{PaTa19}}  & List scheduler & 50 & 100 & $0.103$ \\\cline{2-5}
 & Heuristic & 50 & 100 & $1.58$ \\\hline
Reusch \textit{et al.} \cite{ReCr22}  & SA + list scheduler & 192 & 144 & $1200$ \\\hline
Syed \textit{et al.} \cite{SyAy21b}  & Heuristic & N/A & 500 & $0.41$ \\\hline
Syed \textit{et al.} \cite{SyAy21}  & Heuristic & N/A & 500 & $0.170$ \\\hline
Syed \textit{et al.} \cite{SyAy22}  & Heuristic & N/A & 90 & $\sim 342$ \\\hline
Syed \textit{et al.} \cite{SyAy23}  & Heuristic & N/A & 90 & $\sim 8$ -- 10 \\\hline
Wang \textit{et al.} \cite{WaCh20}  & Heuristic & 30 & 100 & $72$ \\\hline
Yang \textit{et al.} \cite{YaGo22} & Tabu search & N/A & 180 & $\sim 11160$ \\\hline
Yang \textit{et al.} \cite{YaWe22} & Machine learning & 18 & 900 & N/A \\\hline
Yu \textit{et al.} \cite{YuWa20} & Heuristic & 96 & 350 & N/A \\\hline
Zhou \textit{et al.} \cite{ZhWa21} & Heuristic & N/A & 650 & $\sim 2700$ \\\hline
    \end{tabular}
    \caption{Overview of solving times of the respective largest reported problem instance for which a schedule was found. Only research works with heuristic approach and joint routing are included for comparability.}
    \label{tab:runtime-joint-routing-heuristic}
\end{table}
\vspace{3cm}

\vspace*{5cm} 
\section{Publication History}
\label{sec:publication}
We give an overview of the publication history of TSN scheduling.
First, we highlight seminal works from the literature.
Then, we analyze the development of the field with respect to the number of published papers per year.

\subsection{Seminal Works}
Early works about per-flow scheduling in Ethernet networks were presented by Steiner \cite{St10} and Schweissguth \textit{et al.} \cite{ScDa17}.
While these works are not specifically for TSN and abstract on the details of the real-time enhancement for Ethernet. they influenced many later works presented in this survey.
The first works specifically about scheduling in TSN were presented in 2016.
Dürr \textit{et al.} \cite{DuNa16} presented an ILP for no-wait scheduling and identified the problem of guard bands consuming bandwidth. 
Craciunas \textit{et al.} \cite{CrOl16} adapted the work of Steiner \cite{St10} for TSN.
They introduced isolation constraints and incremental scheduling to the domain of TSN.
Gavrilut \textit{et al.} \cite{GaZa17} is the first work which features joint routing and reliability considerations.
Raagard \textit{et al.} \cite{RaPo17} introduced reconfiguration of schedules to TSN scheduling.
Oliver \textit{et al.} \cite{OlCr18} proposed a scheduling approach with limits the number of used GCL entries by computing them in a joint approach with transmission offsets.
All earlier works computed GCLs by a post-processing after scheduling.

\subsection{Published Papers}
\fig{papers} shows the number of papers about TSN scheduling per year.
The first papers about scheduling in TSN were published in 2016. 
The general trend is that the field grows almost monotonously from one year to the next, with only one exception in 2019.
We observe a significant increase in published works since the year 2020.
Given the fast growth of the last 2 years, we expect even more research works about TSN scheduling in the future.
\figeps[0.9\columnwidth]{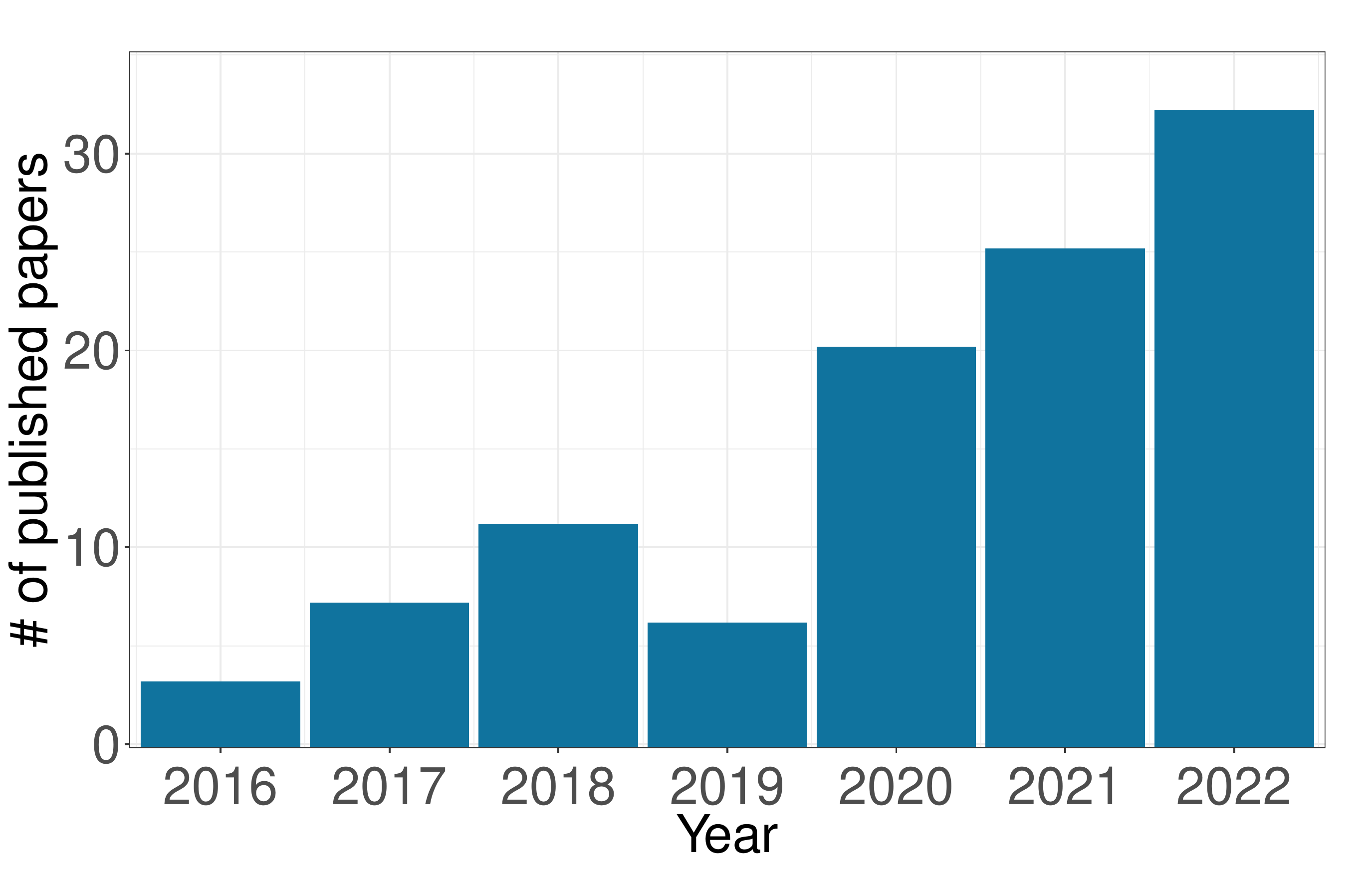}{Number of published research works per year about TSN scheduling. Only works published before June 2022 are counted.}

\section{Future Works}
\label{sec:discussion}
In this section, we discuss the results of the literature study.
First, we suggest improvements for future research works.
Then, we highlight open problems not handled sufficiently so far.

\subsection{Suggestions for Improvement}
The surveyed literature features many high quality research works.
However, there is room for improvement in the presentation of some of these works. 
We suggest improvements in the hope that the overall quality of the TSN scheduling literature can be improved even more in the future.
First, we discuss shortcomings and improvements in the presentation of evaluation methodologies.
Then, we suggest the use of the technical terminology used in Ethernet bridging.

\subsubsection{Evaluation Methodologies}
The scalability of the proposed solutions from the literature was extensively evaluated. 
Unfortunately, the impact of possible additional features and changes in parameters to the various objective functions was mostly ignored so far.
Although scalability is an important property of a scheduling algorithm, it would be interesting to see more evaluations regarding solution quality. 
Most test cases in the literature are synthetically constructed, both network topology and streams.
Even though it is hard to obtain test cases from the industry, let alone publish them, we would like to see more evaluations with realistic instances.
It is not clear whether the proposed algorithms are suitable for large industry-scale instances or how they look like.
It is also extremely difficult to compare the results of different research works as there is no public set of test cases for benchmarks.
Consequently, there is little research work available about which algorithm should be used in which setting.

Unfortunately, it is also hard to assess the significance of evaluation results in some papers for two reasons.
First, the instances solved are not sufficiently described. 
At least the topology and a description of assumed delays, e.g., processing and propagation delay, should be contained in the description of the network.
Important properties of streams like deadlines or periods are often missing.
Second, some evaluations report results for individual problem instances and are thus more of anecdotal character.
There are easy and hard instances for every algorithm.
Comparing multiple approaches on the same selected instances can be useful, but this may have the taste of picking specific instances in support of some conclusion. 
Instead of reporting results for individual instances, average results for multiple instances with the same evaluation setting should be reported.
Another property covered by many evaluations is the schedulability of the respective proposed approach.
These evaluations treat instances as infeasible when no schedule was found before some timeout. 
Thus, comparing the schedulability of two scheduling approaches which support different features is biased, as timeouts do not prove infeasibility.
This may lead to wrong conclusions in favor of some algorithm or model, although schedulability is actually equal.

\subsubsection{Terminology}
Many works surveyed in this paper use a vocabulary loosely related to Ethernet bridging. 
However, the standards and other relevant literature use a specific technical terminology.
We recommend that the scheduling community adopts this jargon.
Readers from adjacent research domains or who have prior knowledge in Ethernet bridging can benefit from a consistent vocabulary.
Therefore, the word \textit{stream} should be used instead of \textit{flow}.
Network devices which send or receive data streams are denoted as \textit{end stations}.
The source end station of a stream is denoted as \textit{talker}, while the destination end station is denoted as \textit{listener}.
Layer 2 switching devices are denoted as \textit{bridges} instead of \textit{switches}. 
Frames are the units of data transmission, as TSN is a layer 2 technology.
Although the meaning of the term \textit{packet} is clear in the context of scheduling, it is technically wrong.
Routing is the process of path computation on layer 3.
Therefore, the term \textit{path selection} is more appropriate in TSN.
However, we note that we used the term \textit{routing} in this survey several times. 
The reason for this is to ensure consistency with the reviewed literature which solves the so-called joint routing problem.

\subsection{Open Problems}
The available literature is comprehensive with regard to solution approaches to the unmodified scheduling problem in TSN.
However, there is still a wide field of relevant aspects which are not yet understood.

\subsubsection{Impact of Guard Bands and GCL Entries}
To the best of our knowledge, the impact of guard bands on bandwidth available to lower-priority traffic was not evaluated in the literature.
Likewise, the impact of available GCL entries on available bandwidth for lower-priority traffic is not investigated in detail.
The evaluations so far suggest that AVB streams benefit from schedules with many holes between TT streams with regard to tardiness.
However, such schedules may need more gate closings and thus guard bands, which reduces the available bandwidth.
It is not clear how AVB and BE traffic can be simultaneously integrated in a unified approach for the scheduling of TT streams.

\subsubsection{Routing and Multicast}
The joint routing and scheduling problem was explored in detail in the literature.
All research works about this topic agree that schedulability benefits from joint routing and scheduling.
However, solving the joint routing problem is significantly harder compared to scheduling with a given routing.
Unfortunately, there is currently no exact and scalable approach known for joint routing.
Additionally, it is not understood which properties a routing should have to benefit schedule synthesis and quality.
TSN supports multicast streams which are relevant in real use cases.
Some of the algorithms presented in research works covered by this survey can handle multicast streams.
However, the literature lacks evaluations and insights about the appropriate integration of multicast streams in a schedule.

\subsubsection{Online Reconfiguration}
There is also little work about online schedule reconfiguration, though it is important for operation.
In some scenarios, e.g., automotive networks, insertion and removal of streams at execution time of the schedule can be important.
So far it is not explored exhaustively what properties a schedule should have such that reconfiguration can by computed efficiently.
Additionally, there are no reconfiguration algorithms for most of the problem extensions from Section \ref{sec:problem_modifications}.

\subsubsection{Queuing and Handling of Non-Determinism}
An important open problem in TSN is sufficient integration of queuing.
Almost all research works use isolation constraints from \cite{CrOl16}, i.e., they do not allow frames of different streams to reside in the same queue at the same time.
However, this is not a requirement of the TAS. 
Some approaches even separate streams by assigning them to different egress queues during scheduling.
The rational of this is to reduce the impact of non-determinism like frame loss. 
Other attempts to reduce the influence of such causes of non-determinism are not yet explored.
The benefits of unrestricted queuing regarding schedulability or solution quality has not yet been evaluated.

Real hardware bridges are subject to non-determinism.
There is jitter in processing delays, and clocks are not exactly synchronized in reality. 
Additionally, frames that are scheduled to arrive approximately at the same time at two ingress ports of the same bridge may cause race conditions, i.e., processing order is not deterministic. 
All research works covered by this survey assume bridges are perfectly deterministic.
Thus, the literature lacks handling of such causes of non-determinism.

\subsubsection{Support of PSFP}
Per-Stream Filtering and Policing (PSFP) is a standard defined in IEEE 802.1Qci \cite{802.1Qci} for filtering and policing in TSN.
Currently, there are no devices available implementing PSFP.
However, filtering and policing could be used to prevent violations of schedules through unexpected packets.
Packets not scheduled, delayed frames, and frames larger than expected can be filtered at execution time of a schedule.
Thus, PSFP requires configuration of filtering entries that need to be derived from the schedule. A joint approach may be needed as PSFP imposes additional restrictions, e.g., the number of available filtering entries will be limited in bridges.

\subsubsection{Use of TSN Mechanisms}
TSN is not limited to scheduled traffic and the TAS.
Other traffic classes may have real-time requirements, but cannot be scheduled as the respective streams are not periodic.
Different traffic classes may have different sets of real-time requirements, e.g., demanding bounded jitter instead of bounded latency.
TSN features more mechanisms which may be applied to fulfill these requirements, such as Asynchronous Traffic Shaping \cite{802.1Qcr} or Cyclic Queuing and Forwarding \cite{802.1Qch}.
A major open problem in TSN is the coexistence of multiple mechanisms and the assignment of streams to them. Input may be a set of streams or traffic rates with their descriptors and real-time requirements, and output is their assignment to appropriate TSN mechanisms together with the complete network configuration. This problem goes far beyond the TSN scheduling problem, but may impose additional constraints on the latter.  
Some requirements can only be fulfilled by scheduling the respective streams and computing GCLs for the TAS.
Others may not even know the traffic streams in advance and can be implemented by the CBS or even simpler mechanisms. As even computing GCLs for the TAS is a challenging task for current state-of-the-art scheduling and optimization algorithms, such a comprehensive approach is currently unreachable. Hopefully, future works will move towards such long term goals and enable users to exploit the full potential of TSN.

\subsubsection{Understanding of the TSN Problem}
So far, scalability analyses have been conducted on special algorithms. However, they do not provide insights in what makes the TSN problem hard. This also pertains to all problem extensions like joint routing and multicast, reliability, robustness, BE or ABE traffic, etc. Moreover, properties of schedules such as tightness or average duration of open periods of the TAS have not yet been investigated. It would be helpful to understand the impact of problem extensions on the structure of schedules in an intuitive way.
A better understanding of extensions and their impact on schedule structure may facilitate the development of heuristic algorithms that solve larger instances of the TSN problem with acceptable quality compared to exact approaches.

\section{Conclusion}
\label{sec:conclusion}
TSN is a set of standards to enable real-time transmission over switched Ethernet networks.
IEEE 802.1Qbv \cite{802.1Qbv} defines traffic scheduling combined with the Time-Aware Shaper (TAS), i.e., transmissions of periodic high-priority streams are scheduled such that packets hardly interfere and that ultra-low latency is achieved. Moreover, the TAS protects scheduled traffic against traffic from other traffic classes. This approach requires the configuration of transmission times for streams at the Talkers (source nodes) as well as the configuration of the TAS on the switches.

In this paper, we first gave an introduction to TSN with focus on traffic scheduling and the TAS.
We defined the ``TSN scheduling problem'' and discussed common extensions such as scheduling with fixed or joint routing, various forms of queuing, support for reliability or lower-priority traffic, or respecting technical restrictions. Some of these extensions lead to optimization problems. We summarized frequently used scheduling and optimization methods to tackle these challenges.
Then we reviewed a large body of literature about the TSN scheduling problem and classified it regarding the mentioned extensions. Subsequently, we analyzed and compared the works with respect to modelling assumptions, scheduling objectives, problem instances, and scalability, and pointed out advances. We tracked seminal works and identified popular publication venues for TSN scheduling. We discussed the area by suggesting improvements and pointing out open problems.

This survey serves researchers to identify the current state of the art and open problems in TSN scheduling.
The many problem extensions suggest that the construction of an efficient scheduling or optimization algorithm which considers all relevant aspects is infeasible. We expect future work to provide a better understanding of the complexity of the TSN scheduling problem to cover more problem extensions while maintaining scalability.

\vspace{3cm}

\subsection*{Acknowledgments}
\noindent The authors thank Manuel Eppler and Lukas Bechtel for valuable input and stimulating discussions.
This work has been supported by the German Federal Ministry of Education and  Research  (BMBF)  under  support  code  16KIS1161 (Collaborative Project KITOS). 
The authors alone are responsible for the content of the paper.

\AtNextBibliography{\small}
\printbibliography

\begin{IEEEbiography}
    [{\includegraphics[width=1in,height=1.25in,clip,keepaspectratio] 
        {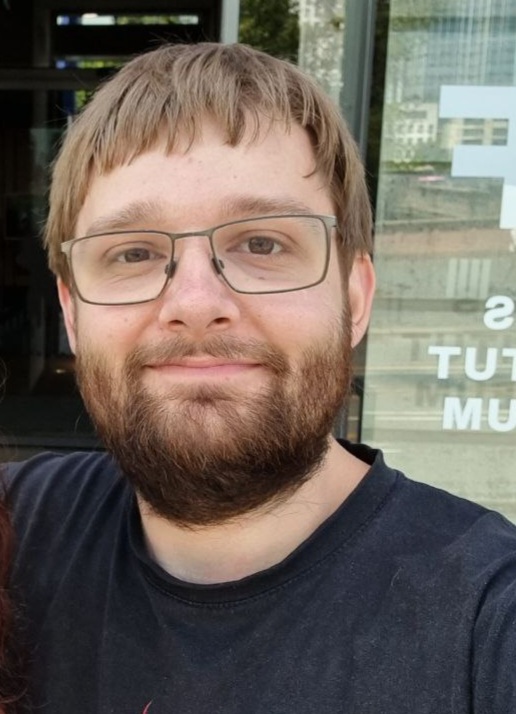}}]{Thomas Stüber}
   is a Ph.D. student at the chair of communication networks of Prof. Dr. habil. Michael Menth at the Eberhard Karls University Tuebingen, Germany. He obtained his master's degree in 2018 and afterwards, became part of the communication networks research group. His research interests include Time-Sensitive Networking (TSN), scheduling, performance evaluation, and operations research. 
\end{IEEEbiography}

\begin{IEEEbiography}
    [{\includegraphics[width=1in,height=1.25in,clip,keepaspectratio] 
        {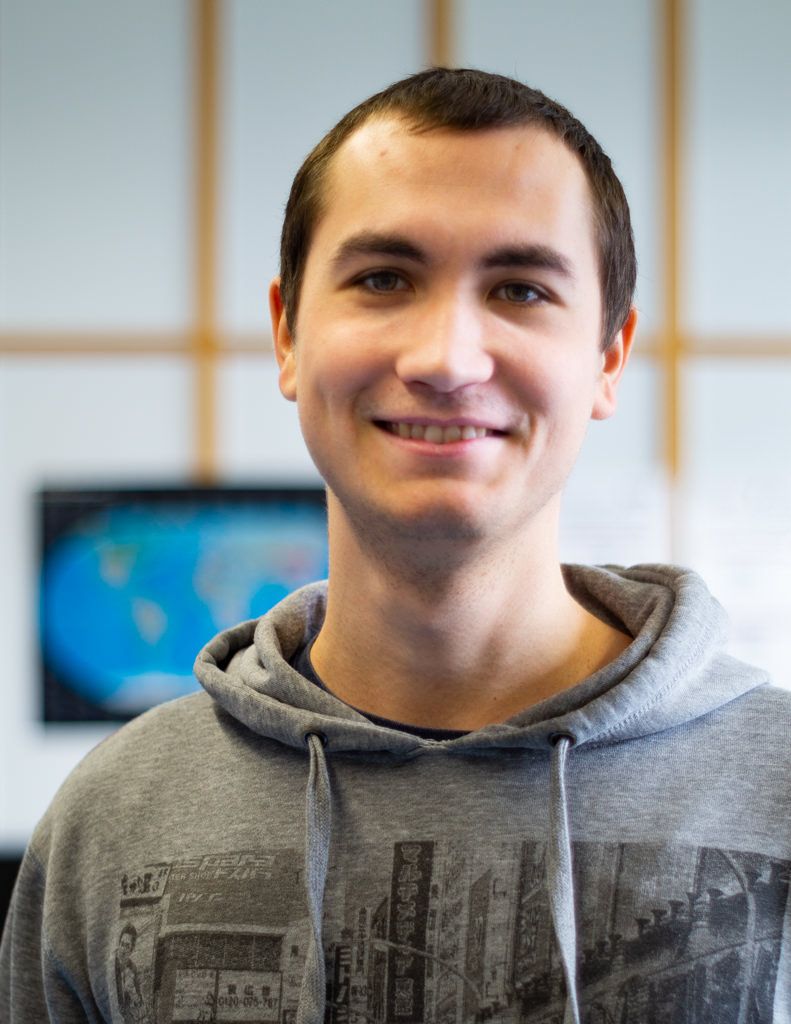}}]{Lukas Osswald}
    is a Ph.D. student at the chair of communication networks of Prof. Dr. habil. Michael Menth at the Eberhard Karls University Tuebingen, Germany. He obtained his master's degree in 2020 and afterwards, became part of the communication networks research group. His research interests include Time-Sensitive Networking (TSN), admission control and network configuration. 
\end{IEEEbiography}

\begin{IEEEbiography}
    [{\includegraphics[width=1in,height=1.25in,clip,keepaspectratio] 
        {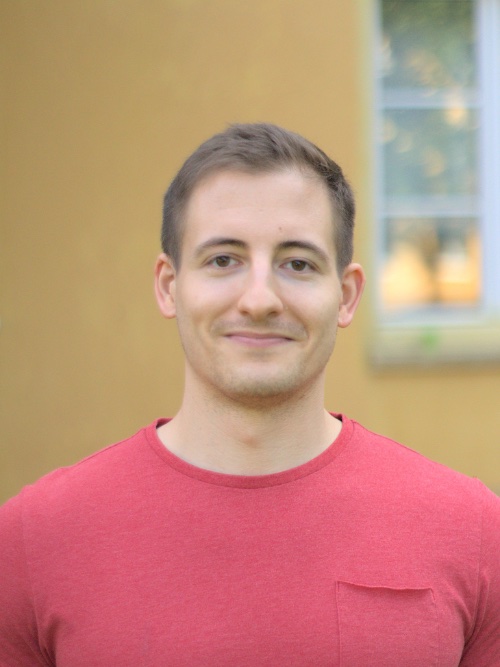}}]{Steffen Lindner}
    is a Ph.D. student at the chair of communication networks of Prof. Dr. habil. Michael Menth at the Eberhard Karls University Tuebingen, Germany. 
    He obtained his master's degree in 2019 and afterwards, became part of the communication networks research group. 
    His research interests include software-defined networking, P4 and congestion management. 
\end{IEEEbiography}

\begin{IEEEbiography}
    [{\includegraphics[width=1in,height=1.25in,clip,keepaspectratio] 
        {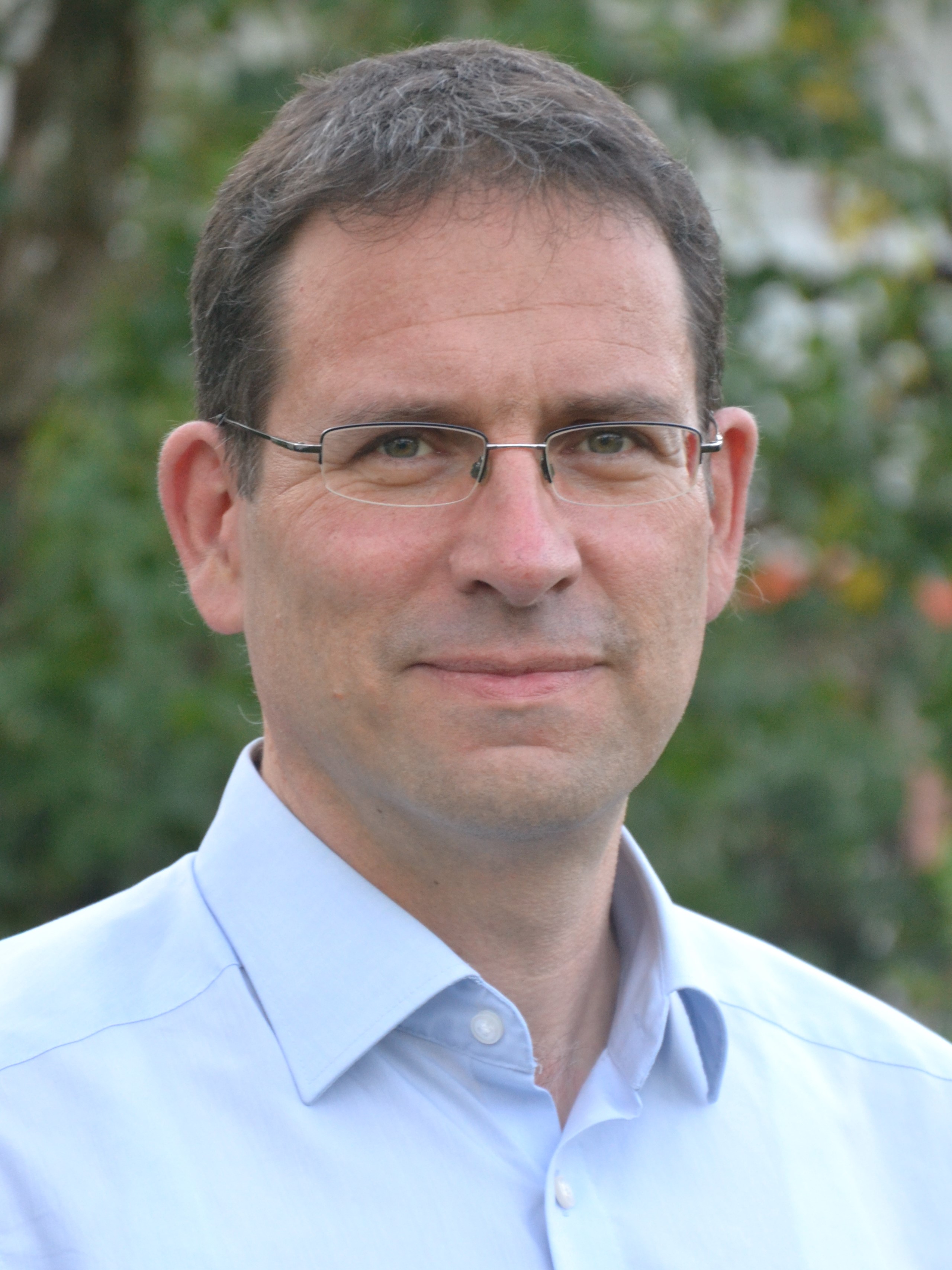}}]{Michael Menth} (Senior Member, IEEE) is professor at the Department
of Computer Science at the University of Tuebingen/Germany and
chairholder of Communication Networks since 2010. He studied,
worked, and obtained diploma (1998), PhD (2004), and habilitation
(2010) degrees at the universities of Austin/Texas, Ulm/Germany,
and Wuerzburg/Germany. His special interests are performance
analysis and optimization of communication networks, resilience and
routing issues, as well as resource and congestion management. His
recent research focus is on network softwarization, in particular
P4-based data plane programming, Time-Sensitive Networking (TSN),
Internet of Things, and Internet protocols. Dr. Menth contributes
to standardization bodies, mainly to the IETF.
\end{IEEEbiography}

\vspace{\fill}

\end{document}